\documentclass[11pt,a4paper]{article}

\usepackage{amssymb}
\usepackage[dvips]{graphicx}
\usepackage{bm}
\usepackage{transparent}

\begin{document}

\title{Three-loop verification of a new algorithm for the calculation of a $\beta$-function in supersymmetric theories regularized by higher derivatives for the case of ${\cal N}=1$ SQED}

\author{
S.S.Aleshin\,${}^a$,
I.S.Durandina\,${}^b$, D.S.Kolupaev\,${}^b$, D.S.Korneev\,${}^c$, M.D.Kuzmichev\,${}^b$,\\ N.P.Meshcheriakov\,${}^b$, S.V.Novgorodtsev\,${}^b$, I.A.Petrov${}^b$, V.V.Shatalova\,${}^b$, I.E.Shirokov\,${}^b$,\\ V.Yu.Shirokova\,${}^b$, K.V.Stepanyantz\,${}^b$
\medskip\\
${}^a${\small{\em Institute for Information Transmission Problems,}}\\
{\small{\em 127051, Moscow, Russia,}}
\medskip
\\
${}^b${\small{\em Moscow State University, Faculty of Physics, Department of Theoretical Physics,}}\\
{\small{\em 119991, Moscow, Russia,}}
\medskip
\\
${}^c${\small{\em RWTH Aachen University, Faculty of Mathematics, Computer Science and Natural Science,}}\\
{\small{\em Institute for Theoretical Particle Physics and Cosmology, 52062, Aachen, Germany}}
}

\maketitle

\begin{abstract}
We verify a recently proposed method for obtaining a $\beta$-function of ${\cal N}=1$ supersymmetric gauge theories regularized by higher derivatives by an explicit calculation. According to this method, a $\beta$-function can be found by calculating specially modified vacuum supergraphs instead of a much larger number of the two-point superdiagrams. The result is produced in the form of a certain integral of double total derivatives with respect to the loop momenta. Here we compare the results obtained for the three-loop $\beta$-function of ${\cal N}=1$ SQED in the general $\xi$-gauge with the help of this method and with the help of the standard calculation. Their coincidence confirms the correctness of the new method and the general argumentation used for its derivation. Also we verify that in the considered approximation the NSVZ relation is valid for the renormalization group functions defined in terms of the bare coupling constant and for the ones defined in terms of the renormalized coupling constant in the HD+MSL scheme, both its sides being gauge-independent.
\end{abstract}

\unitlength=1cm

\section{Introduction}
\hspace*{\parindent}

Investigations of higher order quantum corrections in supersymmetric theories sometimes reveal interesting features of their structure. For example, after the three-loop calculation of Ref. \cite{Avdeev:1980bh} it was natural to suggest that the ${\cal N}=4$ supersymmetric Yang--Mills (SYM) theory is finite in all orders. This statement has soon been proved in Refs. \cite{Grisaru:1982zh,Mandelstam:1982cb,Brink:1982pd,Howe:1983sr}. The three- and four-loop calculations of Refs. \cite{Jack:1996vg,Jack:1996cn,Jack:1998uj,Harlander:2006xq} (see also the review \cite{Mihaila:2013wma}) confirmed existence of a renormalization prescription in which the $\beta$-function of ${\cal N}=1$ supersymmetric theories is related to the anomalous dimension of the matter superfields by the NSVZ equation proposed in \cite{Novikov:1983uc,Jones:1983ip,Novikov:1985rd,Shifman:1986zi} on the base of some general arguments. The calculations of Refs. \cite{Parkes:1985hh,Jack:1996qq} demonstrated existence of a subtraction scheme in which one-loop finite ${\cal N}=1$ supersymmetric theories are also finite in the two- and three-loop approximations. From the other side, the calculation of Ref. \cite{Jack:1996qq} did not confirm the existence of an exact equation for the anomalous dimension of the matter superfields in a special class of ${\cal N}=1$ supersymmetric theories. Calculations made with the higher derivative regularization \cite{Slavnov:1971aw,Slavnov:1972sq} formulated in terms of ${\cal N}=1$ superfields \cite{Krivoshchekov:1978xg,West:1985jx} for the ${\cal N}=1$ supersymmetric electrodynamics (SQED) \cite{Soloshenko:2003nc,Smilga:2004zr} revealed that the integrals giving the $\beta$-function are integrals of total and double total derivatives. Subsequently, it becomes clear that this structure of the loop integrals is a general feature of ${\cal N}=1$ supersymmetric gauge theories, regularized by higher covariant derivatives, \cite{Pimenov:2009hv,Stepanyantz:2011cpt,Stepanyantz:2011bz,Stepanyantz:2012zz,Kazantsev:2014yna,Aleshin:2016yvj,Shakhmanov:2017soc,Kazantsev:2018nbl}. The all-loop proof of this fact has been done in Refs. \cite{Stepanyantz:2011jy,Stepanyantz:2014ima} for the Abelian case and in Ref. \cite{Stepanyantz:2019ihw} for a general ${\cal N}=1$ non-Abelian supersymmetric gauge theory. Such a structure of loop integrals leads to the NSVZ relation between the $\beta$-function and the anomalous dimension of the matter superfields \cite{Novikov:1983uc,Jones:1983ip,Novikov:1985rd,Shifman:1986zi}. For ${\cal N}=1$ SQED with $N_f$ flavors, which will be considered in this paper, the exact NSVZ $\beta$-function takes the form \cite{Vainshtein:1986ja,Shifman:1985fi}

\begin{equation}\label{NSVZ_Bare}
\frac{\beta(\alpha_0)}{\alpha_0^2} = \frac{N_f}{\pi}\Big(1-\gamma(\alpha_0)\Big).
\end{equation}

\noindent
Note that Eq. (\ref{NSVZ_Bare}) is written for the renormalization group functions (RGFs) defined in terms of the bare coupling constant $\alpha_0 = e_0^2/4\pi$. According to \cite{Stepanyantz:2011jy,Stepanyantz:2014ima} the relation (\ref{NSVZ_Bare}) is valid for these functions in the case of using the higher derivative regularization independently of a renormalization prescription.\footnote{Note that with dimensional reduction the integrals for the $\beta$-function are not integrals of total derivatives \cite{Aleshin:2015qqc}. Moreover, the results of Ref. \cite{Aleshin:2016rrr} indicate that RGFs defined in terms of the bare couplings do not satisfy the NSVZ equation for theories regularized by dimensional reduction.} RGFs defined in terms of the renormalized couplings satisfy the NSVZ equation only in a certain class of subtraction schemes, which in the Abelian case has been described in \cite{Goriachuk:2018cac}. This class includes the HD+MSL\footnote{In this scheme the theory is regularized by higher derivatives, and divergences are removed with the help of minimal subtractions of logarithms \cite{Shakhmanov:2017wji,Stepanyantz:2017sqg}. Also the HD+MSL scheme can be constructed by imposing certain boundary conditions on renormalization constants \cite{Kataev:2013eta}.} and on-shell schemes, see Refs. \cite{Kataev:2013eta,Kataev:2013csa} and \cite{Kataev:2019olb}, respectively. However, the $\overline{\mbox{DR}}$ scheme does not belong to it \cite{Jack:1996vg,Jack:1996cn,Jack:1998uj}, so does the MOM scheme \cite{Kataev:2014gxa}.

The factorization into integrals of double total derivatives with the higher derivative regularization also gives the NSVZ-like relations for the renormalization of the photino mass \cite{Hisano:1997ua,Jack:1997pa,Avdeev:1997vx} in softly broken ${\cal N}=1$ SQED \cite{Nartsev:2016nym} and for the $D$-function \cite{Adler:1974gd} in ${\cal N}=1$ SQCD \cite{Shifman:2014cya,Shifman:2015doa}. In both cases the HD+MSL scheme is NSVZ due to the same reasons as for rigid ${\cal N}=1$ SQED \cite{Nartsev:2016mvn,Kataev:2017qvk}.

In the non-Abelian case there are strong indications that the factorization of the loop integrals for the $\beta$-function into integrals of double total derivatives leads to the NSVZ equation for RGFs defined in terms of the bare couplings \cite{Stepanyantz:2016gtk,Stepanyantz:2019lfm}. If this is true, then one of the NSVZ schemes is given by the HD+MSL prescription. However, the all-loop proof of this fact has not yet been finished. Nevertheless, arguments used for proving the factorization into double total derivatives allowed to construct a method of calculating the $\beta$-function for ${\cal N}~=~1$ supersymmetric gauge theories regularized by higher covariant derivatives which essentially simplifies the calculations \cite{Stepanyantz:2019ihw}.\footnote{In the Abelian case considered in this paper it is very similar to the one proposed in Ref. \cite{Smilga:2004zr}.} Although this method allowed making some rather complicated calculations \cite{Kuzmichev:2019ywn,Stepanyantz:2019lyo}, it is desirable to demonstrate that the results obtained with the help of it coincide with the ones found by the standard technique. This is made in this paper for ${\cal N}=1$ SQED, for which we calculate the three-loop $\beta$-function in the general $\xi$-gauge by the method of Ref. \cite{Stepanyantz:2019ihw} and compare the result with the one obtained by the standard calculation.

The paper is organized as follows. In Sect. \ref{Section_N1_SQED} we describe the regularization of ${\cal N}=1$ SQED by higher derivatives and introduce the notations. Next, in Sect. \ref{Section_Renoralization_Of_N1_SQED} we briefly describe the renormalization of this theory and various definitions of RGFs. In Sect. \ref{Section_Two-Loop_Standard} we recall how the $\beta$-function is calculated by standard methods in the two-loop approximation. Here we also describe how the gauge dependent terms cancel each other in the perturbation theory. In Sect.~\ref{Section_Two-Loop_Beta_New} the two-loop result is compared with the one obtained in Ref. \cite{Stepanyantz:2019lyo}. The three-loop $\beta$-function is calculated in Sect. \ref{Section_Three-Loop} using the technique of Ref. \cite{Stepanyantz:2019ihw} in the general $\xi$-gauge. We demonstrate the coincidence of the result with the one obtained earlier by the standard calculation in the Feynman gauge. Also we verify the cancellation of the gauge dependence in both sides of the NSVZ equation (\ref{NSVZ_Bare}).

\section{${\cal N}=1$ SQED regularized by higher derivatives}
\hspace{\parindent}\label{Section_N1_SQED}

It is convenient to write the action of ${\cal N}=1$ SQED in terms of superfields (see, e.g., \cite{Gates:1983nr,West:1990tg,Buchbinder:1998qv}), because in this case ${\cal N}=1$ supersymmetry becomes a manifest symmetry. In this formulation the theory contains two chiral matter superfields $\phi$ and $\widetilde\phi$ interacting with a real gauge superfield $V$ and having opposite charges with respect to the gauge group $U(1)$. We will consider a generalization of this model which contains $N_f$ flavors (numerated by the index $\alpha$). In the massless limit the action for such a theory takes the form

\begin{equation}
S = \frac{1}{4 e_0^2} \mbox{Re} \int d^4x\, d^2\theta\ W^a W_a + \frac{1}{4}\sum\limits_{\alpha=1}^{N_f}\int d^4x\, d^4\theta\, \Big(\phi_\alpha^* e^{2V} \phi_\alpha + \widetilde\phi_\alpha^* e^{-2V} \widetilde\phi_\alpha\Big),
\end{equation}

\noindent
where the chiral superfield $W_a = \bar D^2 D_a V/4$ is a supersymmetric analog of the gauge field strength. The considered model is invariant under the gauge transformations

\begin{equation}\label{SQED_Gauge_Invariance}
V \to V + \frac{i}{2}\big(A^* - A\big);\qquad \phi_\alpha \to e^{iA} \phi_\alpha; \qquad \widetilde\phi_\alpha \to e^{-iA}\widetilde\phi_\alpha
\end{equation}

\noindent
parameterized by a chiral superfield $A$.

To regularize the theory, one can add a higher derivative term $S_\Lambda$ to its action. Then the regularized action $S_{\mbox{\scriptsize reg}} = S + S_\Lambda$ can be written as

\begin{equation}
S_{\mbox{\scriptsize reg}} = \frac{1}{4 e_0^2} \mbox{Re} \int d^4x\, d^2\theta\ W^a R\big(\partial^2/\Lambda^2\big) W_a + \frac{1}{4} \sum\limits_{\alpha=1}^{N_f}\int d^4x\, d^4\theta\, \Big(\phi_\alpha^* e^{2V} \phi_\alpha + \widetilde\phi_\alpha^* e^{-2V} \widetilde\phi_\alpha\Big),
\end{equation}

\noindent
where $R(0)=1$ and $R(x) \to \infty$ at $x\to \infty$. For example, it is possible to choose $R(x)=1+x^n$, where $n$ is a positive integer. A higher derivative regulator $K(x)$ with analogous asymptotics can be inserted into the gauge fixing term, which takes the form

\begin{equation}
S_{\mbox{\scriptsize gf}} = - \frac{1}{32 e_0^2 \xi_0} \int d^4x\, d^4\theta\, D^2 V K\big(\partial^2/\Lambda^2\big) \bar D^2 V,
\end{equation}

\noindent
where $\xi_0$ is the gauge fixing parameter. The Faddeev--Popov ghosts in the Abelian case are not needed.

Calculating the degree of divergence, one can see that the presence of the higher derivative functions $R$ and $K$ removes divergences beyond the one-loop approximation. However, it is well known \cite{Faddeev:1980be} that the one-loop divergences cannot be regularized in this way. For this purpose it is necessary to insert into the generating functional the Pauli--Villars determinants \cite{Slavnov:1977zf}. For the considered theory one can use two (commuting) chiral superfields $\Phi$ and $\widetilde\Phi$ with the action

\begin{equation}\label{Pauli-Villars_Action}
S_\Phi = \frac{1}{4}\int d^4x\, d^4\theta\, \Big(\Phi^* e^{2V} \Phi + \widetilde\Phi^* e^{-2V} \widetilde\Phi\Big) + \Big(\frac{1}{2} M \int d^4x\, d^2\theta\, \widetilde\Phi\, \Phi+\mbox{c.c.}\Big),
\end{equation}

\noindent
where $M=a\Lambda$ and $a$ is a constant which does not depend on $\alpha_0$. The corresponding Pauli--Villars determinant defined as

\begin{equation}
\mbox{Det}(PV,M)^{-1} = \int D\Phi D\widetilde\Phi\, \exp(iS_\Phi)
\end{equation}

\noindent
is a functional of the gauge superfield $V$. For constructing the regularized theory, it should be inserted into the generating functional,

\begin{equation}
Z = \exp(iW) = \int DV D\phi D\widetilde\phi\, \mbox{Det}(PV,M)^{N_f} \exp\Big(iS_{\mbox{\scriptsize reg}} + i S_{\mbox{\scriptsize gf}} + i S_{\mbox{\scriptsize sources}}\Big),
\end{equation}

\noindent
where

\begin{equation}
S_{\mbox{\scriptsize sources}} = \int d^4x\, d^4\theta\, J V + \Big(\sum\limits_{\alpha=1}^{N_f}\int d^4x\, d^2\theta\, \big(j_\alpha \phi_\alpha + \widetilde j_\alpha \widetilde \phi_\alpha\big) +\mbox{c.c.}\Big).
\end{equation}

\noindent
The effective action $\Gamma$ is defined in the standard way as the Legendre transform of the generating functional for the connected Green functions $W$.

\section{Renormalization and RGFs}
\hspace*{\parindent}\label{Section_Renoralization_Of_N1_SQED}

Using the Ward--Slavnov--Taylor identity corresponding to the gauge invariance (\ref{SQED_Gauge_Invariance}) the quadratic part of the effective action for the considered theory can be written in the form

\begin{eqnarray}\label{Two-Point_Functions}
&& \Gamma^{(2)} - S_{\mbox{\scriptsize gf}} = - \frac{1}{16\pi} \int \frac{d^4p}{(2\pi)^4}\, d^4\theta\, V(-p,\theta) \partial^2 \Pi_{1/2} V(p,\theta)\, d^{-1}(\alpha_0, \Lambda/p)\qquad\nonumber\\
&& + \frac{1}{4}\sum\limits_{\alpha=1}^{N_f} \int \frac{d^4q}{(2\pi)^4}\, d^4\theta\, \Big( \phi_\alpha^{*}(-q,\theta)\, \phi_\alpha(q,\theta) + \widetilde\phi_\alpha^{*}(-q,\theta)\, \widetilde\phi_\alpha(q,\theta)\Big)\, G(\alpha_0,\Lambda/q),
\end{eqnarray}

\noindent
where the supersymmetric transversal projection operator is defined as $\partial^2\Pi_{1/2} \equiv - D^a \bar D^2 D_a/8$. Expressions for the functions $d^{-1}$ and $G$ contain ultraviolet divergences. To get rid of them, it is necessary to renormalize the coupling constant, the matter superfields, and the gauge parameter. For this purpose we define the renormalized coupling constant $\alpha(\alpha_0,\Lambda/\mu)$ and the renormalization constant for the matter superfields $Z(\alpha,\Lambda/\mu)$, where $\mu$ is a renormalization point, in such a way that the functions

\begin{equation}
d^{-1}(\alpha_0(\alpha,\Lambda/\mu),\Lambda/p)\qquad \mbox{and}\qquad Z(\alpha,\Lambda/\mu)\, G(\alpha_0(\alpha,\Lambda/q),\Lambda/q)
\end{equation}

\noindent
are finite in the limit $\Lambda\to \infty$. Taking into account that the longitudinal part of the two-point Green function of the gauge superfield does not receive quantum corrections, the gauge parameter is renormalized as $\xi = \xi_0/Z_\alpha$, where $Z_\alpha \equiv \alpha/\alpha_0$.

The $\beta$-function and the anomalous dimension of the matter superfields are standardly defined as

\begin{equation}
\widetilde\beta(\alpha) = \frac{d\alpha}{d\ln\mu}\Big|_{\alpha_0=\mbox{\scriptsize const}};\qquad \widetilde\gamma(\alpha) = \frac{d\ln Z}{d\ln\mu}\Big|_{\alpha_0=\mbox{\scriptsize const}}.
\end{equation}

\noindent
According to \cite{Kataev:2013eta} these RGFs should be distinguished from the ones defined in terms of the bare coupling constant by the prescription

\begin{equation}
\beta(\alpha_0) = \frac{d\alpha_0}{d\ln\Lambda}\Big|_{\alpha=\mbox{\scriptsize const}};\qquad \gamma(\alpha_0) = -\frac{d\ln Z}{d\ln\Lambda}\Big|_{\alpha=\mbox{\scriptsize const}}.
\end{equation}

\noindent
Up to the renaming of arguments, both these definitions of RGFs give the same functions in the HD+MSL scheme,

\begin{equation}
\widetilde\beta(\alpha)\Big|_{\mbox{\scriptsize HD+MSL}} = \beta(\alpha_0 \to \alpha);\qquad \widetilde\gamma(\alpha)\Big|_{\mbox{\scriptsize HD+MSL}} = \gamma(\alpha_0 \to \alpha).
\end{equation}

\noindent
(By definition, in the HD+MSL scheme the theory is regularized by higher derivatives and divergences are removed by minimal subtractions of logarithms, when renormalization constants include only powers of $\ln\Lambda/\mu$ and all finite constants are set to 0.)

For calculating RGFs defined in terms of the bare coupling constant one can use the equations relating them to the functions $d^{-1}-\alpha_0^{-1}$ and $G$ defined by Eq. (\ref{Two-Point_Functions}),

\begin{equation}\label{RGFs_Bare_From_Green_Functions}
\frac{\beta(\alpha_0)}{\alpha_0^2} = \frac{d}{d\ln\Lambda}\Big(d^{-1} - \alpha_0^{-1}\Big)\bigg|_{\alpha=\mbox{\scriptsize const}, p = 0};\qquad \gamma(\alpha_0) = \frac{d\ln G}{d\ln \Lambda}\bigg|_{\alpha=\mbox{\scriptsize const}, q = 0}.
\end{equation}

\noindent
These equalities follow from the finiteness of the functions $d^{-1}$ and $ZG$ expressed in terms of the renormalized coupling constant. The conditions $p=0$ and $q=0$ are needed for removing terms proportional to powers of $p/\Lambda$ or $q/\Lambda$.

\section{Two-loop $\beta$-function: the standard calculation}
\hspace{\parindent}\label{Section_Two-Loop_Standard}

Usually, to calculate a $\beta$-function, one should consider superdiagrams with two external lines of the gauge superfield $V$. (Note that in the Abelian case there is no need to use the background superfield method.) Calculating these superdiagrams we find the function $d^{-1}-\alpha_0^{-1}$, which encodes quantum corrections to the coupling constant. Then the $\beta$-function defined in terms of the bare coupling constant can be found using the first equation in (\ref{RGFs_Bare_From_Green_Functions}).

\begin{figure}[h]
\begin{picture}(0,2.3)
\put(4.5,0.4){\includegraphics[scale=0.152]{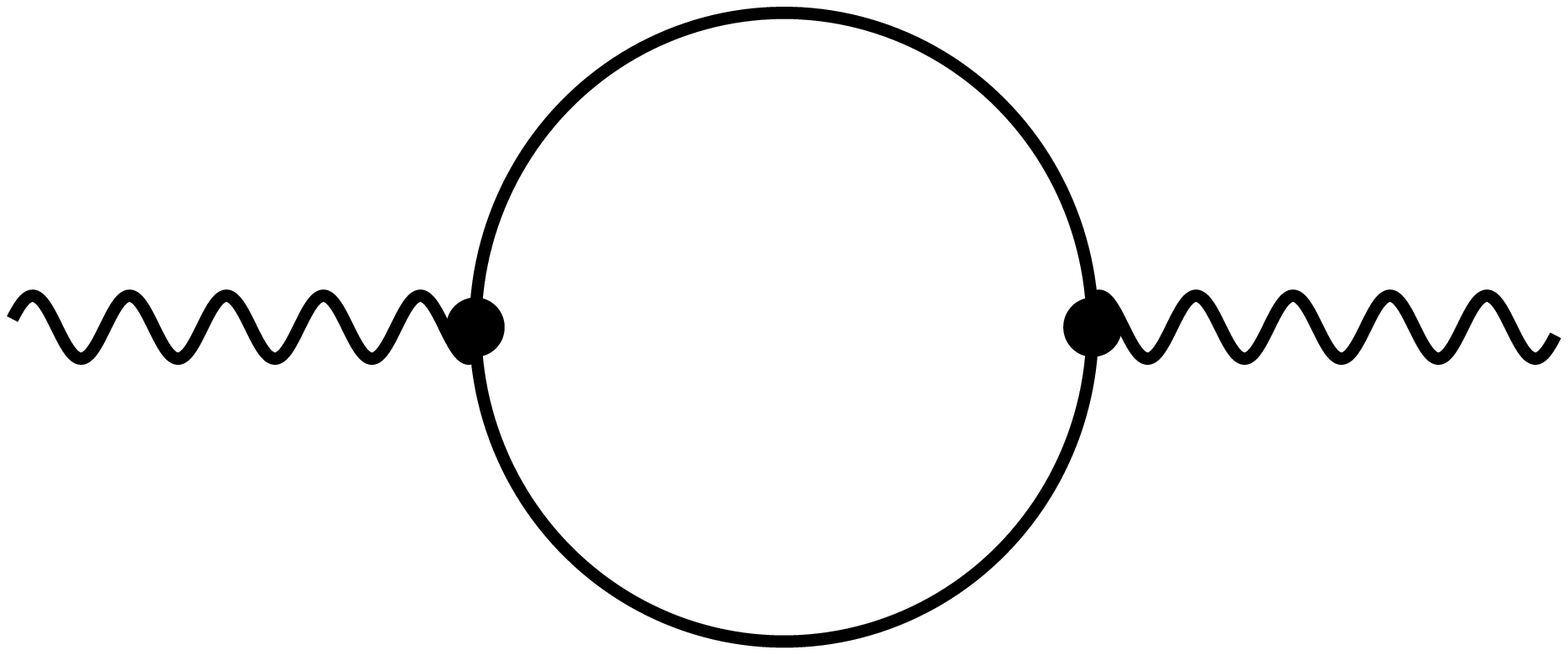}}
\put(4.5,1.7){B1}
\put(8.5,0.1){\includegraphics[scale=0.152]{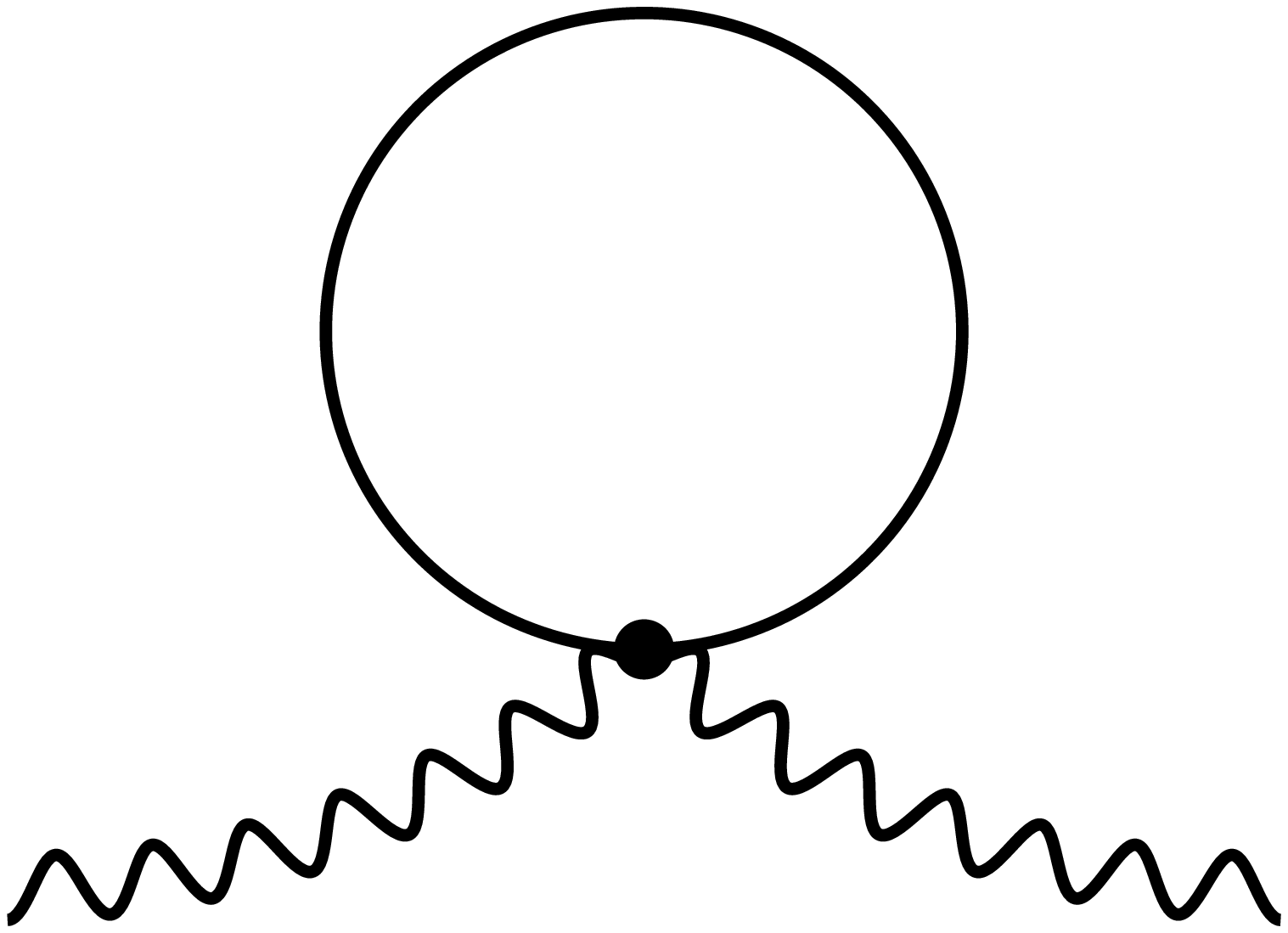}}
\put(8.5,1.7){B2}
\end{picture}
\caption{One-loop superdiagrams contributing to the $\beta$-function of ${\cal N}=1$ SQED.}\label{Figure_One_Loop}
\end{figure}

\begin{figure}[h]
\begin{picture}(0,4.8)
\put(0,2.9){\includegraphics[scale=0.152]{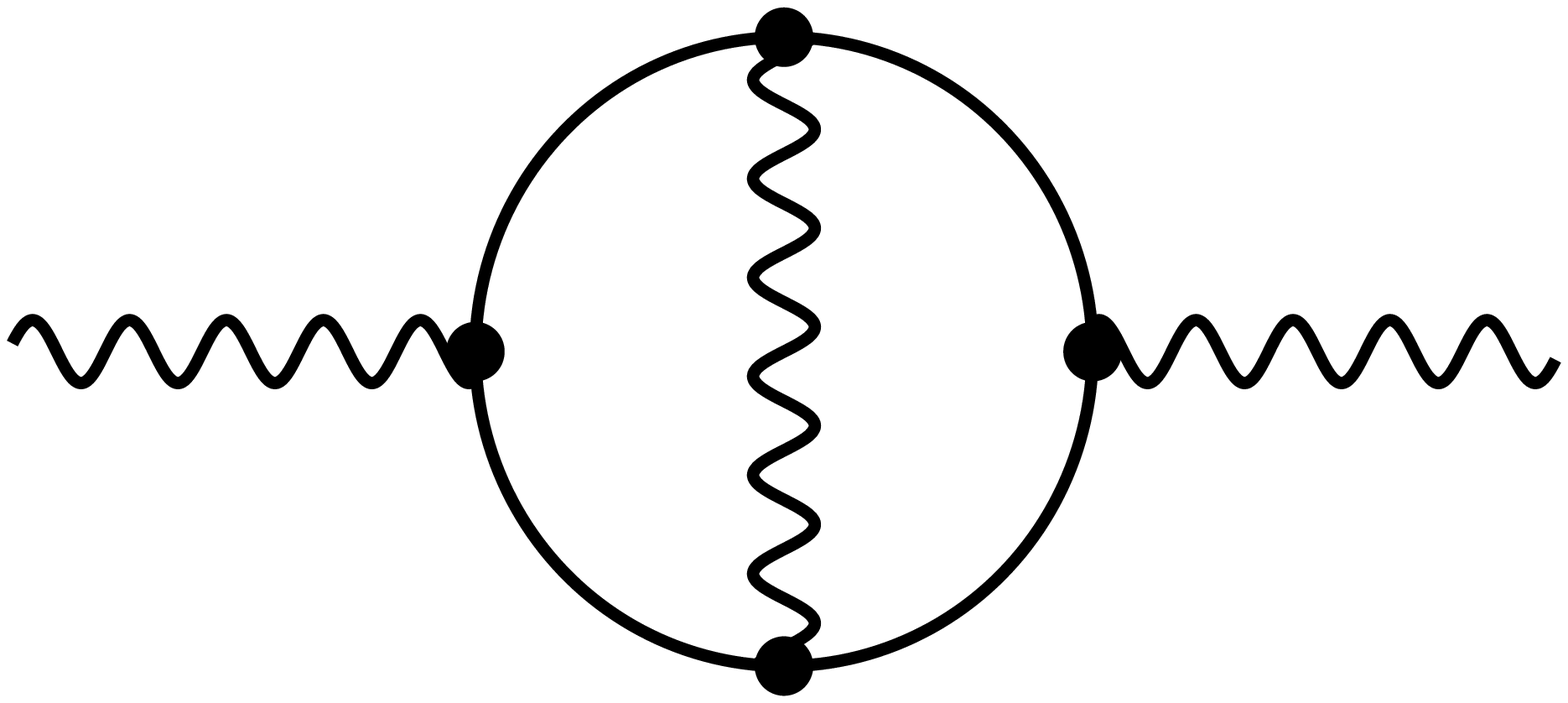}}
\put(0,4.2){B3}
\put(3.25,2.9){\includegraphics[scale=0.152]{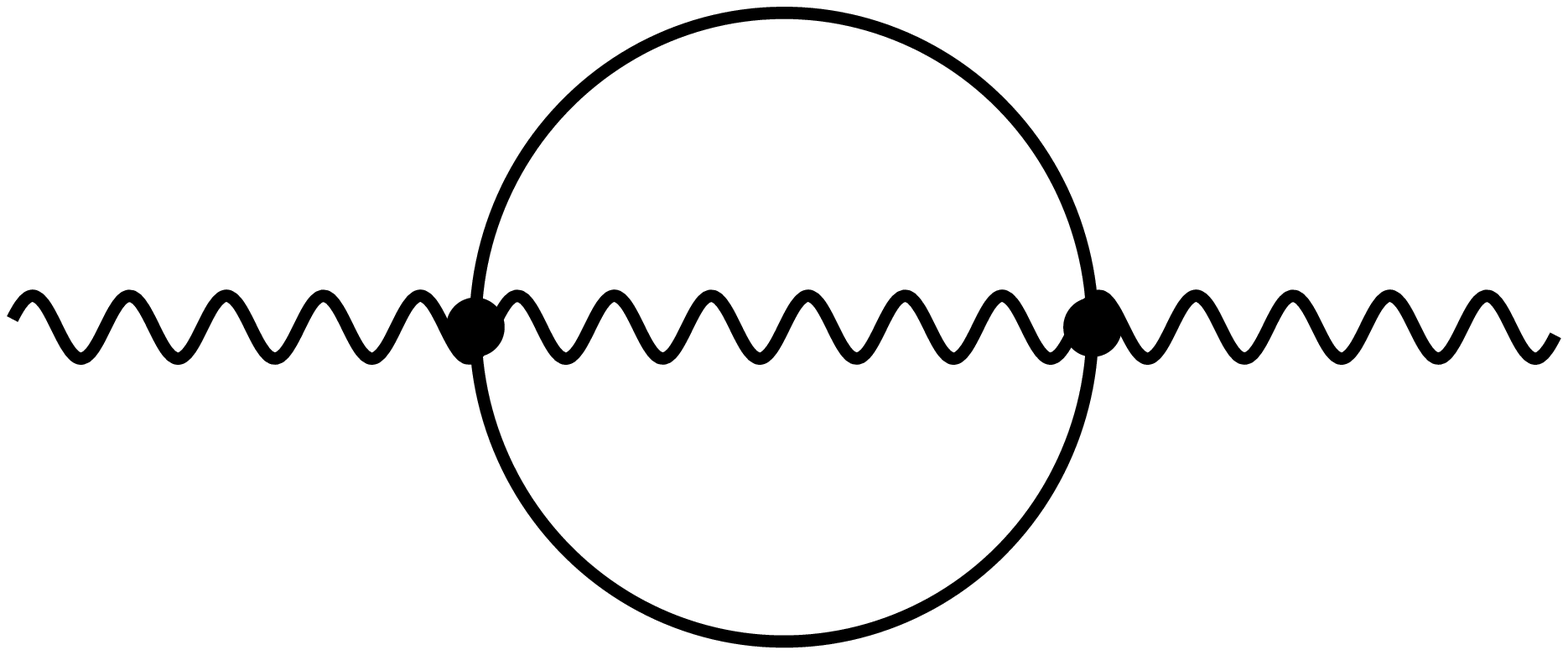}}
\put(3.25,4.2){B4}
\put(6.5,2.9){\includegraphics[scale=0.152]{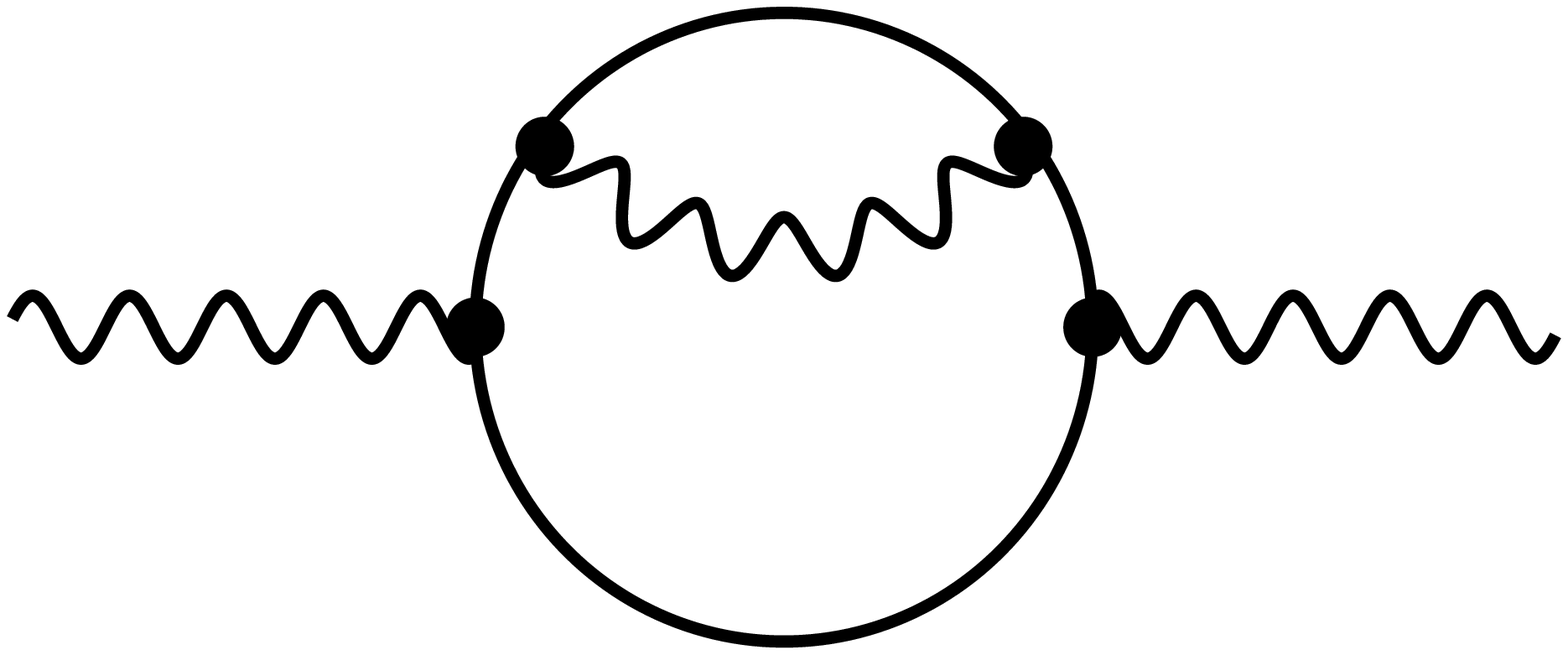}}
\put(6.5,4.2){B5}
\put(9.75,2.9){\includegraphics[scale=0.152]{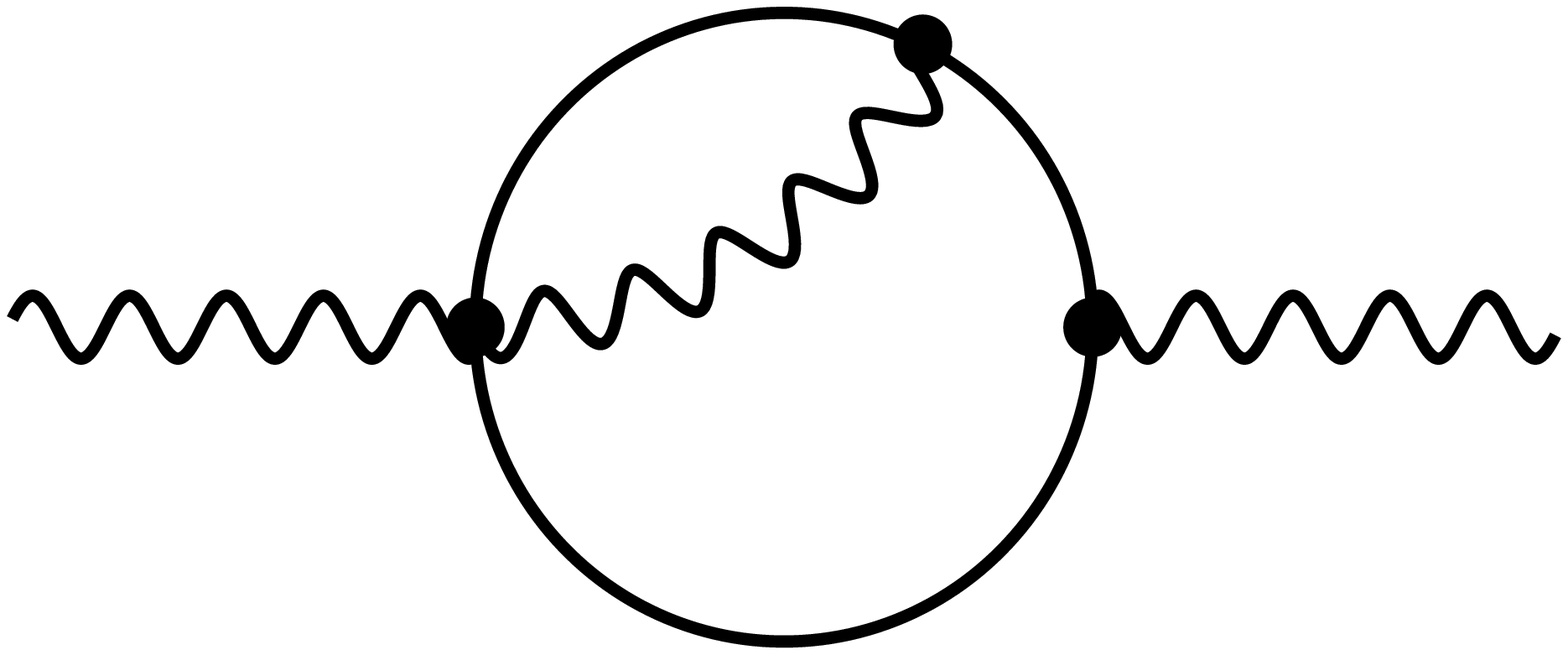}}
\put(9.75,4.2){B6}
\put(13,2.9){\includegraphics[scale=0.152]{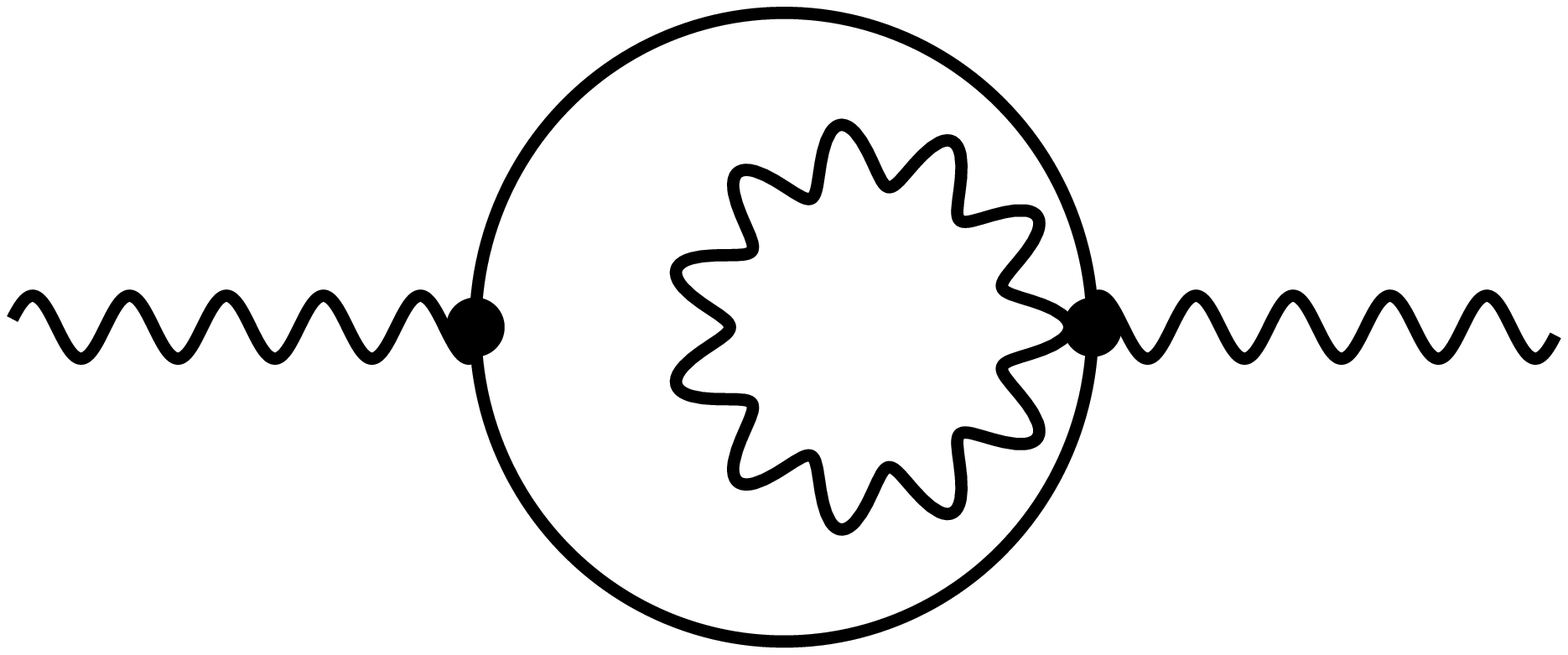}}
\put(13.2,4.2){B7}
\put(0,0.4){\includegraphics[scale=0.152]{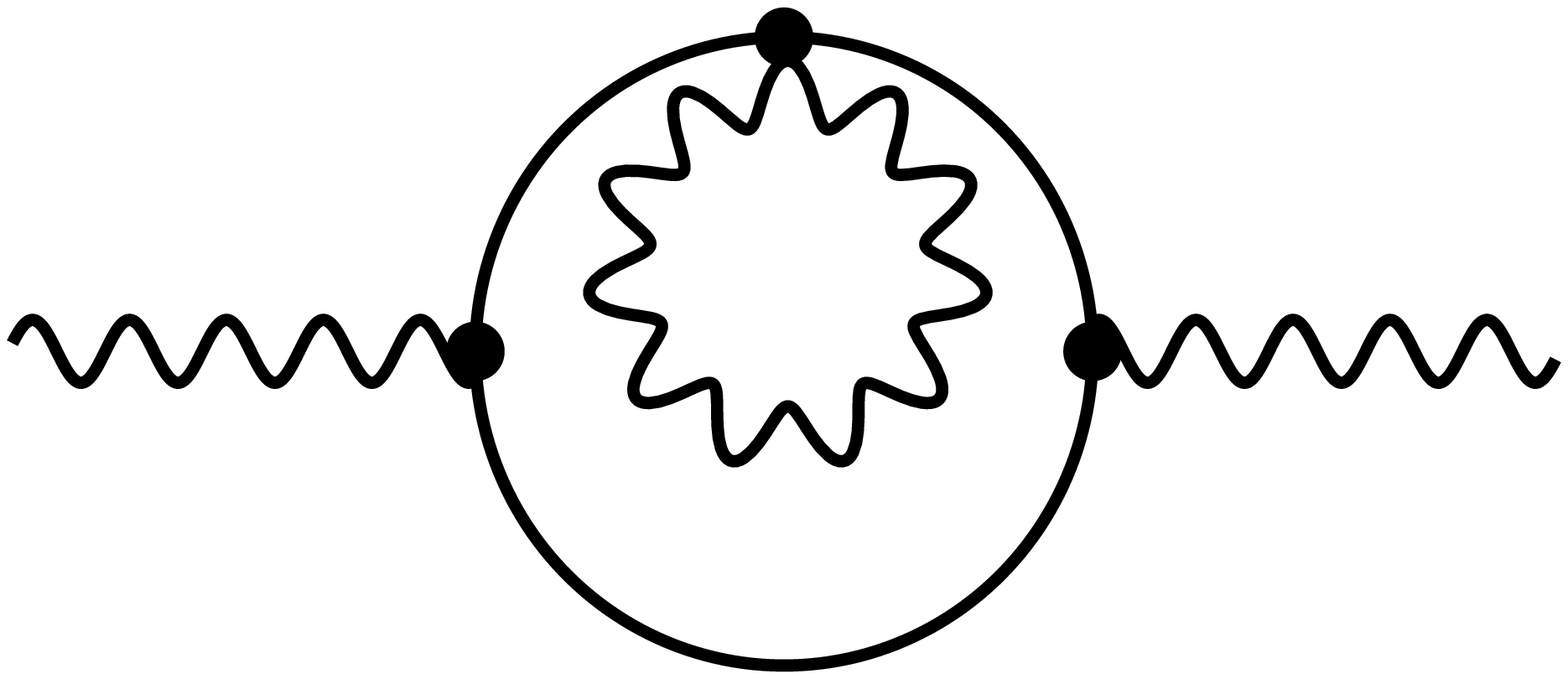}}
\put(0,1.7){B8}
\put(3.55,0.2){\includegraphics[scale=0.152]{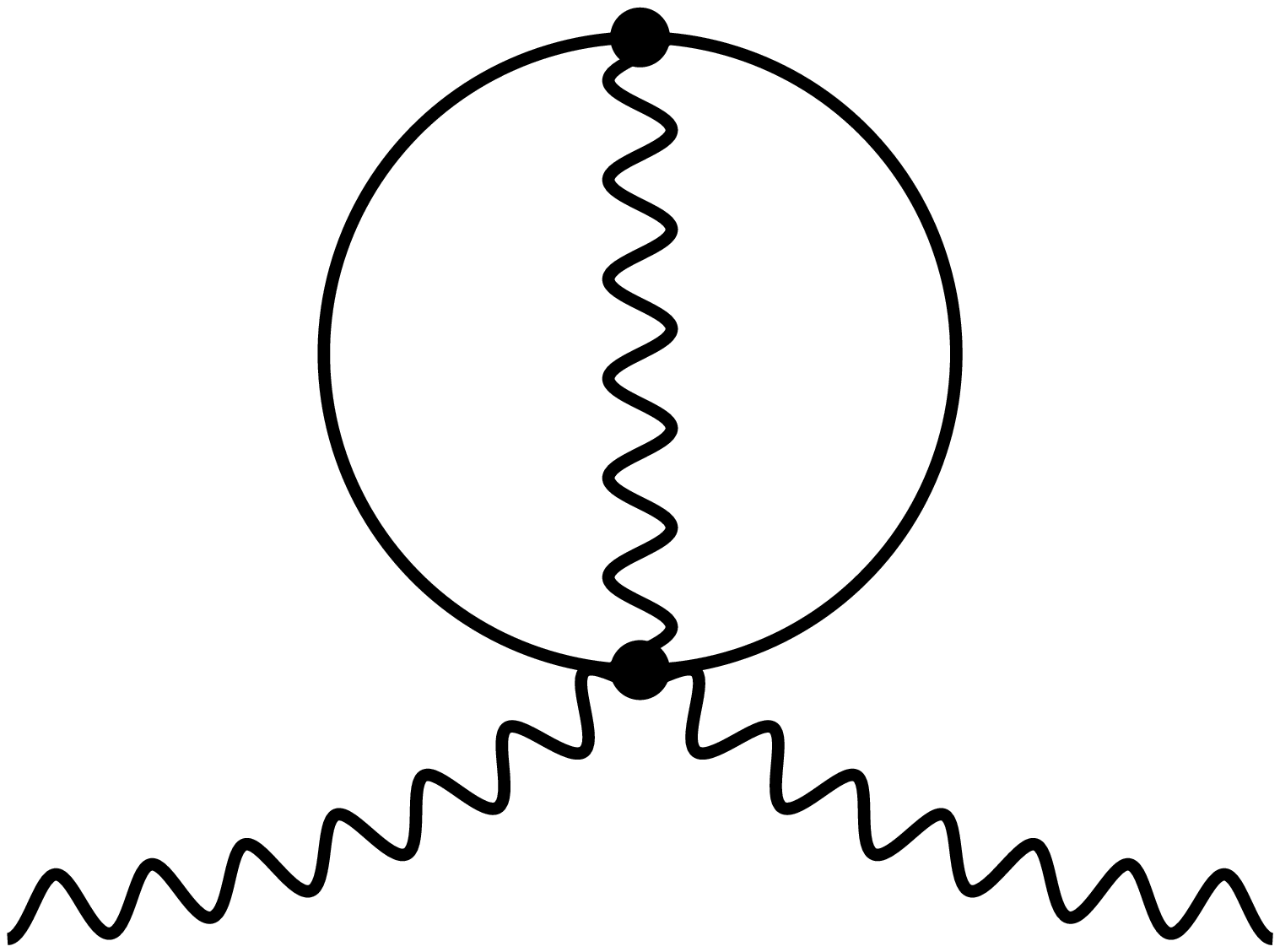}}
\put(3.35,1.7){B9}
\put(6.8,0.2){\includegraphics[scale=0.152]{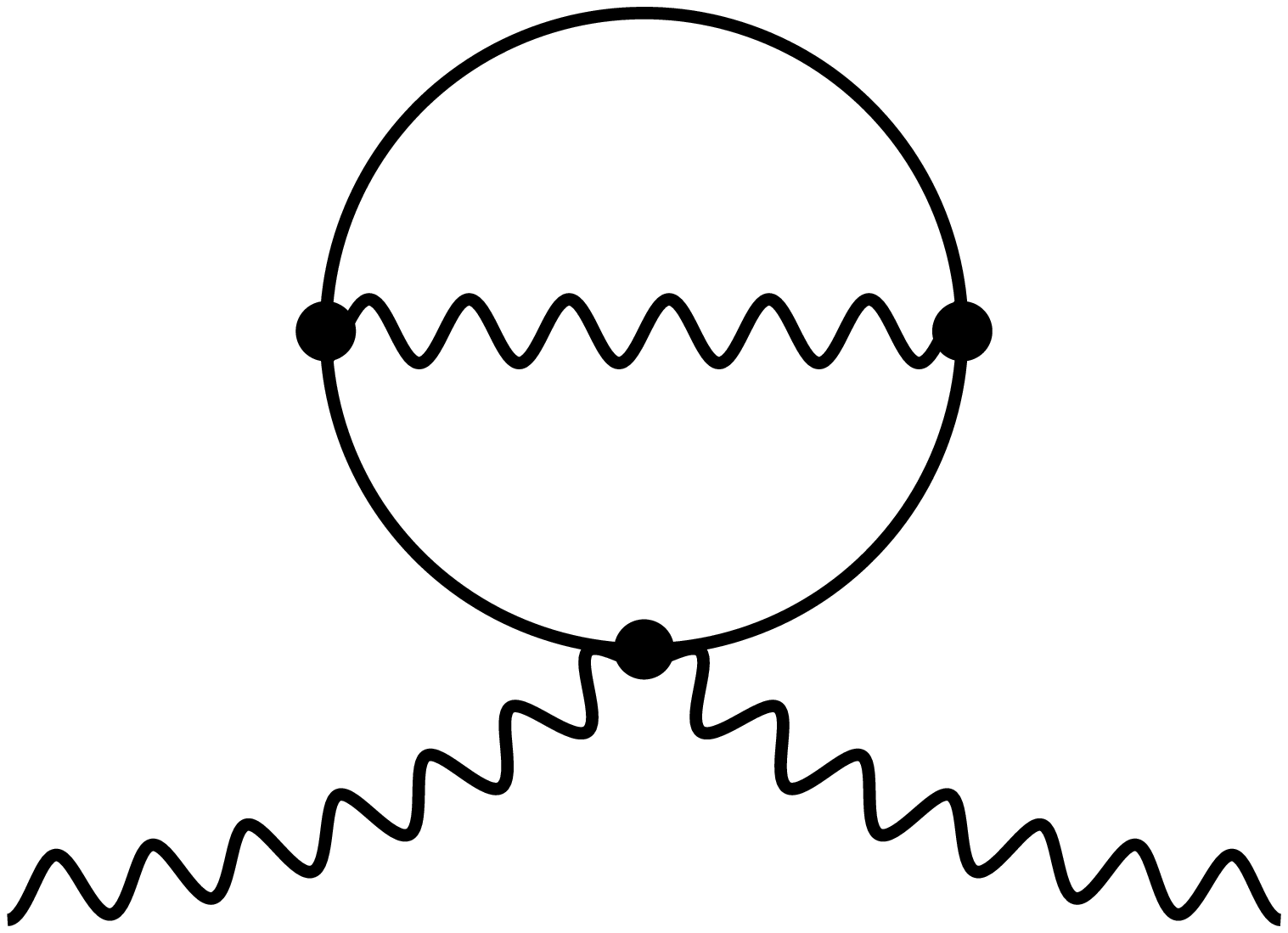}}
\put(6.6,1.7){B10}
\put(10.05,0.2){\includegraphics[scale=0.152]{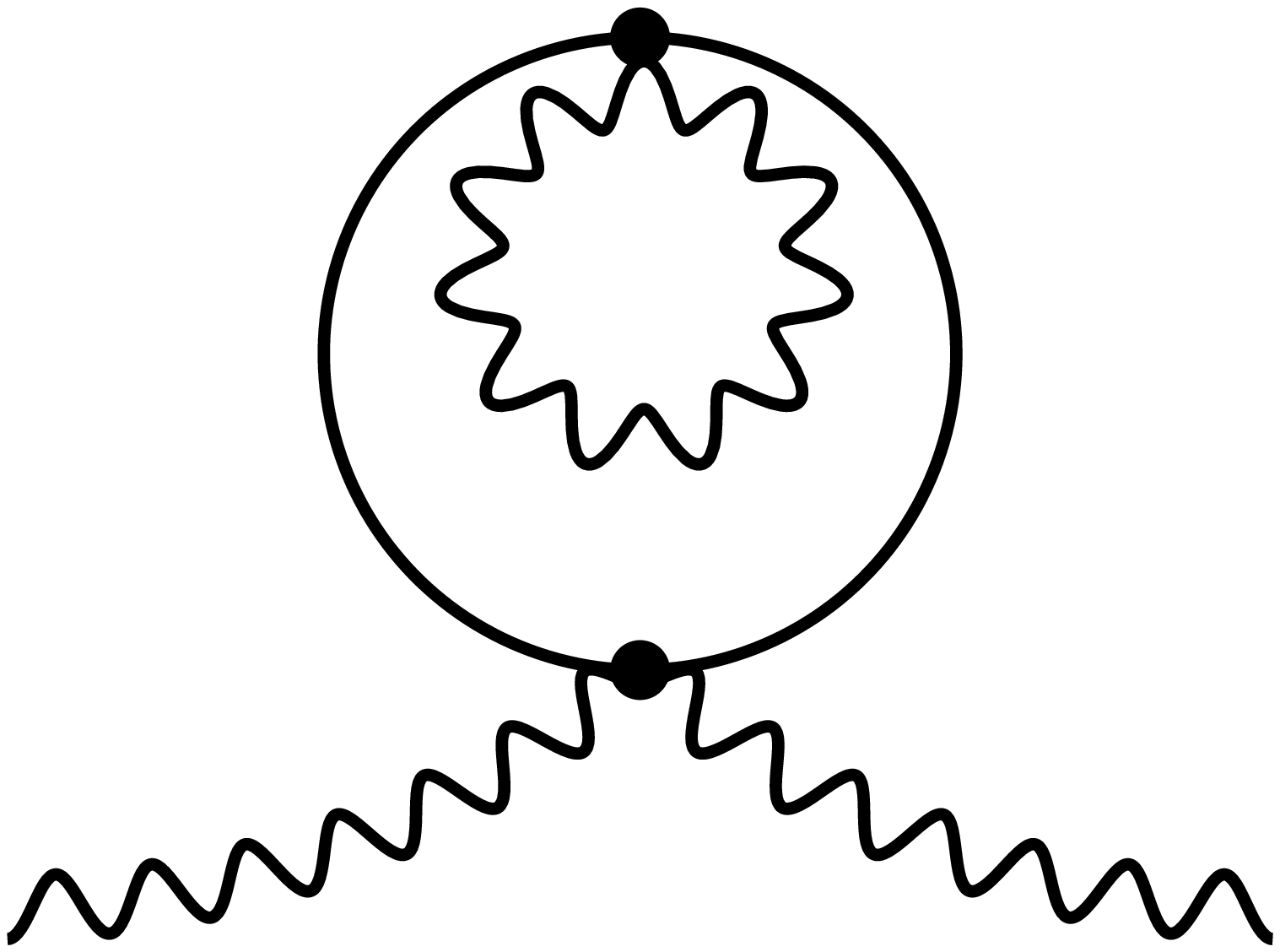}}
\put(9.85,1.7){B11}
\put(13.3,0.2){\includegraphics[scale=0.152]{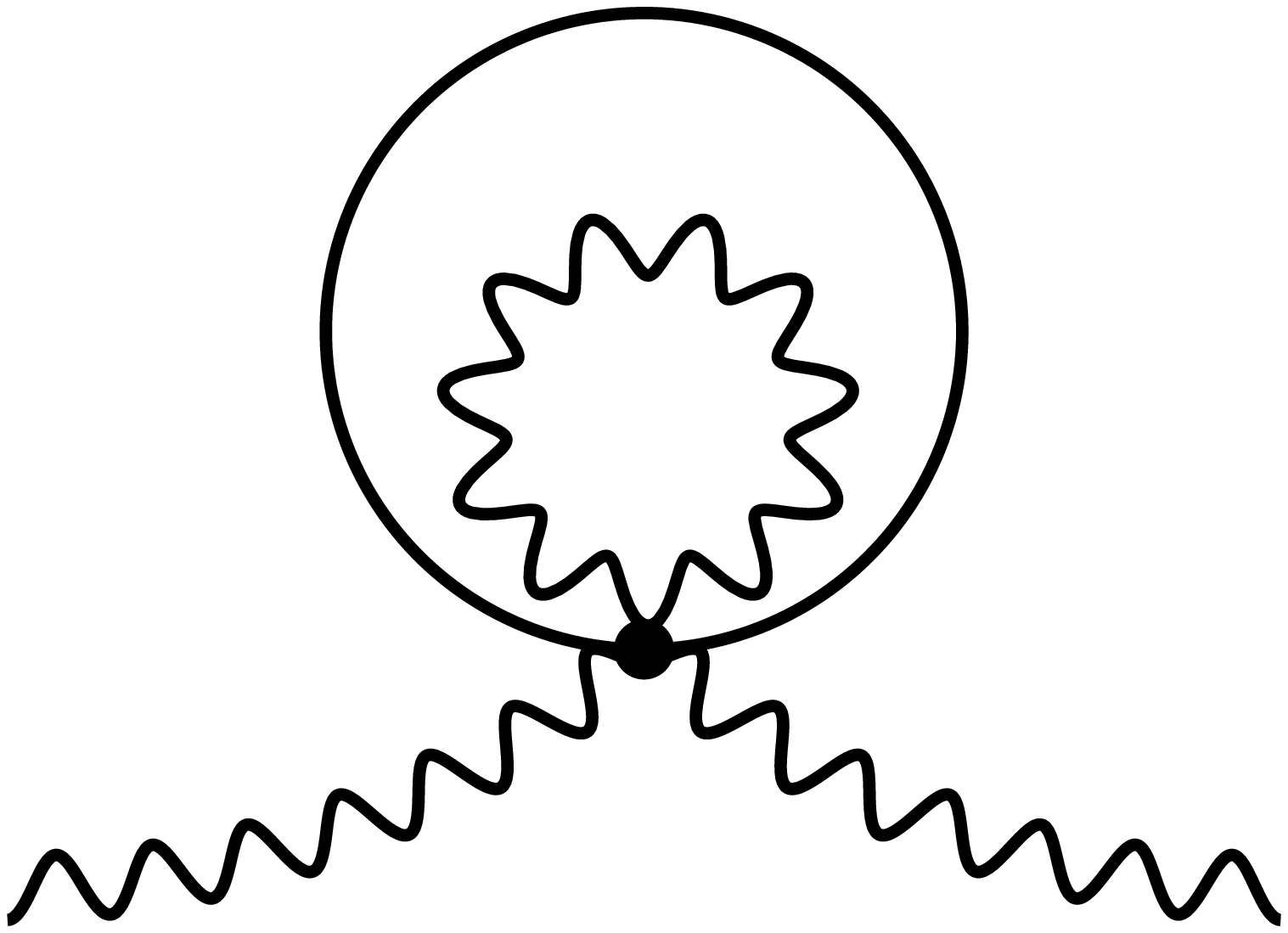}}
\put(13.1,1.7){B12}
\end{picture}
\caption{The superdiagrams contributing to the two-loop $\beta$-function of ${\cal N}=1$ SQED.}\label{Figure_Two_Point_V_Superdiagrams}
\end{figure}

The one- and two-loop superdiagrams contributing to the $\beta$-function are presented in Figs.~\ref{Figure_One_Loop} and \ref{Figure_Two_Point_V_Superdiagrams}, respectively. Propagators of the matter and Pauli--Villars superfields are denoted by solid lines. They are proportional to

\begin{eqnarray}
&& P(\phi_{\alpha,x}, \phi^*_{\beta,y}) = P(\widetilde\phi_{\alpha,x},\widetilde\phi^*_{\beta,y}) = \delta_{\alpha\beta}\frac{\bar D_x^2 D_y^2}{4\partial^2} \delta^8_{xy};\qquad\quad P(\Phi_x,\widetilde\Phi_y) = \frac{M \bar D^2}{\partial^2+M^2}\delta^8_{xy};
\qquad\nonumber\\
&& P(\Phi_x^*, \widetilde\Phi_y^*) = \frac{M D^2}{\partial^2+M^2}\delta^8_{xy};\qquad\qquad\ P(\Phi_x, \Phi^*_y) = P(\widetilde\Phi_x, \widetilde\Phi^*_y) = \frac{\bar D_x^2 D_y^2}{4(\partial^2+M^2)} \delta^8_{xy}.
\end{eqnarray}

\noindent
Propagators of the gauge superfield $V$ are denoted by wavy lines. They are proportional to

\begin{equation}\label{Gauge_Propagator}
P(V_x,V_y) = 2e_0^2 \bigg[ -\frac{1}{R\partial^2} + \frac{1}{16\partial^4}\Big(\frac{\xi_0}{K} - \frac{1}{R}\Big)\Big(\bar D^2 D^2 + D^2 \bar D^2\Big) \bigg]\delta^8_{xy}
\end{equation}

\noindent
and depend on the gauge parameter $\xi_0$. Taking into account that in the Abelian case the gauge fixing action does not produce vertices containing this parameter, we see that all gauge dependence comes from the gauge propagators. This implies that the one-loop contribution to the $\beta$-function is gauge independent, because it does not contain gauge propagators.

\begin{figure}[h]
\begin{picture}(0,2.2)
\put(7.0,0.2){\includegraphics[scale=0.15,clip]{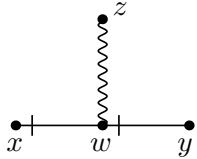}}
\put(8.35,1.8){$z$}
\put(6.95,0){$x$} \put(8.05,0){$w$} \put(9.2,0){$y$}
\end{picture}
\caption{The triple gauge-matter vertex is a key element for understanding how the gauge dependence disappears in the massless case in the perturbation theory.}\label{Figure_Vertex}
\end{figure}

It is also possible to verify that all gauge dependent terms present in the two-loop supergraphs B3 --- B12 cancel each other. To see this, we use the following method. Let us first consider the superdiagrams presented in Fig. \ref{Figure_Two_Point_V_Superdiagrams} with a loop of the usual (massless) matter superfields $\phi$ and $\widetilde\phi$. Some of them contain triple vertices without external lines, such as the one presented in Fig. \ref{Figure_Vertex}. Note that no other propagators are attached to the point $w$, while various internal and external lines can be attached to the points $x$, $y$, and $z$.

The gauge propagator $P(V_z,V_w)$ in the subdiagram presented in Fig. \ref{Figure_Vertex} can be naturally split into the three parts, see Eq. (\ref{Gauge_Propagator}). The first one does not contain supersymmetric covariant derivatives and the gauge parameter. It exactly coincides with the propagator in the Feynman gauge $\xi_0=1$ in the case $K(x)=R(x)$. The second and third parts contain the terms proportional to $D_w^2 \bar D_w^2 \delta^8_{zw}$ and $\bar D_w^2 D_w^2 \delta^8_{zw}$, respectively. They include all gauge dependent terms in the considered propagator. A part of the subdiagram coming from the second part of the propagator is proportional to

\begin{equation}\label{Identity1}
\int d^8w\, \frac{D_w^2 \bar D_w^2}{16\partial^4}\Big(\frac{\xi_0}{K} - \frac{1}{R}\Big)\delta^8_{wz}\cdot \frac{\bar D_x^2 D_x^2}{4\partial^2}\delta_{xw}^8\cdot \frac{\bar D_w^2
D_w^2}{4\partial^2}\delta_{wy}^8 = - \frac{D_y^2 \bar D_y^2}{16\partial^4}\Big(\frac{\xi_0}{K} - \frac{1}{R}\Big)\delta^8_{yz}\cdot \frac{\bar D_x^2 D_x^2}{\partial^2}\delta_{xy}^8.
\end{equation}

\begin{figure}[h]
\begin{picture}(0,2.3)
\put(0.5,0.2){\includegraphics[scale=0.15,clip]{subdiagram1.eps}}
\put(1.85,1.8){$z$}
\put(0.3,1.0){$D_w^2\bar D_w^2$}
\put(0.45,0){$x$} \put(1.55,0){$w$} \put(2.7,0){$y$}
\put(3.2,1.0){\vector(1,0){1}}
\put(4.5,0.2){\includegraphics[scale=0.15,clip]{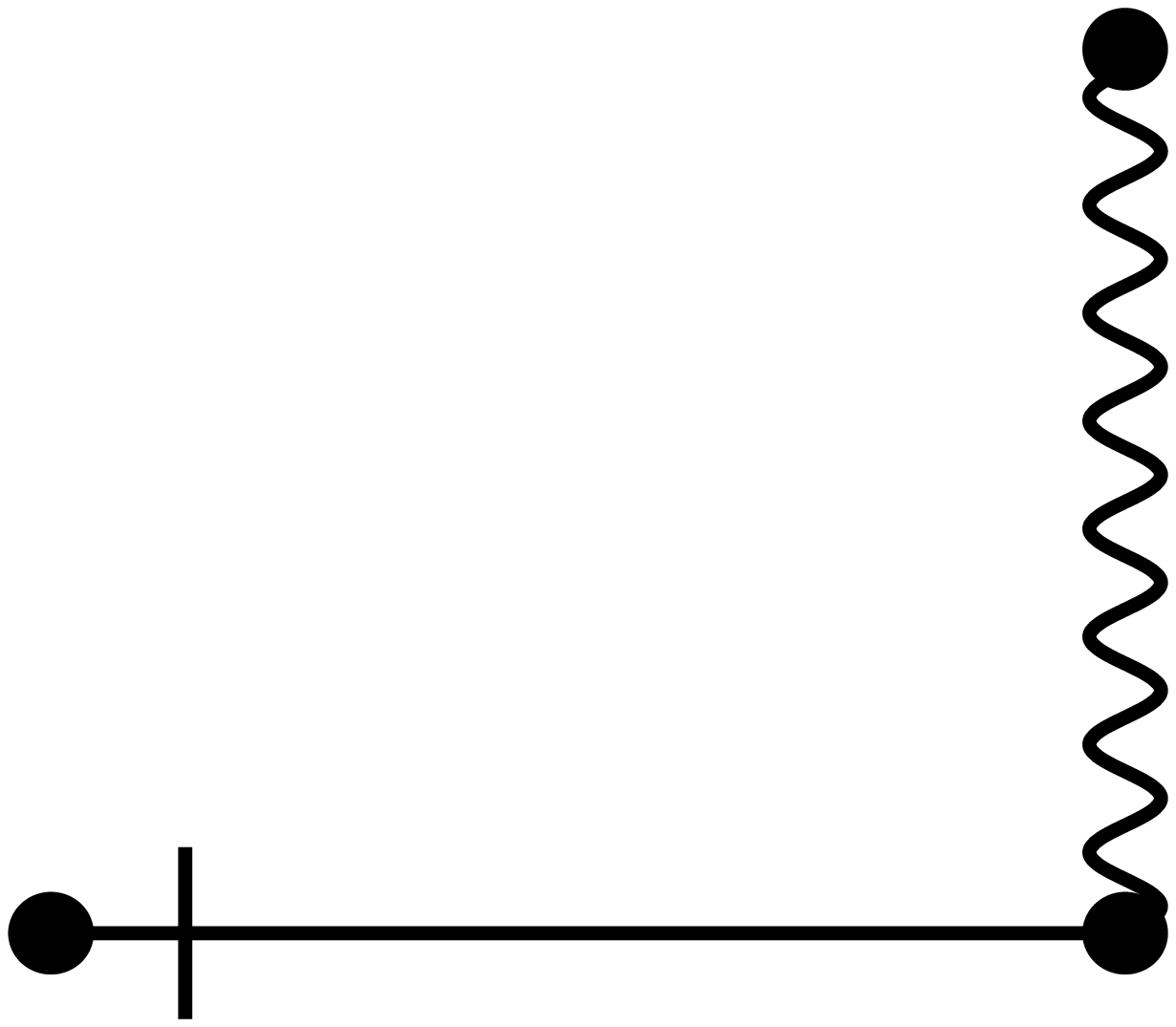}}
\put(6.4,1.8){$z$}
\put(4.95,1.0){$D_y^2\bar D_y^2$}
\put(4.45,0){$x$} \put(6.2,0){$y$}
\put(9,0.2){\includegraphics[scale=0.15,clip]{subdiagram1.eps}}
\put(10.35,1.8){$z$}
\put(8.8,1.0){$\bar D_w^2 D_w^2$}
\put(8.95,0){$x$} \put(10.1,0){$w$} \put(11.3,0){$y$}
\put(11.9,1.0){\vector(1,0){1}}
\put(13.5,0.2){\includegraphics[scale=0.15,clip]{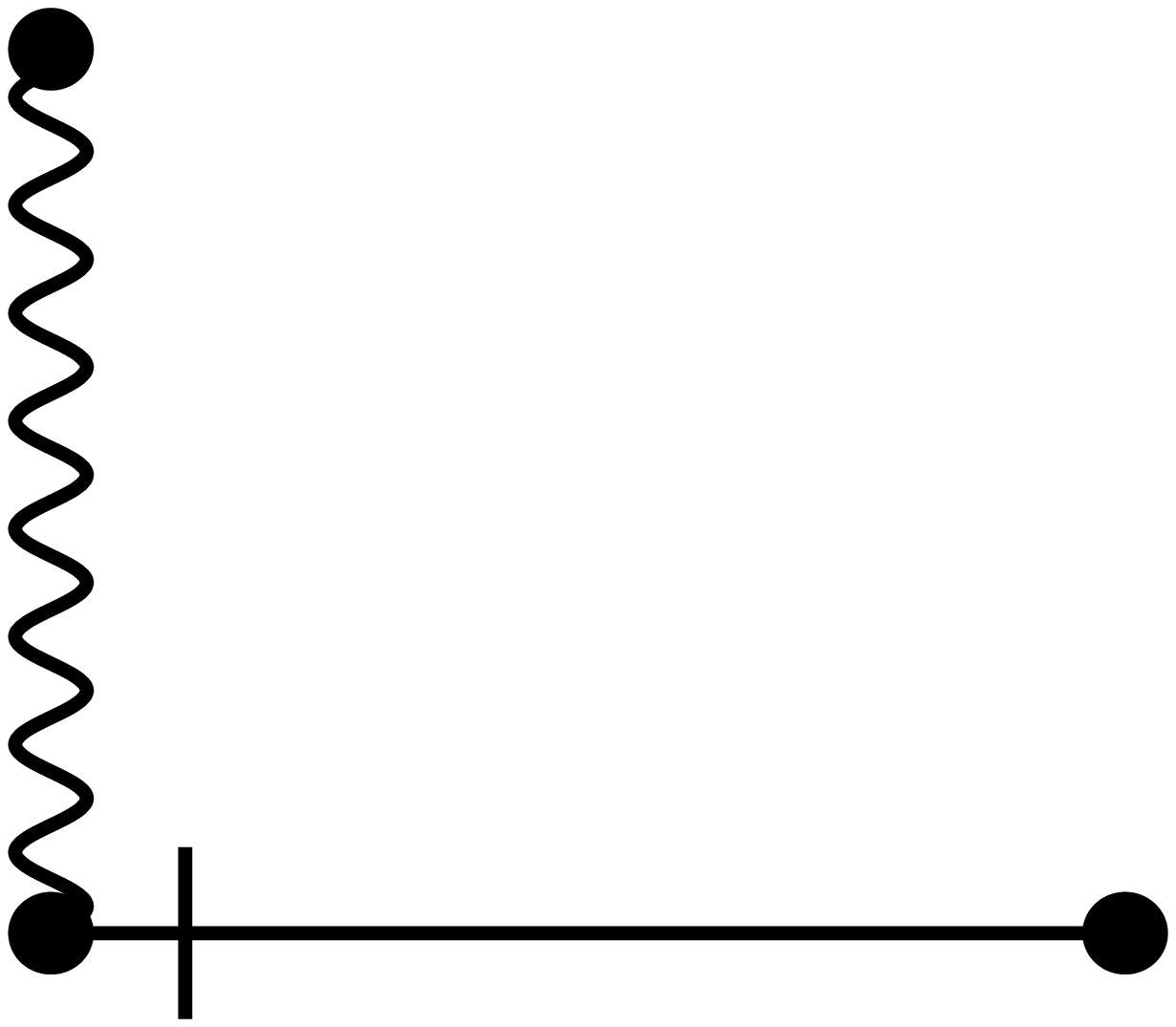}}
\put(13.7,1.8){$z$}
\put(13.8,1.0){$\bar D_x^2 D_x^2$}
\put(13.45,0){$x$} \put(15.2,0){$y$}
\end{picture}
\caption{Graphical form of the identities (\ref{Identity1}) and (\ref{Identity2}) needed for summing the gauge dependent parts of various supergraphs.}\label{Figure_Subdiagrams}
\end{figure}

\noindent
The equality can easily be verified with the help of integrating by parts with respect to the derivative $D_w^2$ and using the identity $D^2 \bar D^2 D^2 = -16 D^2 \partial^2$. The result has a simple graphical interpretation, which is presented in Fig. \ref{Figure_Subdiagrams} on the left. We see that effectively the considered subdiagram is reduced to another subdiagram which does not contain the triple gauge-matter vertex.

Similarly, for the remaining part of the propagator one can prove the equality

\begin{equation}\label{Identity2}
\int d^8w\, \frac{\bar D_w^2 D_w^2}{16\partial^4}\Big(\frac{\xi_0}{K} - \frac{1}{R}\Big)\delta^8_{wz}\cdot \frac{\bar D_x^2 D_x^2}{4\partial^2}\delta_{xw}^8\cdot \frac{\bar D_w^2 D_w^2}{4\partial^2}\delta_{wy}^8 = - \frac{\bar D_x^2 D_x^2}{16\partial^4}\Big(\frac{\xi_0}{K} - \frac{1}{R}\Big)\delta^8_{xz}\cdot \frac{\bar D_x^2 D_x^2}{\partial^2}\delta_{xy}^8,
\end{equation}

\noindent
which has an analogous graphical interpretation, see the right part of Fig. \ref{Figure_Subdiagrams}. Similar graphical identities for the superdiagrams with the massive Pauli--Villars propagators are more complicated. We present them in Appendix \ref{Appendix_Identities}.

Using the identities (\ref{Identity1}) and (\ref{Identity2}) and the analogous identities presented in Appendix \ref{Appendix_Identities} we can reduce gauge dependent terms in diagrams with triple gauge-matter vertices to the ones in diagrams which do not contain them. Certainly, it is necessary to take care of the overall numerical coefficients. We have obtained that, as a result of the above described procedure, all gauge dependent terms in the two-loop $\beta$-function cancel each other, so that the result in the general $\xi$-gauge coincides with the one in the Feynman gauge. For the considered regularization it can be found, e.g., in Ref. \cite{Stepanyantz:2012zz} and has the form

\begin{eqnarray}\label{2Loop_Beta}
&&\hspace*{-6mm} \frac{\beta(\alpha_0)}{\alpha_0^2} = 2\pi N_f \frac{d}{d\ln\Lambda}
\int \frac{d^4Q}{(2\pi)^4} \frac{\partial^2}{\partial Q^\mu \partial Q_\mu}
\frac{\ln(Q^2+M^2)}{Q^2} + 4\pi N_f \frac{d}{d\ln\Lambda} \int \frac{d^4Q}{(2\pi)^4} \frac{d^4K}{(2\pi)^4} \frac{e_0^2}{K^2 R_K}
\nonumber\\
&&\hspace*{-6mm} \times  \frac{\partial^2}{\partial Q^\mu \partial Q_\mu} \bigg(\frac{1}{Q^2 (K+Q)^2} -
\frac{1}{(Q^2+M^2)((K+Q)^2 + M^2)}\bigg) + O(e_0^4) = \frac{N_f}{\pi}\Big(1 + \frac{\alpha_0}{\pi} + O(\alpha_0^2)\Big),\nonumber\\
\end{eqnarray}

\noindent
where $R_K\equiv R(K^2/\Lambda^2)$. (In our notation Euclidean momenta are denoted by capital letters.)

Taking into account that the two-loop $\beta$-function is scheme independent, for an arbitrary renormalization prescription we obtain the well-known result \cite{Jones:1974pg}

\begin{equation}
\frac{\widetilde\beta(\alpha)}{\alpha^2} =  \frac{N_f}{\pi}\Big(1 + \frac{\alpha}{\pi} + O(\alpha^2)\Big).
\end{equation}

\section{Two-loop $\beta$-function: a new method of calculation}
\hspace*{\parindent}\label{Section_Two-Loop_Beta_New}

\begin{figure}[h]
\begin{picture}(0,2.5)
\put(0.5,0){\includegraphics[scale=0.183]{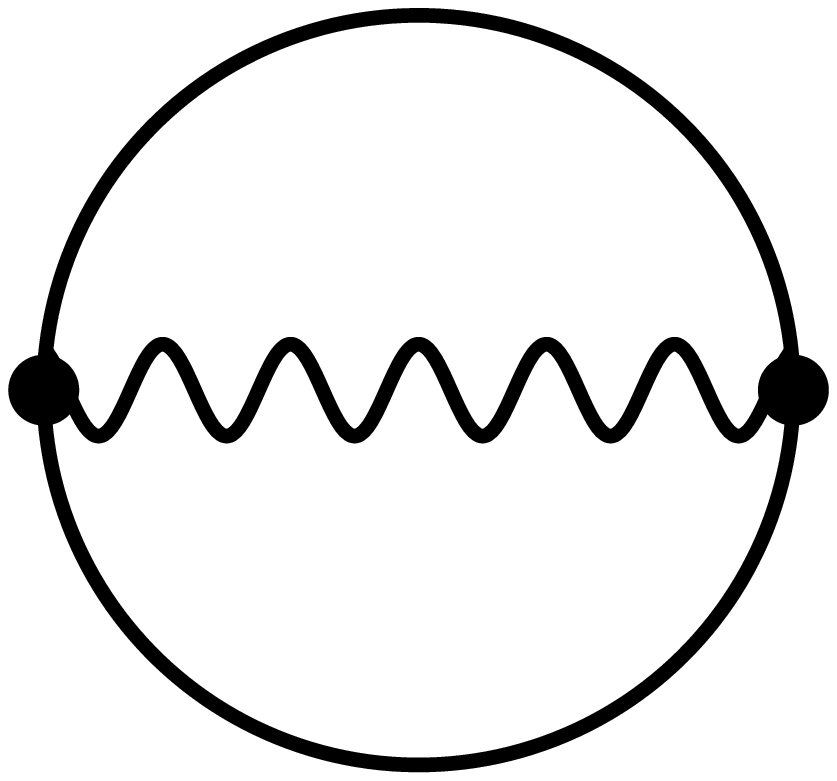}} \put(8.9,0){\includegraphics[scale=0.209]{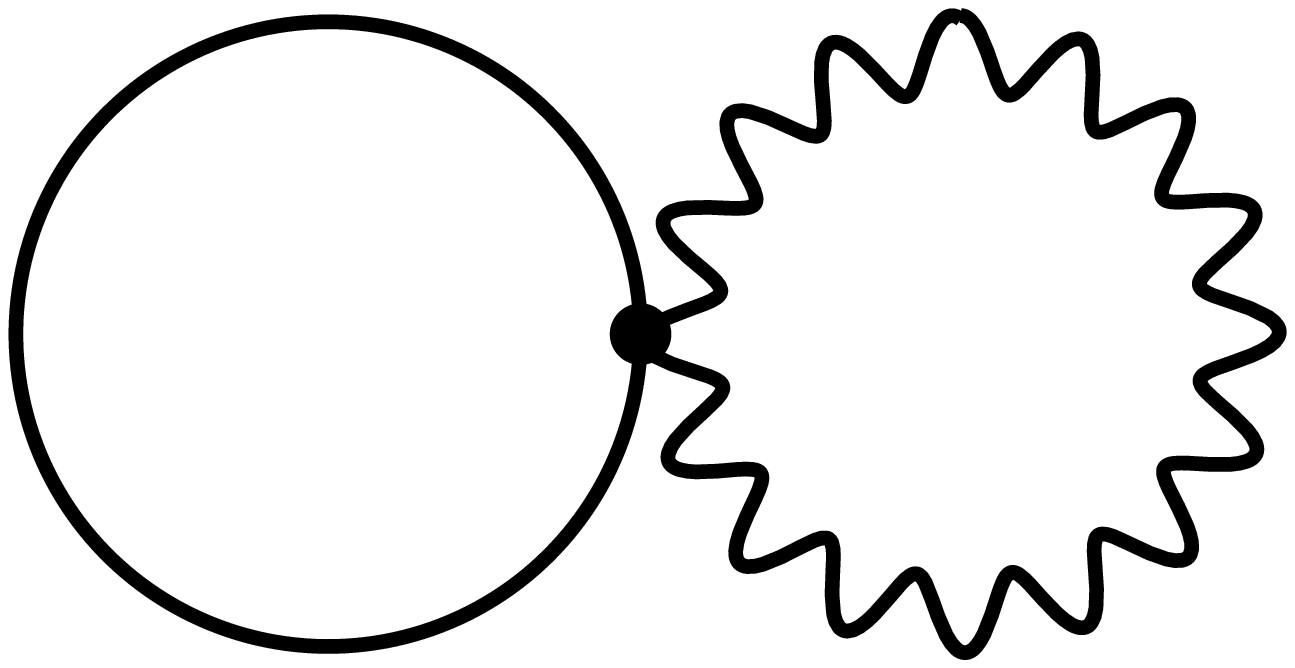}}
\put(0.2,1.5){S1} \put(8.6,1.5){S2}
\put(2.5,0.7){$\to$\ \ B3,\ B4,\ B5,\ B6,\ B9,\ B10}
\put(12.1,0.7){$\to$\ \ B7,\ B8,\ B11,\ B12}
\end{picture}
\caption{The two-loop contribution to the $\beta$-function can be obtained by considering two supergraphs presented in this figure.}\label{Figure_Vacuum_Supergraph}
\end{figure}

According to Ref. \cite{Stepanyantz:2019ihw} a $\beta$-function of ${\cal N}=1$ supersymmetric gauge theories regularized by higher covariant derivatives can be obtained in a much simpler way. In particular, to find the two-loop contribution to the $\beta$-function, instead of calculating the superdiagrams presented in Fig. \ref{Figure_Two_Point_V_Superdiagrams} one can consider only two supergraphs depicted in Fig. \ref{Figure_Vacuum_Supergraph}. The superdiagrams presented in Fig. \ref{Figure_Two_Point_V_Superdiagrams} are obtained by attaching to them two external gauge lines in all possible ways. In Fig. \ref{Figure_Vacuum_Supergraph} we also point which superdiagrams in Fig. \ref{Figure_Two_Point_V_Superdiagrams} are obtained from each supergraph in this way.

According to Refs. \cite{Stepanyantz:2019ihw,Kuzmichev:2019ywn}, to calculate a part of the $\beta$-function corresponding to a certain vacuum supergraph in ${\cal N}=1$ supersymmetric gauge theories regularized by higher derivatives, one should use the following algorithm:

1. It is necessary to insert $\theta^4 (v^B)^2$ into an arbitrary point of the supergraph, where $v^B$ (with $B=1,\ldots,r$) are functions which slowly decrease at a very large scale $R\to \infty$ and $\theta^4\equiv \theta^a \theta_a \bar\theta^{\dot b}\bar\theta_{\dot b}$.

2. Next, one should calculate the supergraph using the $D$-algebra and omit all terms suppressed by inverse powers of $R$.

3. Let us mark $L$ propagators with independent Euclidean momenta $Q^\mu_i$. Let $a_i$ and $b_i$ be indices corresponding to their beginnings and endings, respectively. Then in the integrand of the momentum integral the product $\prod_{i=1}^L \delta_{a_i}^{b_i}$ coming from the marked propagators should be replaced by the differential operator

\begin{equation}\label{Differential_Operator}
\sum\limits_{k,l=1}^L \prod\limits_{i\ne k,l} \delta_{a_i}^{b_i} (T^A)_{a_k}{}^{b_k} (T^A)_{a_l}{}^{b_l} \frac{\partial^2}{\partial Q^\mu_k \partial Q^{\mu}_l},
\end{equation}

\noindent
where $(T^A)_a{}^b$ are the generators of the gauge group in a relevant representation.

4. Finally, one should apply to the result the operator

\begin{equation}
-\frac{2\pi}{r{\cal V}_4} \frac{d}{d\ln\Lambda},
\end{equation}

\noindent
where $r$ denotes a dimension of the gauge group, and the constant

\begin{equation}
{\cal V}_4 \equiv \int d^4x\, \big(v^B\big)^2
\end{equation}

\noindent
is of the order $R^4\to \infty$.

Using this algorithm the two-loop $\beta$-function (in the general $\xi$-gauge) has been calculated for a general ${\cal N}=1$ supersymmetric gauge theory regularized by higher covariant derivatives in Ref. \cite{Stepanyantz:2019lyo}. This theory can contain Yukawa couplings $\lambda_0^{ijk}$ and one more higher derivative regulator $F(x)$. ${\cal N}=1$ SQED considered in this paper is a particular case of this theory which corresponds to

\begin{equation}\label{Replacement}
r\to 1;\qquad C(R)_i{}^j\to \delta_\alpha^\beta
\left(\begin{array}{cc}1 & 0\\
0 & 1
\end{array}\right);\qquad C_2 \to 0;\qquad \lambda_0^{ijk} \to 0;\qquad F(x) \to 1.
\end{equation}

\noindent
(In the non-Abelian case $C(R)_i{}^j\equiv (T^A T^A)_i{}^j$ and $C_2\delta^{AB} = f^{ACD} f^{BCD}$, where $T^A$ are the generators of the representation for the matter superfields and $f^{ABC}$ are the gauge group structure constants.)

Thus, it is possible to derive the two-loop $\beta$-function by the above described method by making the replacement (\ref{Replacement}) in the result of Ref. \cite{Stepanyantz:2019lyo}. Then we obtain

\begin{equation}
\frac{\beta(\alpha_0)}{\alpha_0^2} - \frac{\beta_{\mbox{\scriptsize 1-loop}}(\alpha_0)}{\alpha_0^2} = \Delta_{\mbox{\scriptsize matter}}\Big(\frac{\beta}{\alpha_0^2}\Big) + O(\alpha_0^2),
\end{equation}

\noindent
where

\begin{eqnarray}
&& \Delta_{\mbox{\scriptsize matter}}\Big(\frac{\beta}{\alpha_0^2}\Big) = 4\pi N_f \frac{d}{d\ln\Lambda}\int \frac{d^4Q}{(2\pi)^4} \frac{\partial^2}{\partial Q^\mu \partial Q_\mu} \int \frac{d^4K}{(2\pi)^4} \frac{e_0^2}{K^2 R_K} \left\{\frac{1}{Q^2 (Q+K)^2} \right. \qquad\nonumber\\
&& \left. - \frac{1}{\big(Q^2 + M^2\big) \big((K+Q)^2 + M^2\big)} \right\}.
\end{eqnarray}

\noindent
We see that this expression coincides with the two-loop contribution to the $\beta$-function obtained by the standard method, see Eq. (\ref{2Loop_Beta}). (This contribution is proportional to $e_0^2$.) Also it agrees with the result of Ref. \cite{Smilga:2004zr} if the difference in the regularizations is taken into account. Thus, in the two-loop approximation both methods give the same results.

\section{Three-loop $\beta$-function}
\hspace*{\parindent}\label{Section_Three-Loop}

To obtain the three-loop $\beta$-function of ${\cal N}=1$ SQED regularized by higher derivatives, it is necessary to consider the supergraphs presented in Fig. \ref{Figure_3Loop_Supergraphs}. The corresponding contribution to the $\beta$-function can be obtained by the method described in Sect. \ref{Section_Two-Loop_Beta_New}. For ${\cal N}=1$ SQED this method can be reformulated as follows:

\begin{figure}[h]
\begin{picture}(0,5)
\put(0.6,4.2){S3}
\put(0.6,2.6){\includegraphics[scale=0.2,clip]{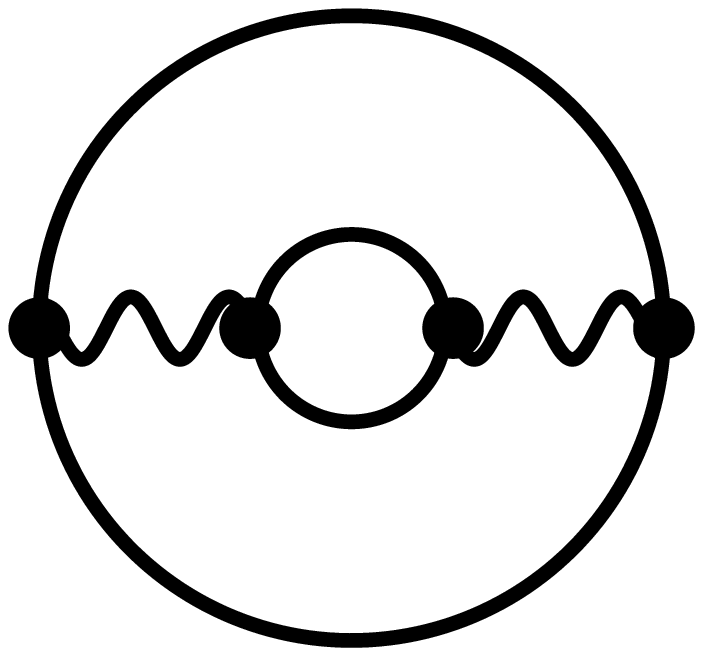}}
\put(2.9,4.2){S4}
\put(2.9,2.6){\includegraphics[scale=0.2,clip]{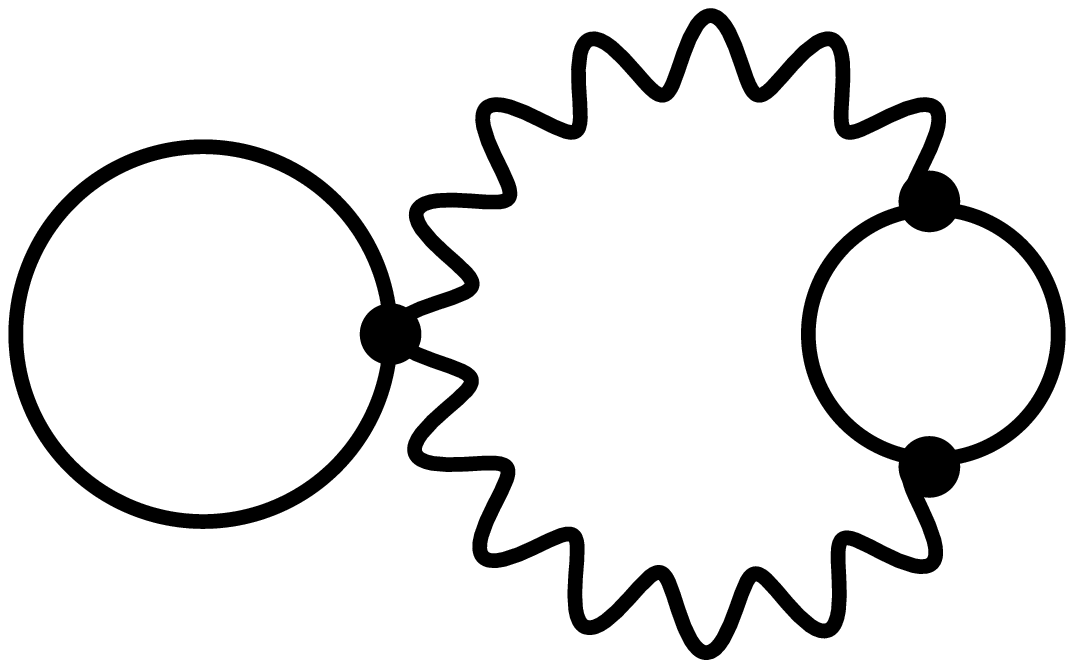}}
\put(5.8,4.2){S5}
\put(5.8,2.6){\includegraphics[scale=0.21,clip]{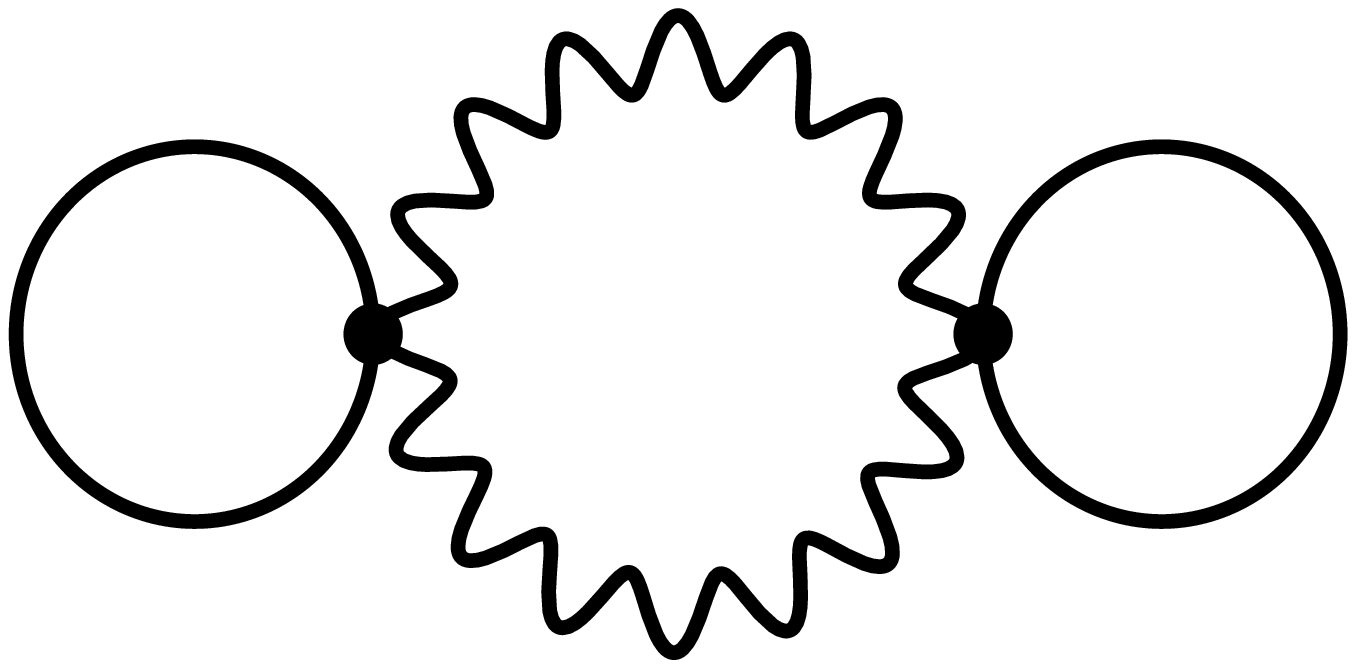}}
\put(9.2,4.2){S6}
\put(9.5,2.6){\includegraphics[scale=0.2,clip]{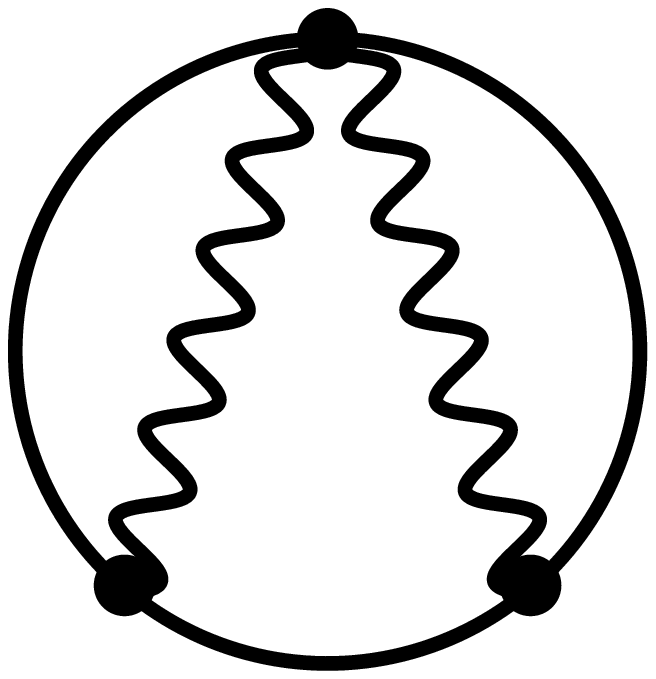}}
\put(11.5,4.2){S7}
\put(11.8,2.6){\includegraphics[scale=0.2,clip]{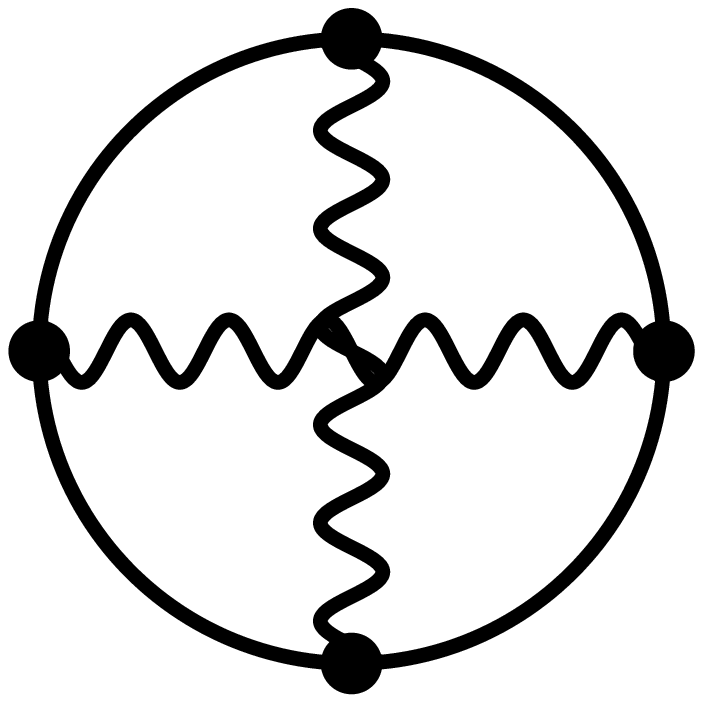}}
\put(13.8,4.2){S8}
\put(14.1,2.6){\includegraphics[scale=0.2,clip]{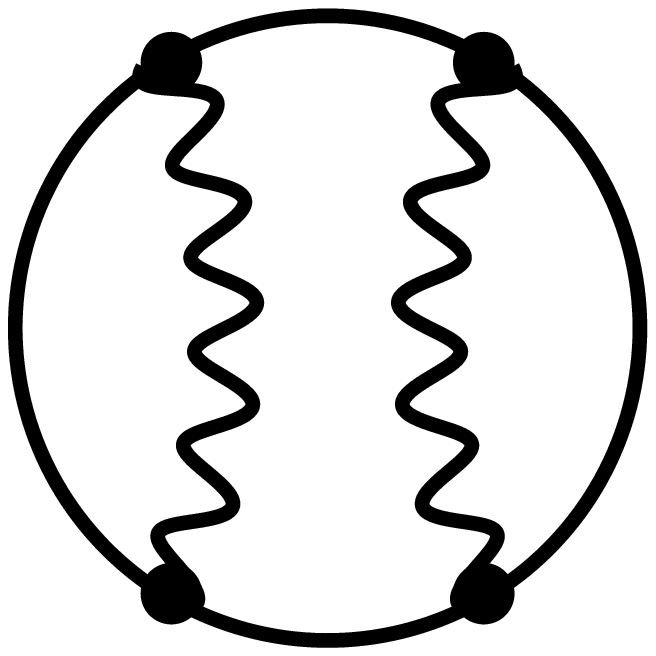}}

\put(0.6,1.6){S9}
\put(0.7,0){\includegraphics[scale=0.24,clip]{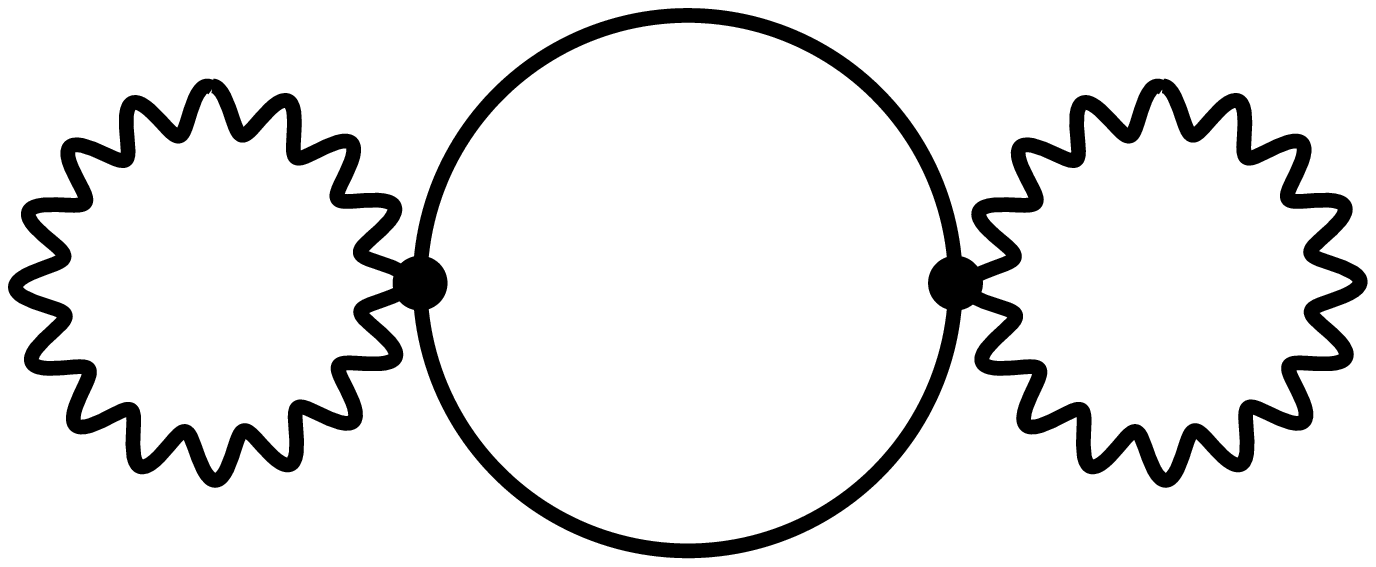}}
\put(4.8,1.6){S10}
\put(4.8,0){\includegraphics[scale=0.2,clip]{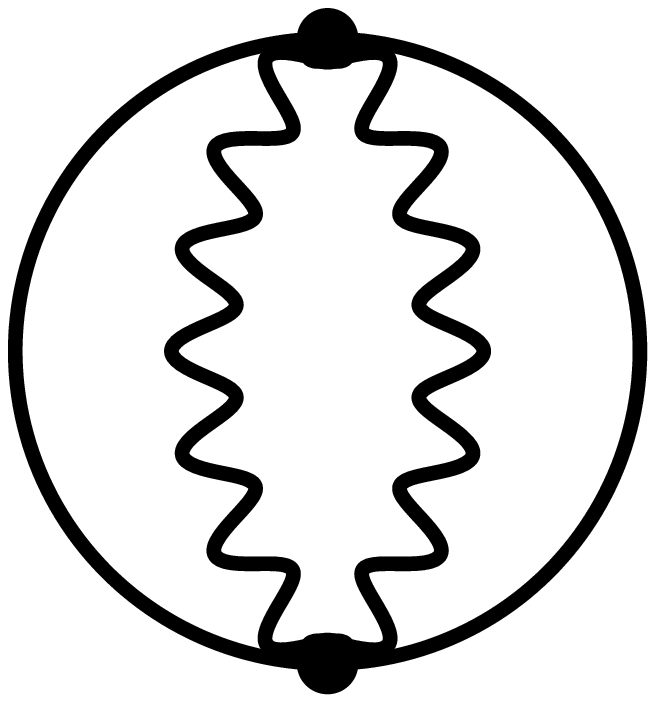}}
\put(6.9,1.6){S11}
\put(6.9,0.02){\includegraphics[scale=0.197,clip]{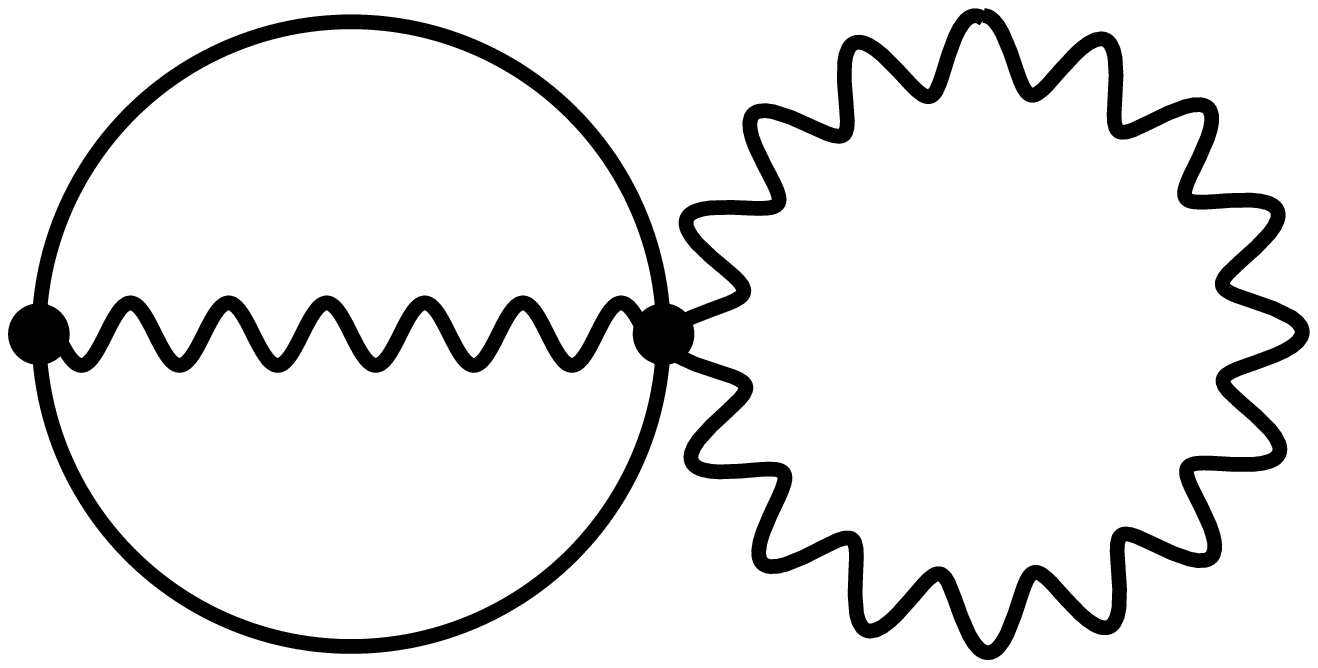}}
\put(10.3,1.6){S12}
\put(10.3,0){\includegraphics[scale=0.2,clip]{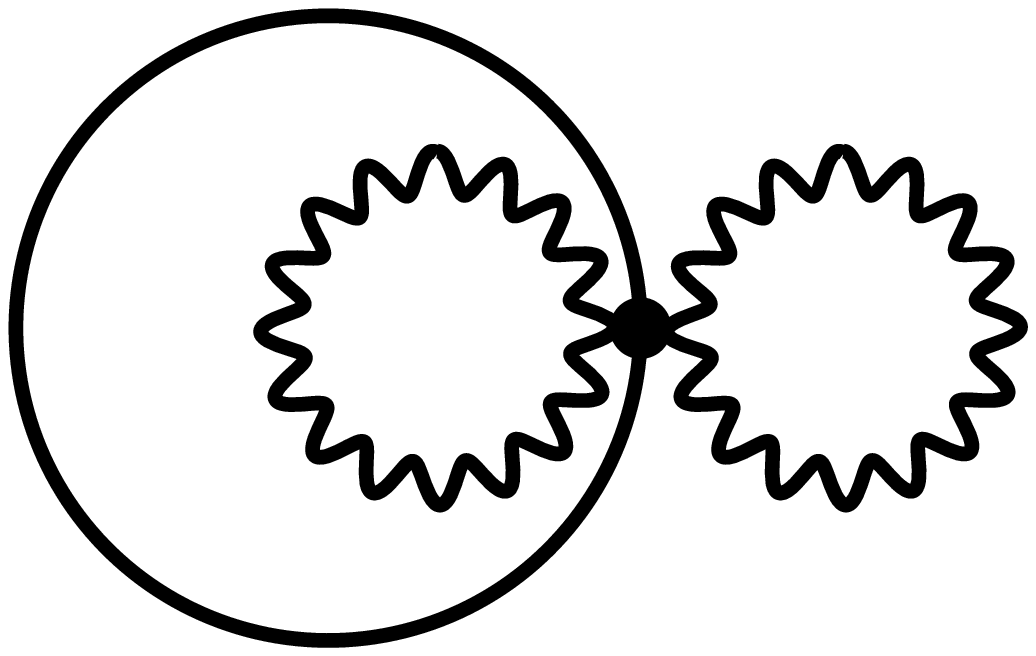}}
\put(13.2,1.6){S13}
\put(13.2,0){\includegraphics[scale=0.195,clip]{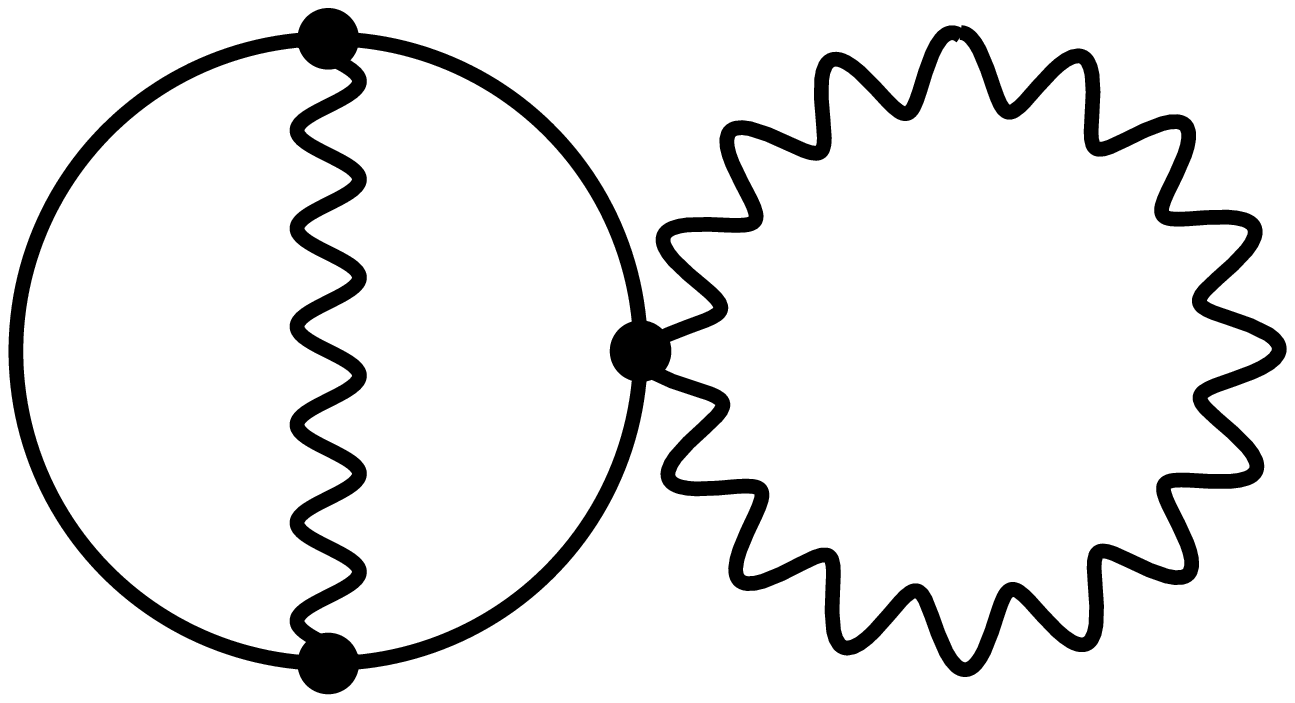}}
\end{picture}
\caption{The supergraphs generating the three-loop contribution to the $\beta$-function of ${\cal N}=1$ SQED.}\label{Figure_3Loop_Supergraphs}
\end{figure}

1. First, we calculate a supergraph with an insertion of $\theta^4 (v^B)^2$. Let $L$ and $M$ be numbers of loops and matter loops in this supergraph, respectively.

2. Because the gauge group is $U(1)$, the generator of the adjoint representation vanishes. This implies that the propagators of the gauge superfield do not contribute to the operator (\ref{Differential_Operator}). The generator of the representation for the matter superfields is

\begin{equation}
T^A\ \to\ \delta_\alpha^\beta \left(\begin{array}{cc}
1 & 0\\
0 & -1
\end{array}\right),
\end{equation}

\noindent
where $1$ and $-1$ correspond to $\phi$ and $\widetilde\phi$, respectively. Therefore, the operator which should be applied to the integrand takes the form

\begin{equation}
\sum\limits_{i=1}^M \frac{\partial^2}{\partial Q_{\mu i}^2},
\end{equation}

\noindent
where the index $i$ numerates the {\it matter} loops.

3. Taking into account that $r=\mbox{dim}\, U(1)=1$, the contribution to the function $\beta(\alpha_0)/\alpha_0^2$ corresponding to the considered supergraph is obtained by differentiating the result with respect to $\ln\Lambda$ and multiplying it to the factor $-2\pi/{\cal V}_4$.

The results for the three-loop supergraphs presented in Fig. \ref{Figure_3Loop_Supergraphs} calculated by this method are collected in Appendix \ref{Appendix_Beta_Supergraphs}. Note that in the general $\xi$-gauge it is necessary to take into account the supergraphs S5, S9 --- S13, which vanish in the Feynman gauge. The sum of all supergraphs can be written in the form

\begin{eqnarray}\label{3Loop_Beta}
&&\hspace*{-6mm} \frac{\beta(\alpha_0)}{\alpha_0^2} = 2\pi N_f \frac{d}{d\ln\Lambda}
\int \frac{d^4Q}{(2\pi)^4} \frac{\partial^2}{\partial Q^\mu \partial Q_\mu}
\frac{\ln(Q^2+M^2)}{Q^2} + 4\pi N_f \frac{d}{d\ln\Lambda} \int \frac{d^4Q}{(2\pi)^4} \frac{d^4K}{(2\pi)^4}
\nonumber\\
&&\hspace*{-6mm} \times \frac{e_0^2}{K^2 R_K^2} \frac{\partial^2}{\partial Q^\mu \partial Q_\mu} \bigg(\frac{1}{Q^2 (Q+K)^2} -
\frac{1}{(Q^2+M^2)((Q+K)^2 + M^2)}\bigg) \bigg[ R_K - 2 N_f e_0^2 \nonumber\\
&&\hspace*{-6mm} \times \int \frac{d^4L}{(2\pi)^4}\, \bigg(\frac{1}{L^2 (L+K)^2} - \frac{1}{(L^2+M^2)
((L+K)^2+M^2)} \bigg)\bigg] \nonumber\\
&&\hspace*{-6mm} + 8\pi N_f \frac{d}{d\ln\Lambda} \int \frac{d^4Q}{(2\pi)^4}
\frac{d^4K}{(2\pi)^4} \frac{d^4L}{(2\pi)^4} \frac{e_0^4}{K^2 R_K L^2 R_L} \frac{\partial^2}{\partial Q^\mu \partial Q_\mu}\bigg\{\bigg( \frac{1}{Q^2(Q+K)^2(Q+L)^2}
\nonumber\\
&&\hspace*{-6mm} - \frac{K^2}{Q^2 (Q+K)^2 (Q+L)^2 (Q+K+L)^2} \bigg) - \bigg(- \frac{K^2 + M^2}{((Q+K)^2+M^2)((Q+L)^2+M^2)}
\nonumber\\
&&\hspace*{-6mm} \times \frac{1}{(Q^2+M^2)((Q+K+L)^2+M^2)} + \frac{1}{(Q^2+M^2) ((Q+K)^2+M^2)((Q+L)^2+M^2)}
\nonumber\\
&&\hspace*{-6mm} - \frac{2M^2}{(Q^2+M^2)^2((Q+K)^2+M^2) ((Q+L)^2+M^2)} \bigg)\bigg\} + O(e_0^6).\vphantom{\frac{1}{2}}
\end{eqnarray}

We see that all gauge dependent terms cancel each other, so that the result is independent of the gauge parameter $\xi_0$, although expressions for separate supergraphs depend on it. The result coincides with one found earlier (see, e.g., \cite{Stepanyantz:2012zz}\footnote{In \cite{Stepanyantz:2012zz} $N_f=1$ and there are $n$ Pauli--Villars determinants. To obtain Eq. (\ref{3Loop_Beta}) one should introduce the factors $N_f$ for each matter loop and consider a particular case $c_I \to 1$, $M_I \to M$.}) in the Feynman gauge by direct summation of the supergraphs with two external lines of the gauge superfield. This confirms the correctness of the technique used for the calculation. Moreover, using the method described in Sect. \ref{Section_Two-Loop_Standard} we have calculated the sum of all gauge dependent terms in the three-loop supergraphs with two external gauge lines which do not contain Pauli--Villars loops and demonstrated that it really vanishes.

\begin{figure}[h]
\begin{picture}(0,9.7)
\put(0.6,7.5){\includegraphics[scale=0.16,clip]{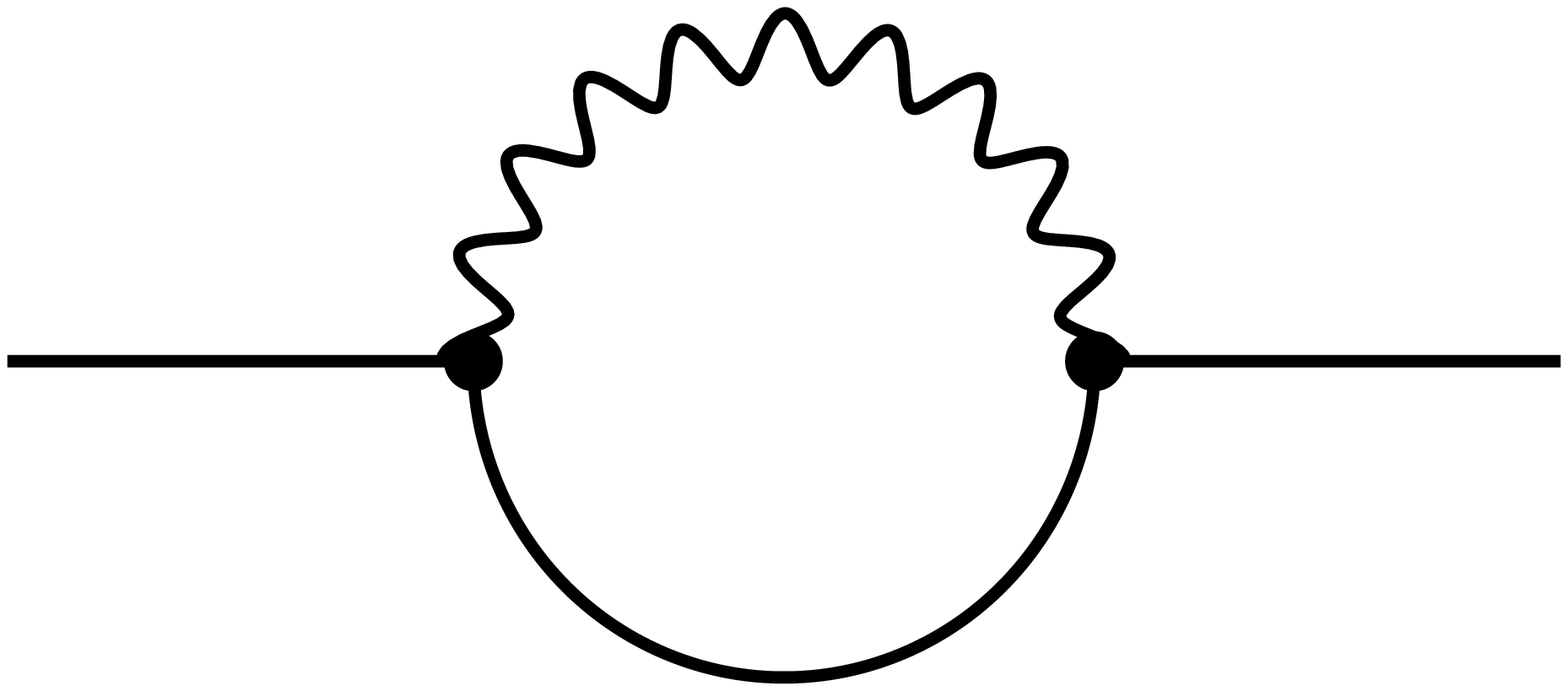}}
\put(0.6,9.15){A1}
\put(4.9,7.5){\includegraphics[scale=0.15,clip]{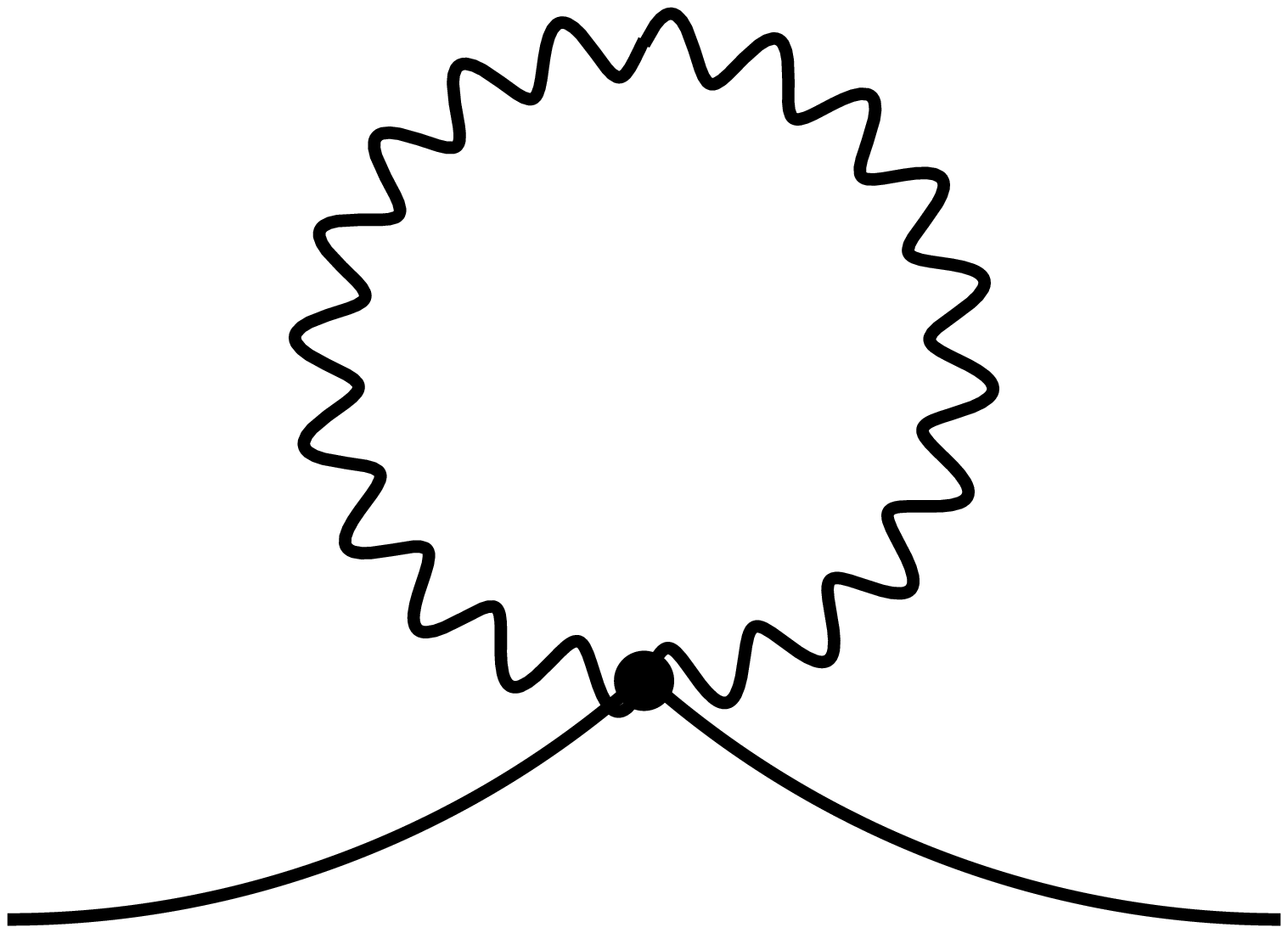}}
\put(4.9,9.15){A2}
\put(8.6,7.5){\includegraphics[scale=0.16,clip]{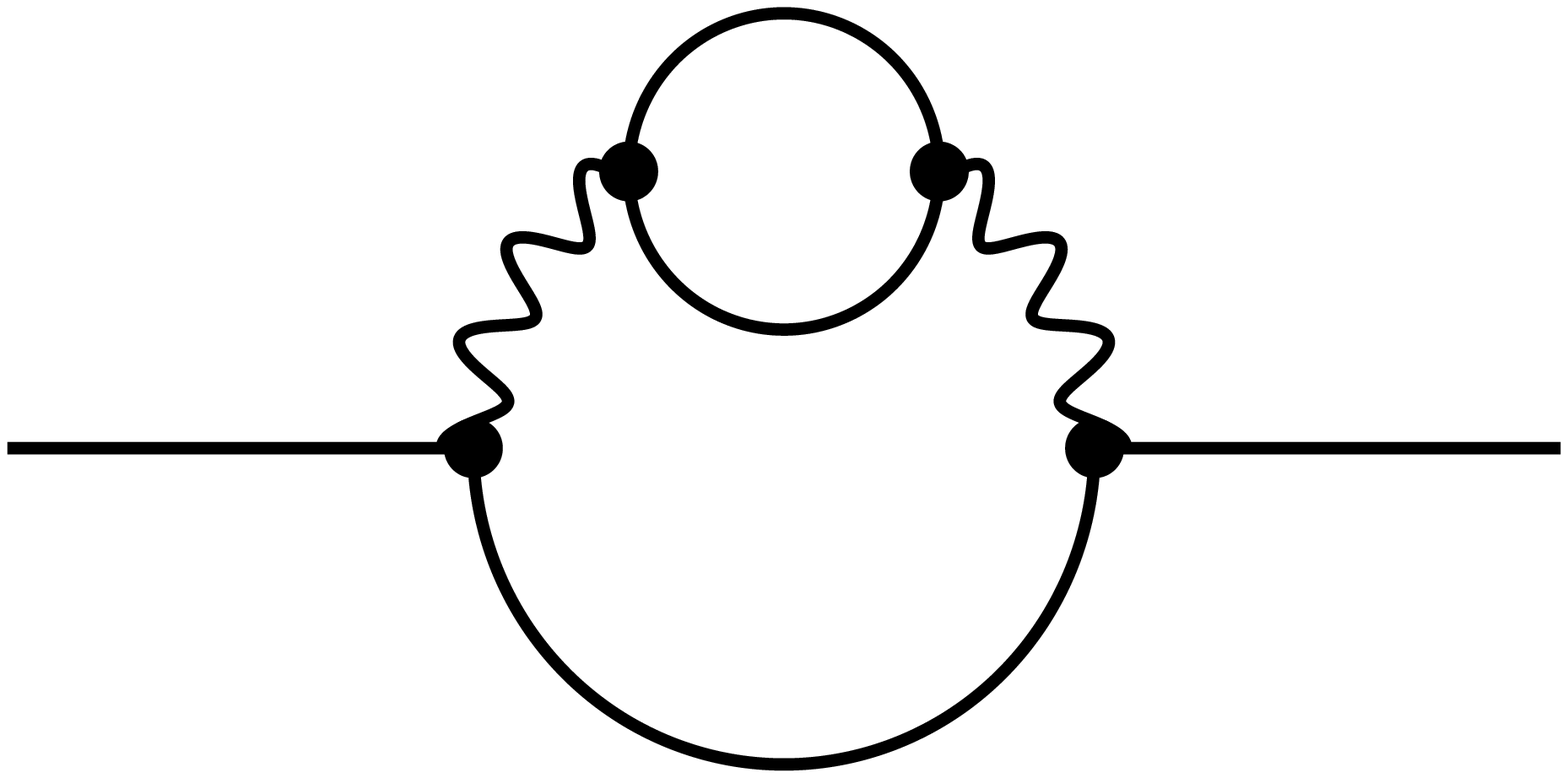}}
\put(8.6,9.15){A3}
\put(12.6,7.5){\includegraphics[scale=0.16,clip]{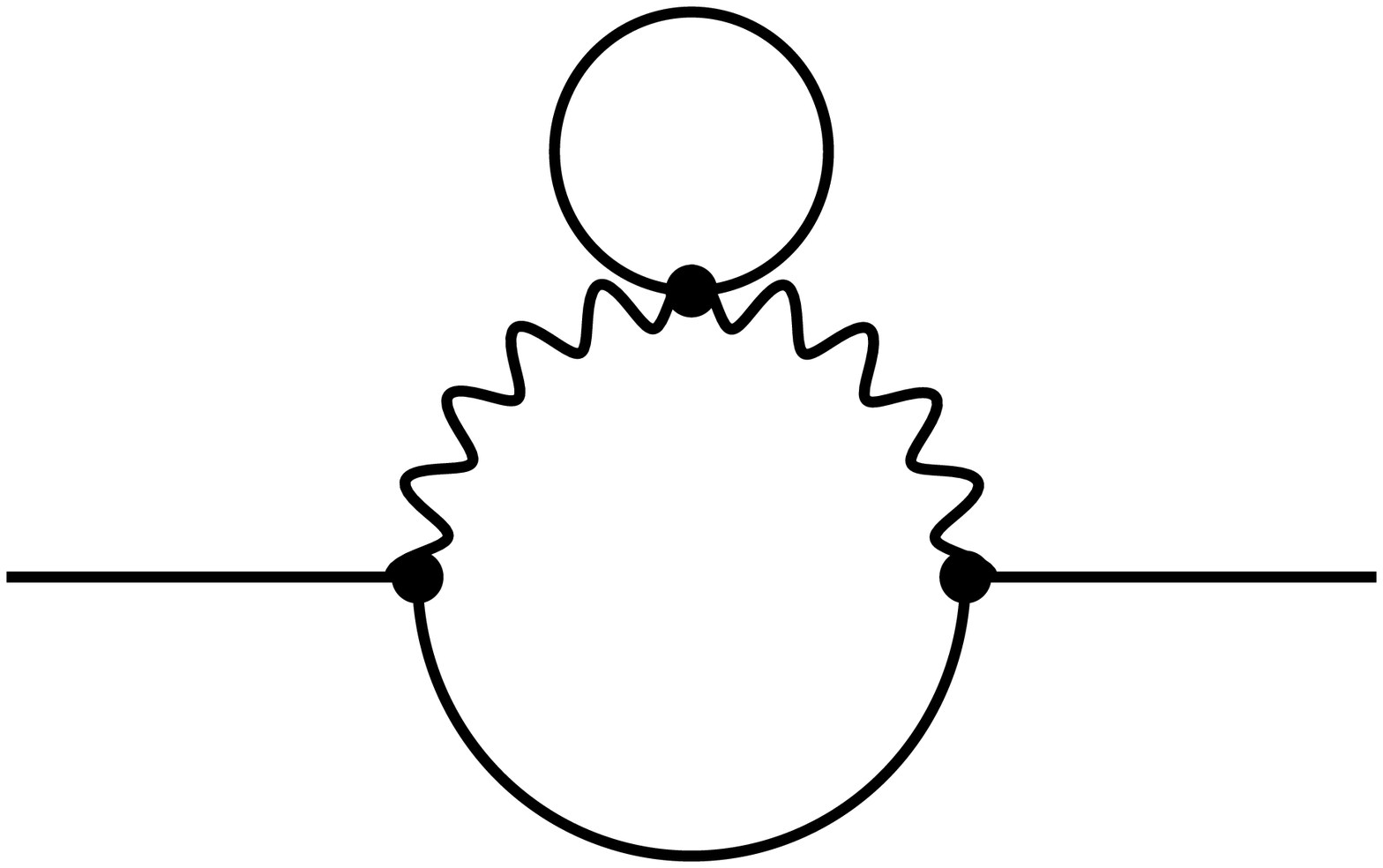}}
\put(12.6,9.15){A4}

\put(0.6,5){\includegraphics[scale=0.15,clip]{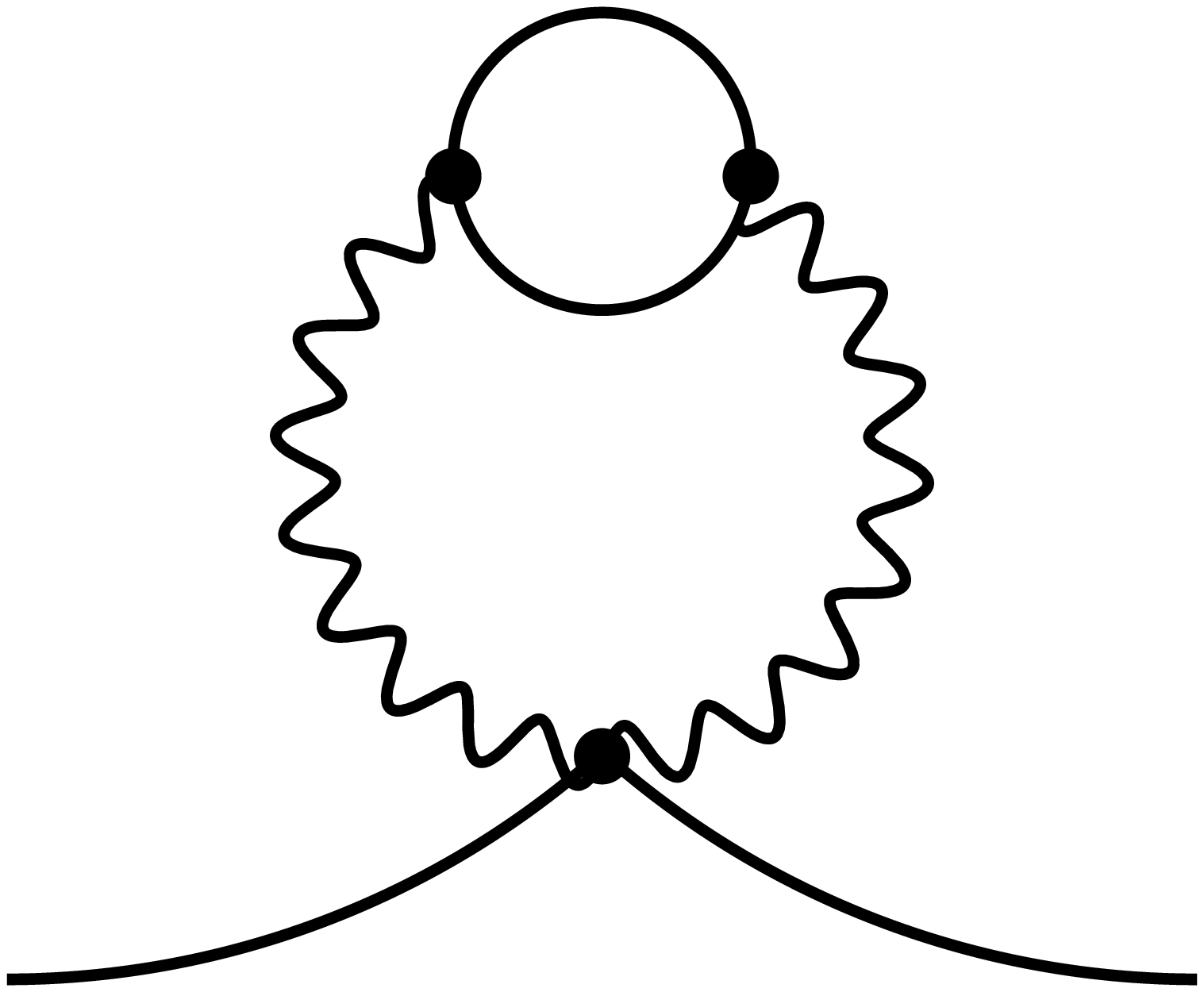}}
\put(0.6,6.7){A5}
\put(4.2,5){\includegraphics[scale=0.15,clip]{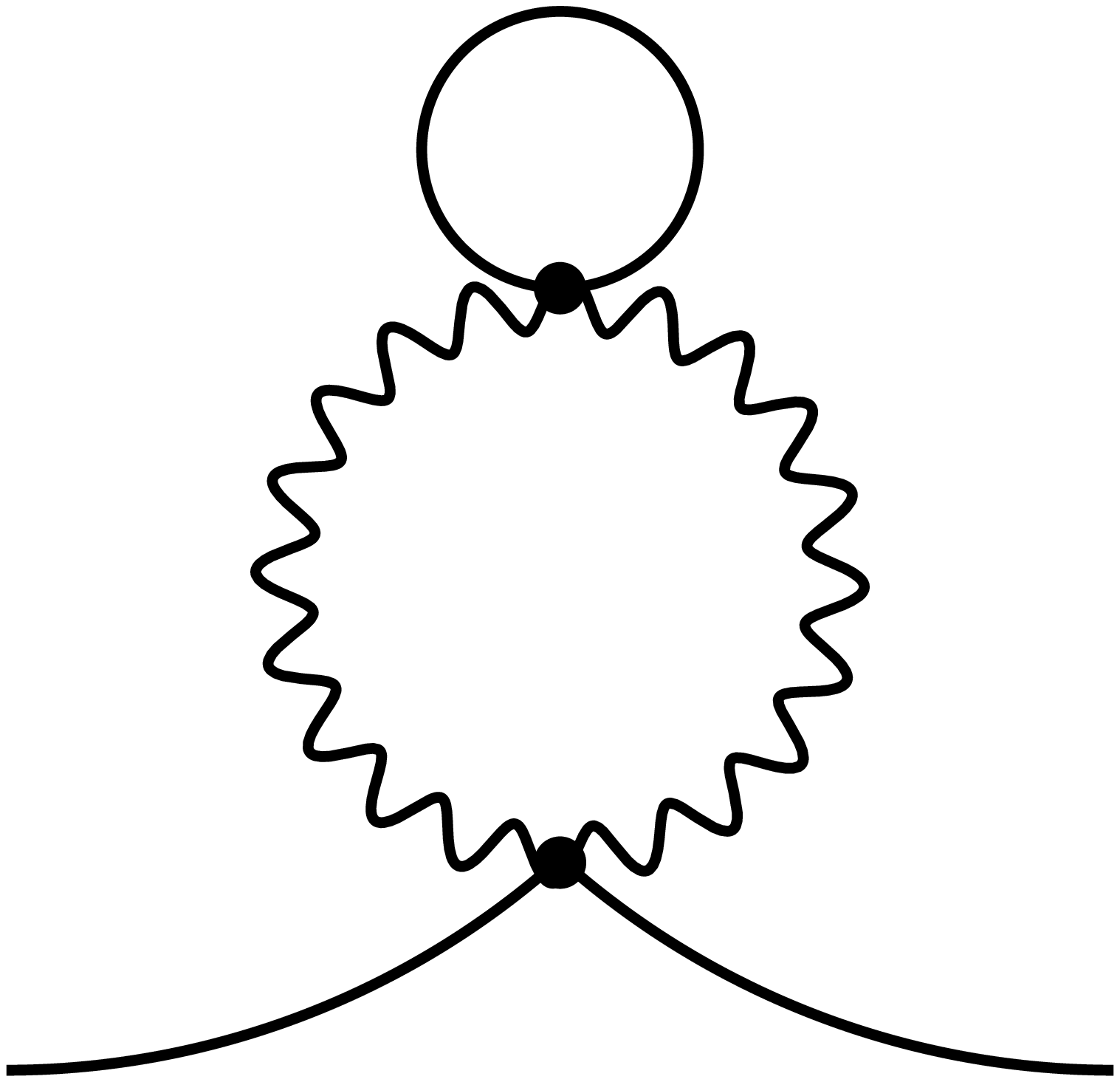}}
\put(4.2,6.7){A6}
\put(7.4,5){\includegraphics[scale=0.22,clip]{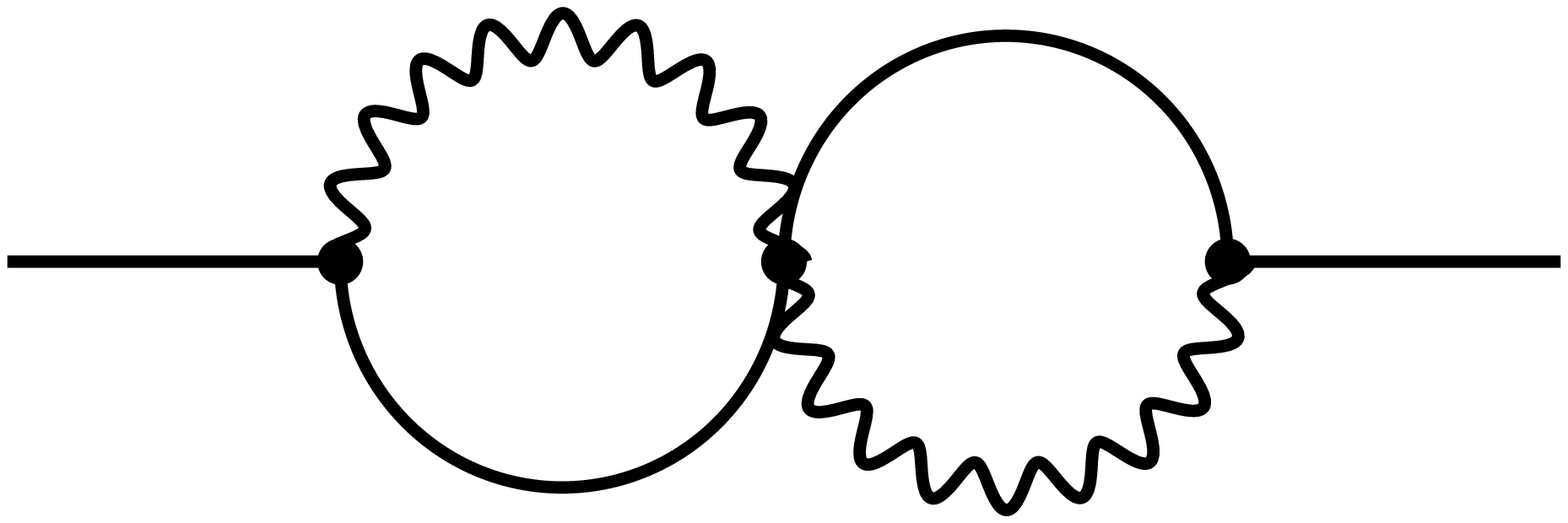}}
\put(7.4,6.7){A7}
\put(12.6,5){\includegraphics[scale=0.16,clip]{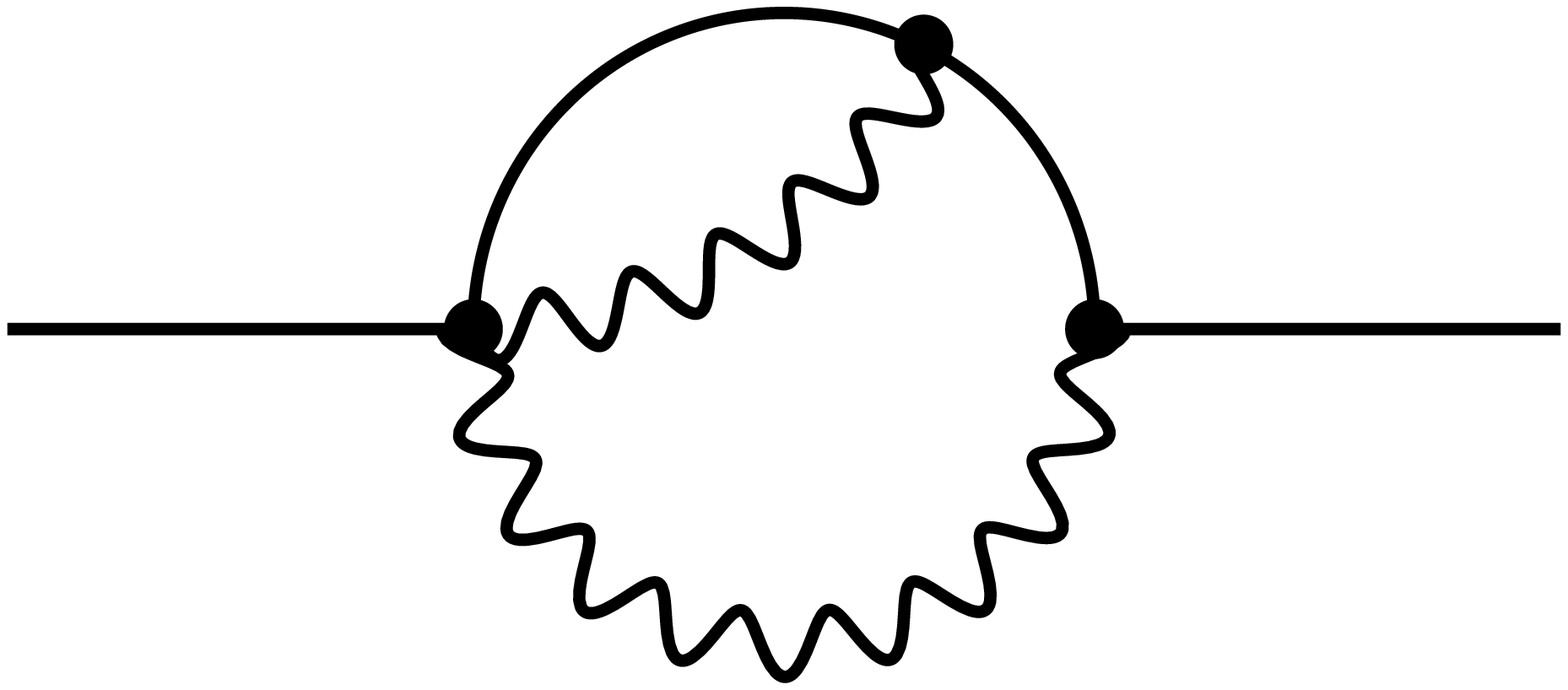}}
\put(12.6,6.7){A8}

\put(0.6,2.5){\includegraphics[scale=0.16,clip]{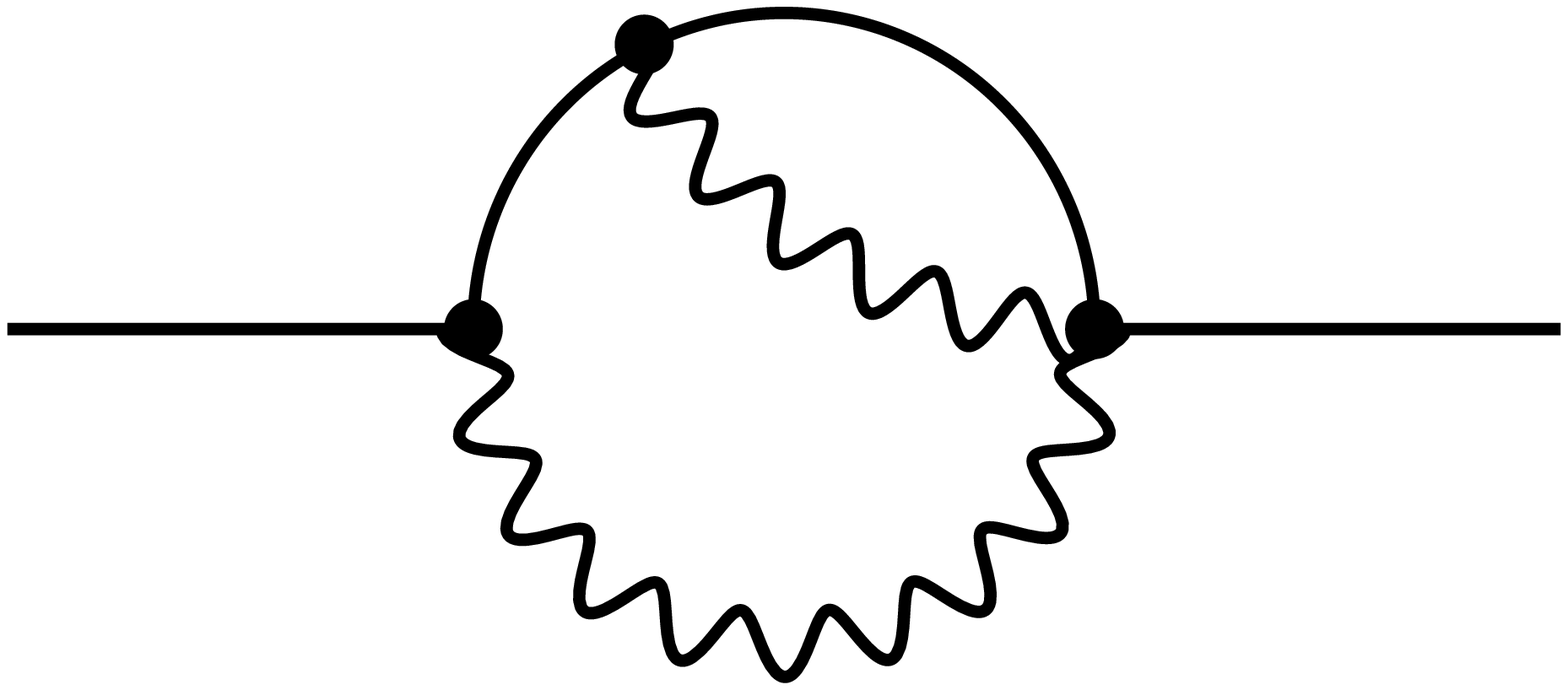}}
\put(0.6,4.15){A9}
\put(4.6,2.5){\includegraphics[scale=0.16,clip]{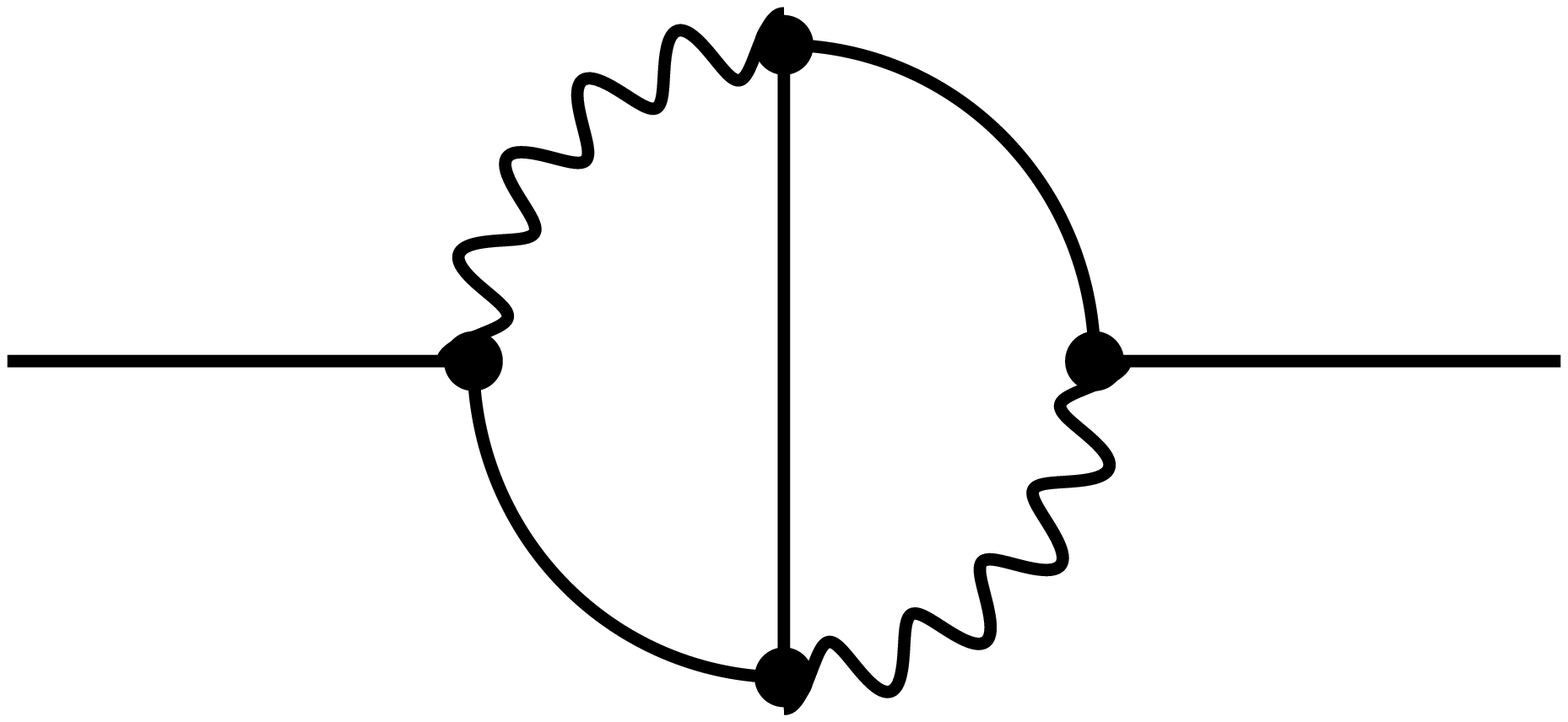}}
\put(4.6,4.15){A10}
\put(8.6,2.5){\includegraphics[scale=0.16,clip]{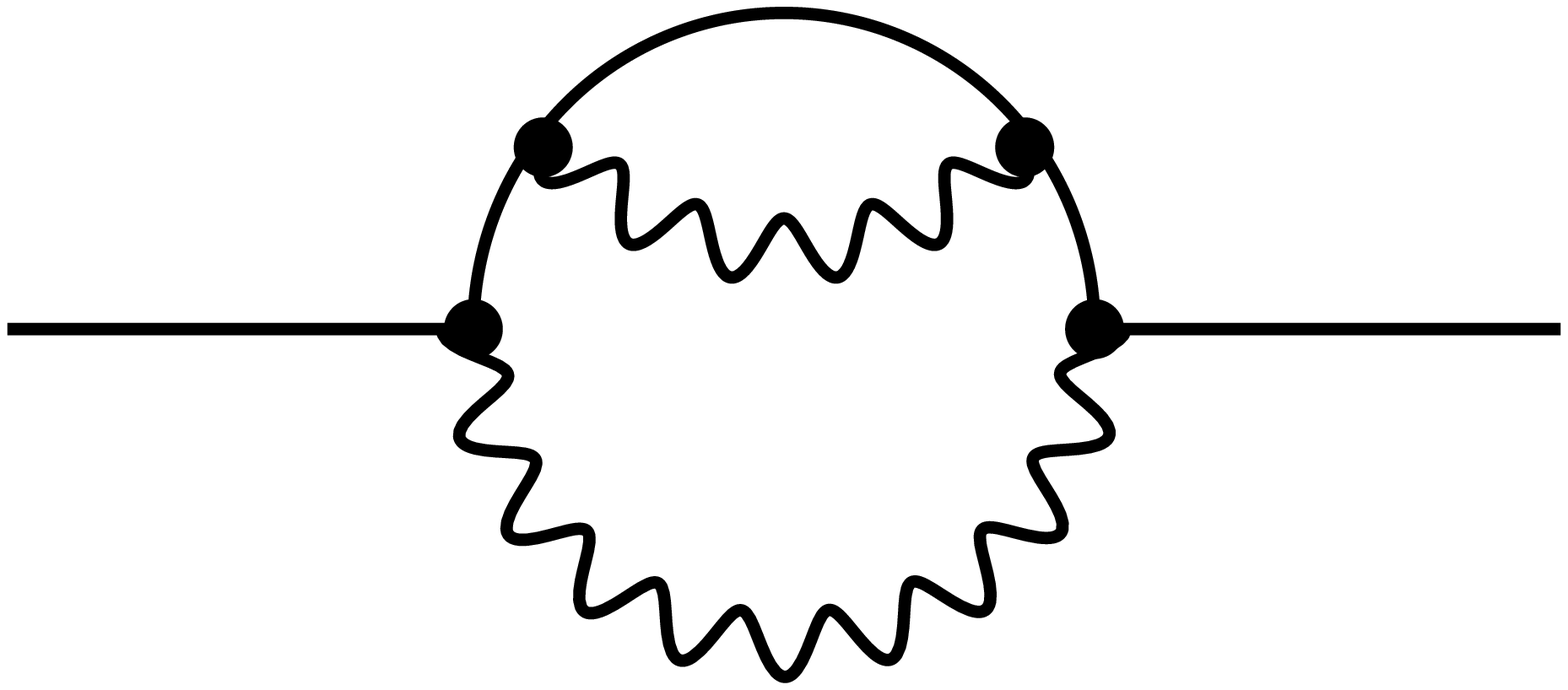}}
\put(8.6,4.15){A11}
\put(12.6,2.5){\includegraphics[scale=0.16,clip]{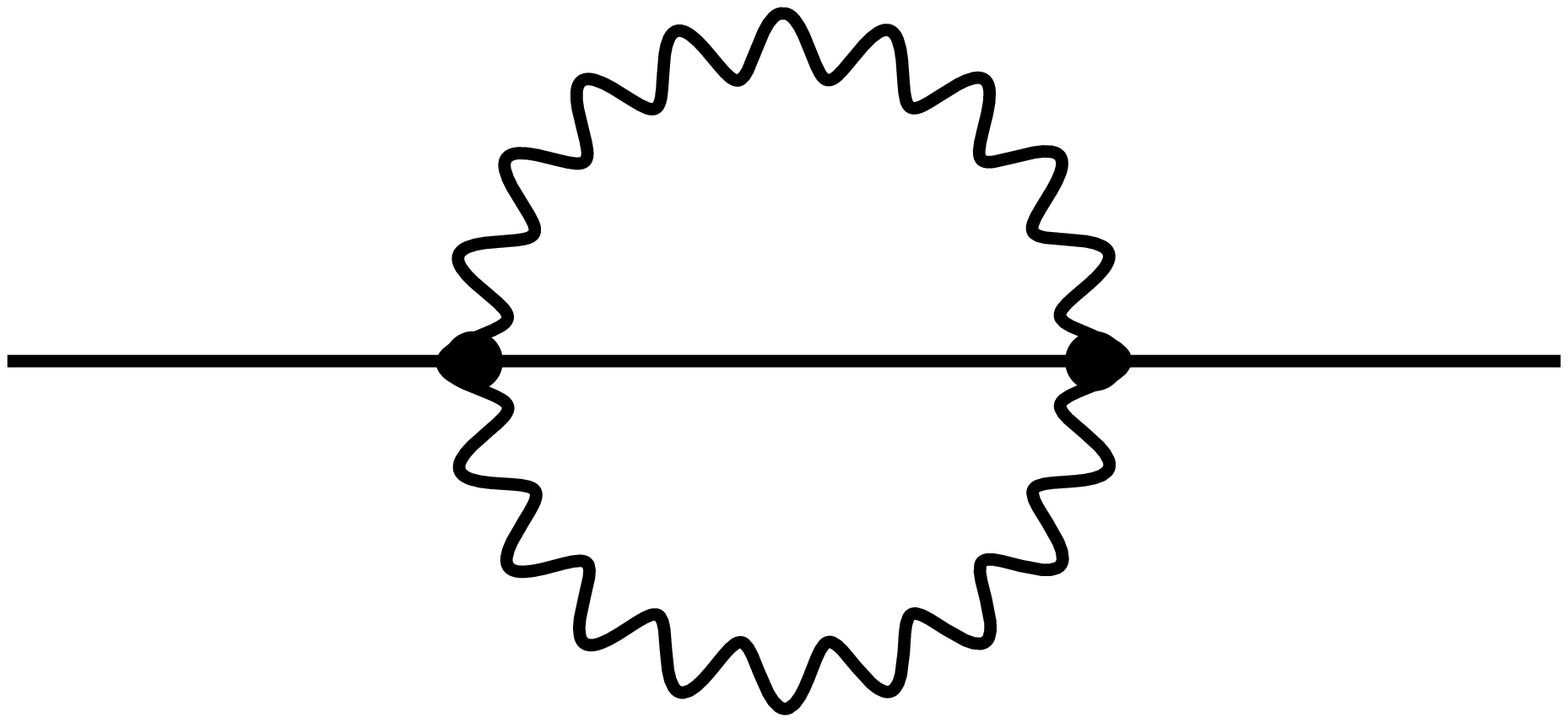}}
\put(12.6,4.15){A12}

\put(0.6,0){\includegraphics[scale=0.16,clip]{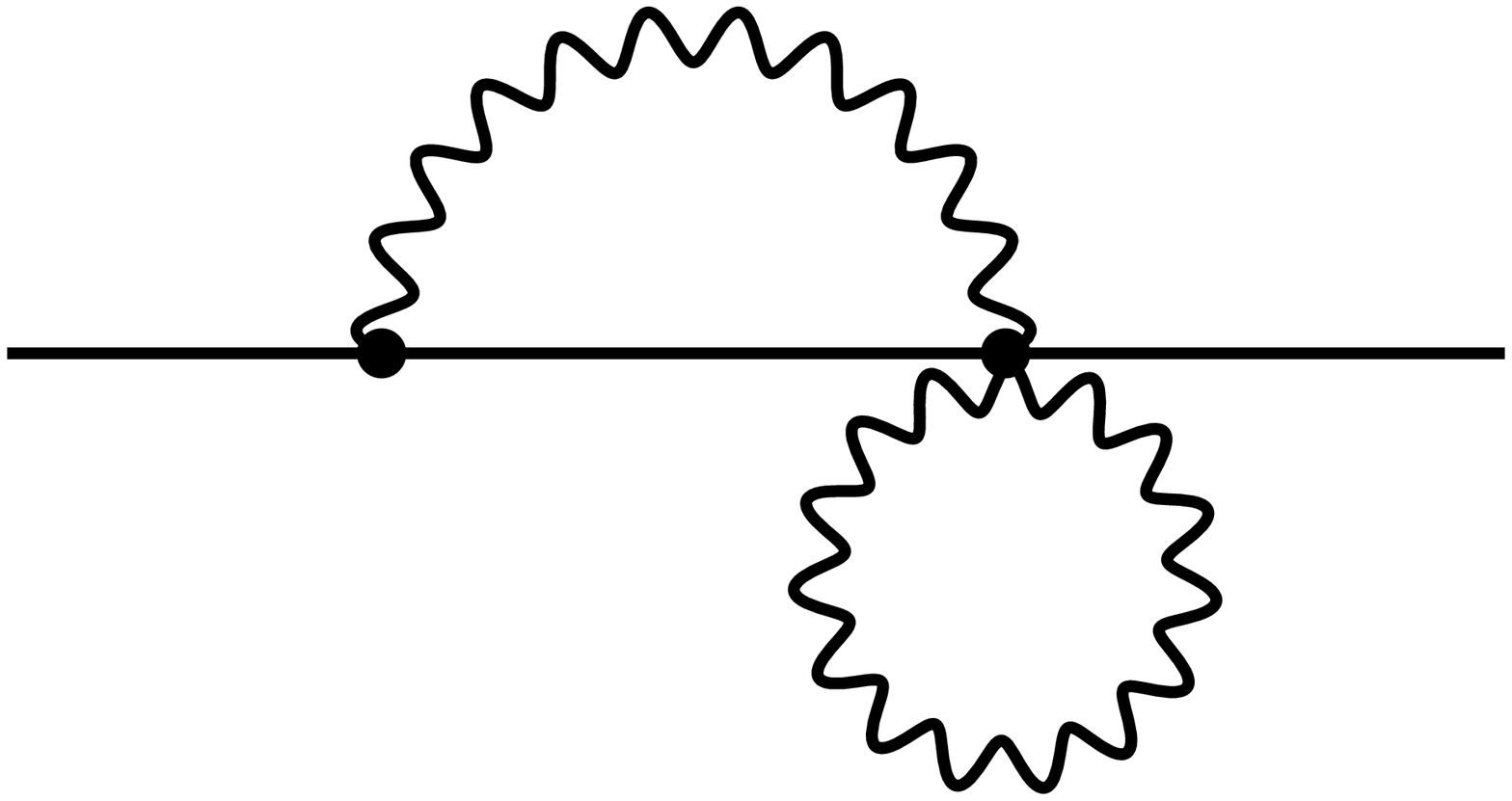}}
\put(0.6,1.7){A13}
\put(4.6,0){\includegraphics[scale=0.16,clip]{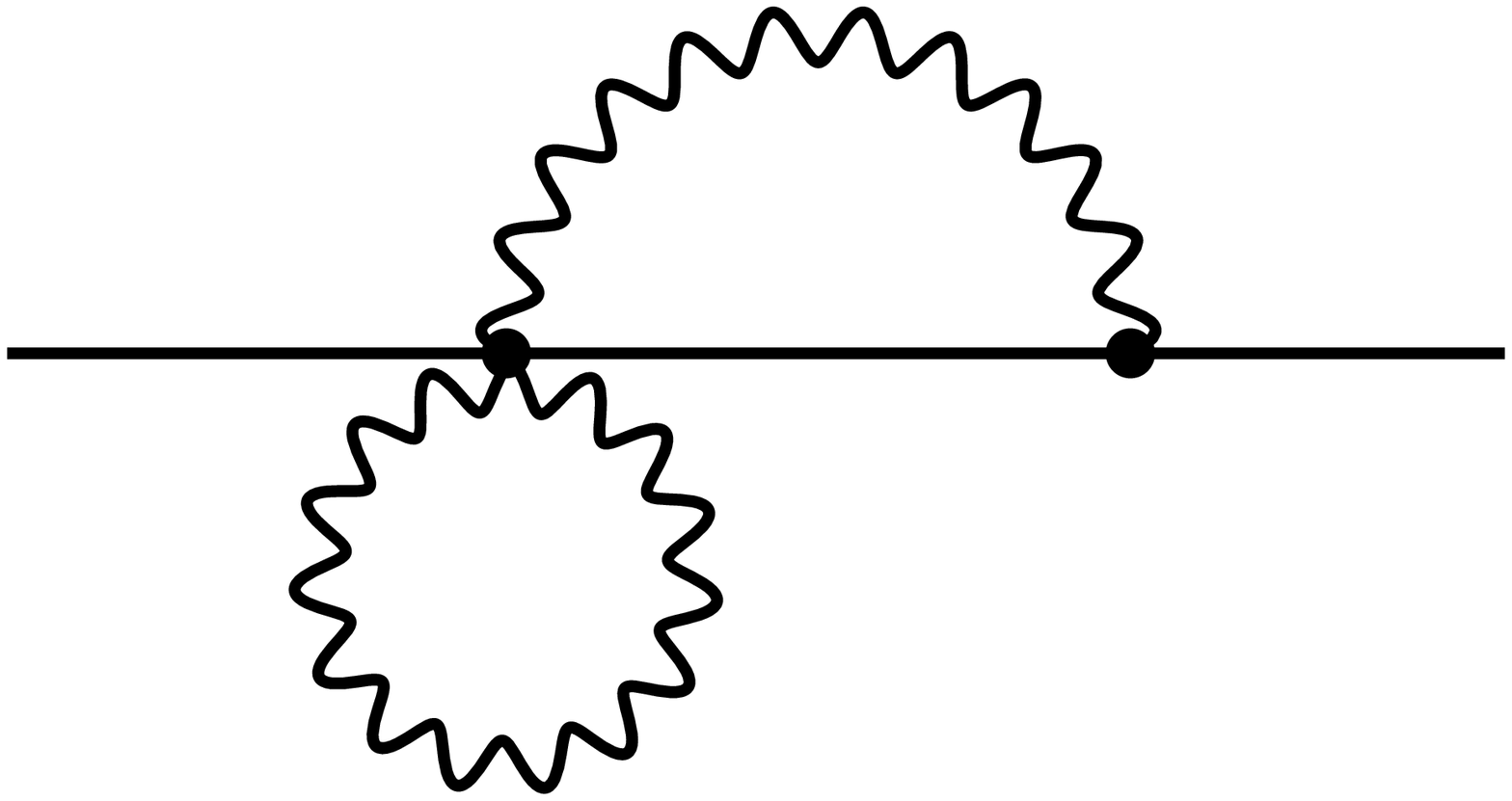}}
\put(4.6,1.7){A14}
\put(8.8,-0.2){\includegraphics[scale=0.14,clip]{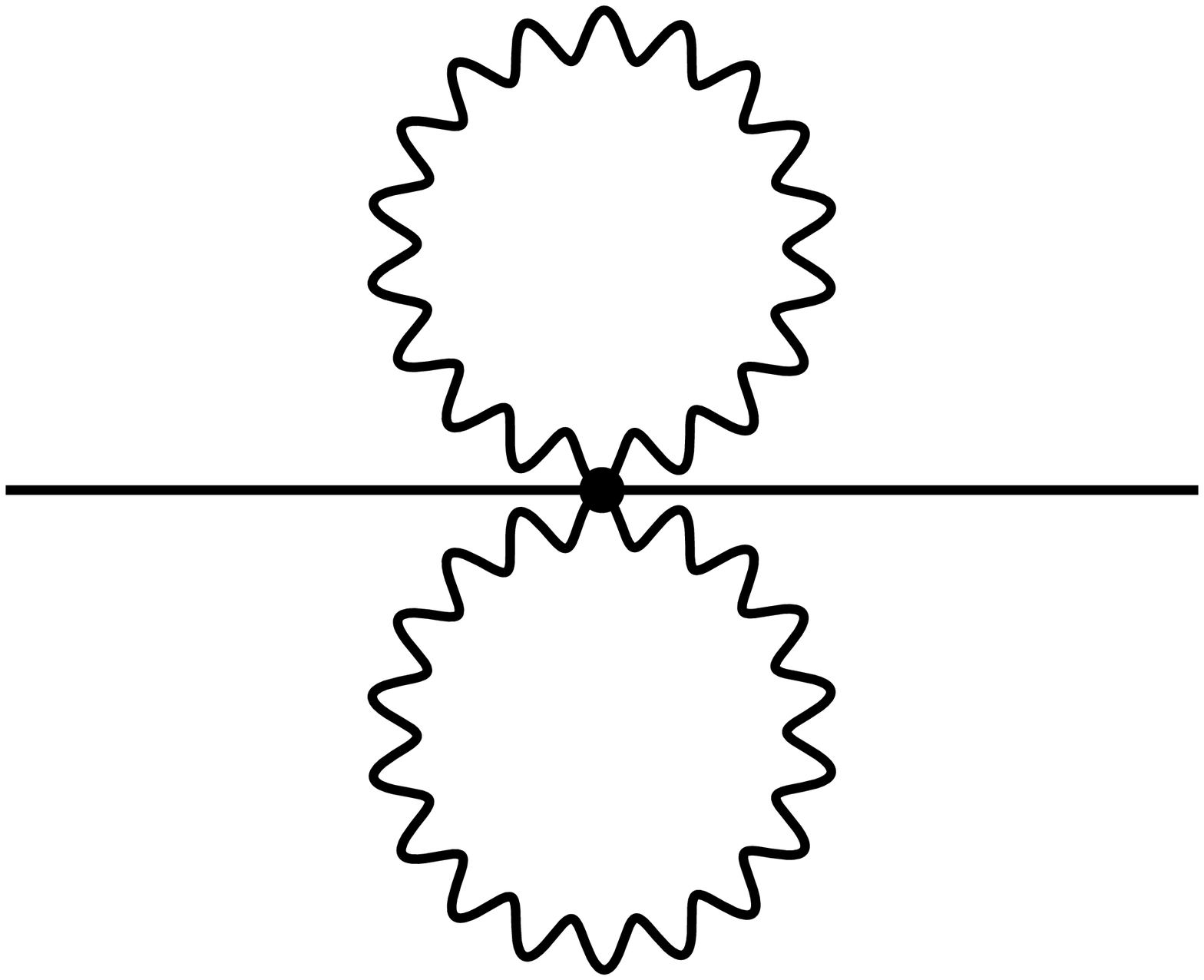}}
\put(8.6,1.7){A15}
\put(12.6,-0.02){\includegraphics[scale=0.16,clip]{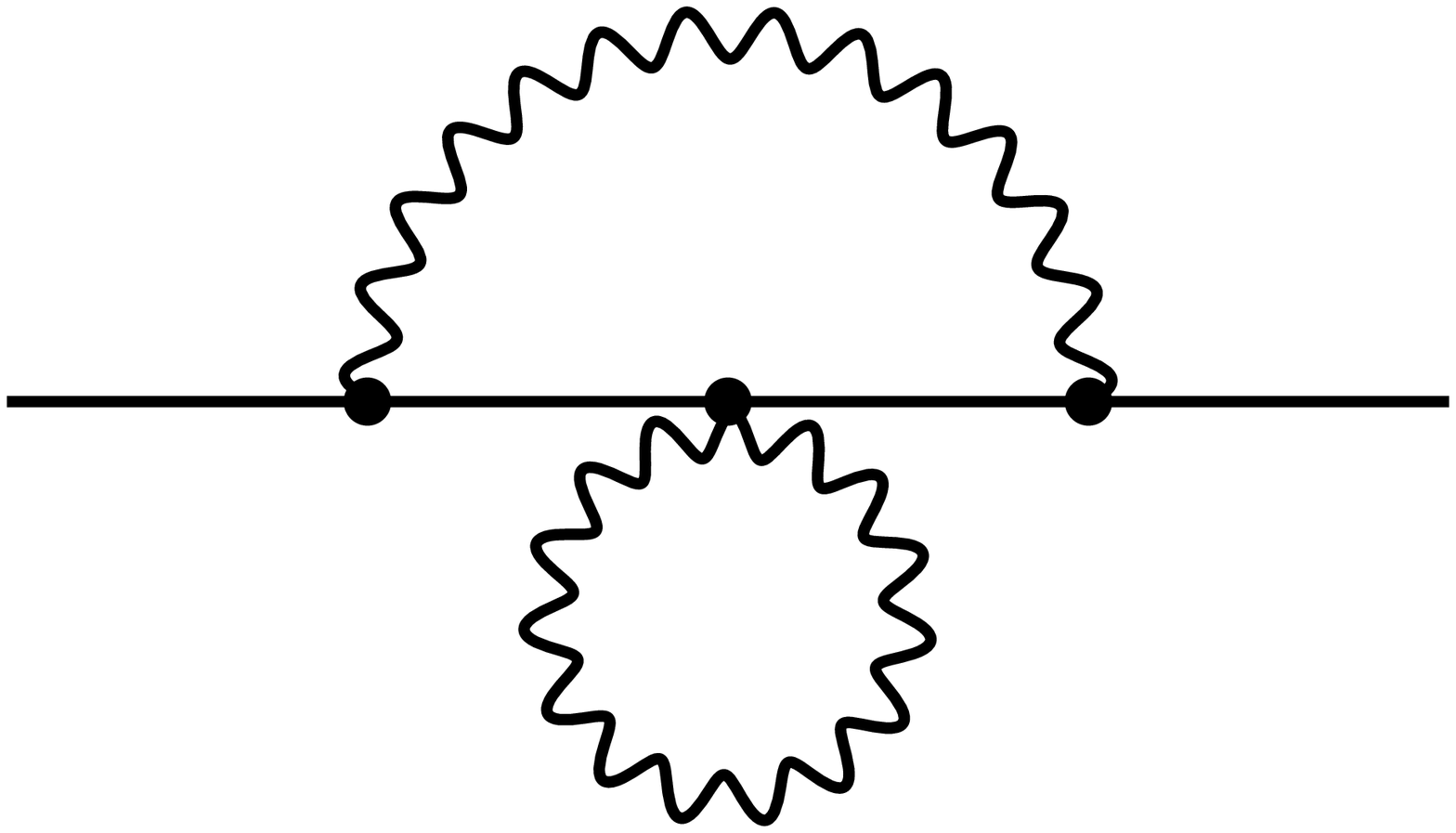}}
\put(12.6,1.7){A16}
\end{picture}
\caption{The one- and two-loop supergraphs contributing to the function $G$. The left and right external lines correspond to $\phi^*$ and $\phi$ (or $\widetilde\phi^*$ and $\widetilde\phi$), respectively.} \label{Figure_Two-Point_Matter}
\end{figure}

By construction, all expressions for the supergraphs S1 --- S13 presented in Appendix \ref{Appendix_Beta_Supergraphs} are integrals of double total derivatives. According to the general argumentation of Ref. \cite{Smilga:2004zr}, calculating the integral over $d^4Q$ we should obtain certain contributions to the function $\ln G$, where $G$ was defined by Eq. (\ref{Two-Point_Functions}).\footnote{The generalization of this statement to the non-Abelian case has been constructed in Ref. \cite{Stepanyantz:2016gtk}.} The relevant (one- and two-loop) supergraphs contributing to the function $G$ are presented in Fig. \ref{Figure_Two-Point_Matter},

\begin{equation}
G \equiv 1+\Delta G = 1 + \sum\limits_{i=1}^{16} \Delta_{\mbox{\scriptsize A}i} G + O(\alpha_0^3).
\end{equation}

\noindent
Cutting a matter line in the vacuum supergraphs we obtain either 1PI supergraph present in Fig.~\ref{Figure_Two-Point_Matter} or a supergraph which is not 1PI. In the latter case (namely, for the supergraphs S8, S9, and S13) it is necessary to cut one more matter line and obtain two 1PI supergraphs which are present in Fig.~\ref{Figure_Two-Point_Matter}. They correspond to the second term in the series expansion of $\ln G = \Delta G - (\Delta G)^2/2 +\ldots$ The expressions for $\Delta_{\mbox{\scriptsize A}i}G$ (for the vanishing external momentum) are listed in Appendix \ref{Appendix_Delta_G}.

After calculating the expressions presented in Appendix \ref{Appendix_Beta_Supergraphs} and comparing them with the superdiagrams presented in Fig. \ref{Figure_Two-Point_Matter} we have obtained the relations

\begin{eqnarray}
&&\hspace*{-6mm} \Delta_{\mbox{\scriptsize S1}}\Big(\frac{\beta}{\alpha_0^2}\Big) = - \frac{N_f}{\pi} \frac{d}{d\ln\Lambda} \Delta_{\mbox{\scriptsize A1}} G\Big|_{q=0};\quad\ \
\Delta_{\mbox{\scriptsize S2}}\Big(\frac{\beta}{\alpha_0^2}\Big) = - \frac{N_f}{\pi} \frac{d}{d\ln\Lambda} \Delta_{\mbox{\scriptsize A2}} G \Big|_{q=0};\nonumber\\
&&\hspace*{-6mm} \Delta_{\mbox{\scriptsize S3}}\Big(\frac{\beta}{\alpha_0^2}\Big) = - \frac{N_f}{\pi} \frac{d}{d\ln\Lambda} \Delta_{\mbox{\scriptsize A3}} G\Big|_{q=0};\quad\ \
\Delta_{\mbox{\scriptsize S4}}\Big(\frac{\beta}{\alpha_0^2}\Big) = - \frac{N_f}{\pi} \frac{d}{d\ln\Lambda} \Big(\Delta_{\mbox{\scriptsize A4}} G + \Delta_{\mbox{\scriptsize A5}} G\Big)\Big|_{q=0};\nonumber\\
&&\hspace*{-6mm} \Delta_{\mbox{\scriptsize S5}}\Big(\frac{\beta}{\alpha_0^2}\Big) = - \frac{N_f}{\pi} \frac{d}{d\ln\Lambda} \Delta_{\mbox{\scriptsize A6}} G\Big|_{q=0};\quad\ \
\Delta_{\mbox{\scriptsize S6}}\Big(\frac{\beta}{\alpha_0^2}\Big) = - \frac{N_f}{\pi} \frac{d}{d\ln\Lambda} \Big(\Delta_{\mbox{\scriptsize A7}} G + \Delta_{\mbox{\scriptsize A8}} G + \Delta_{\mbox{\scriptsize A9}} G\Big)\Big|_{q=0};\nonumber\\
&&\hspace*{-6mm} \Delta_{\mbox{\scriptsize S7}}\Big(\frac{\beta}{\alpha_0^2}\Big) = - \frac{N_f}{\pi} \frac{d}{d\ln\Lambda} \Delta_{\mbox{\scriptsize A10}} G\Big|_{q=0};\quad\
\Delta_{\mbox{\scriptsize S8}}\Big(\frac{\beta}{\alpha_0^2}\Big) = - \frac{N_f}{\pi} \frac{d}{d\ln\Lambda} \Big(\Delta_{\mbox{\scriptsize A11}} G - \frac{1}{2} \big(\Delta_{\mbox{\scriptsize A1}} G\big)^2\Big)\Big|_{q=0};
\nonumber\\
&&\hspace*{-6mm} \Delta_{\mbox{\scriptsize S9}}\Big(\frac{\beta}{\alpha_0^2}\Big) = - \frac{N_f}{\pi} \frac{d}{d\ln\Lambda} \Big( - \frac{1}{2} \big(\Delta_{\mbox{\scriptsize A2}} G\big)^2\Big)\Big|_{q=0};\qquad\quad\ \
\Delta_{\mbox{\scriptsize S10}}\Big(\frac{\beta}{\alpha_0^2}\Big) = - \frac{N_f}{\pi} \frac{d}{d\ln\Lambda} \Delta_{\mbox{\scriptsize A12}} G\Big|_{q=0};\nonumber\\
&&\hspace*{-6mm} \Delta_{\mbox{\scriptsize S11}}\Big(\frac{\beta}{\alpha_0^2}\Big) = - \frac{N_f}{\pi} \frac{d}{d\ln\Lambda} \Big(\Delta_{\mbox{\scriptsize A13}} G + \Delta_{\mbox{\scriptsize A14}} G\Big)\Big|_{q=0};\qquad
\Delta_{\mbox{\scriptsize S12}}\Big(\frac{\beta}{\alpha_0^2}\Big) = - \frac{N_f}{\pi} \frac{d}{d\ln\Lambda} \Delta_{\mbox{\scriptsize A15}} G\Big|_{q=0};\nonumber\\
&&\hspace*{-6mm} \Delta_{\mbox{\scriptsize S13}}\Big(\frac{\beta}{\alpha_0^2}\Big) = - \frac{N_f}{\pi} \frac{d}{d\ln\Lambda} \Big(\Delta_{\mbox{\scriptsize A16}} G - \Delta_{\mbox{\scriptsize A1}} G\cdot \Delta_{\mbox{\scriptsize A2}} G\Big)\Big|_{q=0}.
\end{eqnarray}

\noindent
The sum of these relations gives the equation

\begin{equation}
\frac{\beta(\alpha_0)}{\alpha_0^2} - \frac{\beta_{\mbox{\scriptsize 1-loop}}(\alpha_0)}{\alpha_0^2} = - \frac{N_f}{\pi} \frac{d\ln G}{d\ln\Lambda}\bigg|_{q=0} + O(\alpha_0^3) = -\frac{N_f}{\pi}\gamma(\alpha_0) + O(\alpha_0^3),
\end{equation}

\noindent
which is valid even at the level of loop integrals. Substituting the well-known expression for the one-loop $\beta$-function $\beta_{\mbox{\scriptsize 1-loop}}(\alpha_0) = N_f \alpha_0^2/\pi$ we obtain the NSVZ equation for RGFs defined in terms of the bare coupling constant,

\begin{equation}
\frac{\beta(\alpha_0)}{\alpha_0^2} = \frac{N_f}{\pi}\Big(1-\gamma(\alpha_0)\Big) + O(\alpha_0^3).
\end{equation}

\noindent
As we discussed above, both sides of this equation are gauge independent and coincide with the corresponding expressions in the Feynman gauge. The integrals giving the anomalous dimension of the matter superfields defined in terms of the bare coupling constant in the case of using the higher derivative regularization are given by Eq. (\ref{Two-Loop_Anomalous_Dimension_Integral}) in Appendix \ref{Appendix_Delta_G}. They have already been taken in Ref. \cite{Soloshenko:2003sx} (for $N_f=1$ and $R(x)=1+x^n$), where the calculation was done in the Feynman gauge. For arbitrary $N_f$ the result can be found, e.g., in \cite{Kataev:2013csa,Kataev:2014gxa}. For the regularization used in this paper with $R(x)=1+x^n$ it can be written as

\begin{equation}\label{Two-Loop_Anomalous_Dimension_Result}
\gamma(\alpha_0) = -\frac{\alpha_0}{\pi} + \frac{\alpha_0^2}{\pi^2} \Big(N_f \ln a + N_f + \frac{1}{2}\Big) + O(\alpha_0^3),
\end{equation}

\noindent
where $a=M/\Lambda$ is a regularization parameter. This implies that the $\beta$-function defined in terms of the bare coupling constant takes the form

\begin{equation}
\beta(\alpha_0) = \frac{\alpha_0^2 N_f}{\pi}\Big[1+\frac{\alpha_0}{\pi} + \frac{\alpha_0^2}{\pi^2} \Big( -N_f \ln a - N_f - \frac{1}{2}\Big) + O(\alpha_0^3)\Big].
\end{equation}

For completeness, we also present the expressions for RGFs defined in terms of the renormalized coupling constant in the HD+MSL scheme,

\begin{eqnarray}
&& \widetilde\gamma(\alpha) = -\frac{\alpha}{\pi} + \frac{\alpha^2}{\pi^2} (N_f \ln a + N_f + \frac{1}{2}) + O(\alpha^3);\\
&& \widetilde\beta(\alpha) = \frac{\alpha^2 N_f}{\pi}\Big[1+\frac{\alpha}{\pi} + \frac{\alpha^2}{\pi^2} \Big( -N_f \ln a - N_f - \frac{1}{2}\Big) + O(\alpha^3)\Big].\quad
\end{eqnarray}

\noindent
The expressions for RGFs obtained with other renormalization prescriptions can be found in Refs. \cite{Kataev:2013csa,Kataev:2014gxa}.

Thus, we see that the algorithm proposed in \cite{Stepanyantz:2019ihw} really gives the correct result for the three-loop $\beta$-function of the theory under consideration. This confirms its correctness and a possibility to use it for doing more complicated calculations.

\section*{Conclusion}
\hspace*{\parindent}

In this paper we have compared two different methods for calculating the $\beta$-function of ${\cal N}=1$ supersymmetric gauge theories regularized by higher covariant derivatives. Namely, we considered ${\cal N}=1$ SQED as an example and obtained its three-loop $\beta$-function (defined in terms of the bare coupling constant) in a general $\xi$-gauge using the technique of Ref. \cite{Stepanyantz:2019ihw}. In the Abelian case it is similar to the one proposed in \cite{Smilga:2004zr}.  The algorithm described in \cite{Kuzmichev:2019ywn,Stepanyantz:2019lyo} essentially simplifies the calculations, because one should deal with only (specially modified) vacuum supergraphs. In contrast, the standard method requires calculating a large number of superdiagrams with two external lines of the gauge superfield $V$. After this, one should sum the results, verify the cancellation of noninvariant terms, and write the result in the form of an integral of double total derivatives. In the Feynman gauge this has been done earlier (see, e.g., \cite{Stepanyantz:2012zz} and references therein).

The results for the three-loop $\beta$-function obtained by both methods coincided.\footnote{Within the standard technique the sum of the three-loop superdiagrams containing the Pauli--Villars loops has been found only in the Feynman gauge, because the calculation for an arbitrary $\xi_0$ is very cumbersome.} This confirms that the new technique works correctly and justifies a possibility of its using for more complicated calculations in higher orders. Also this can be considered as a check of the general argumentation of Ref. \cite{Stepanyantz:2019ihw} used for the all-loop proof of the factorization of integrals giving the $\beta$-function into integrals of double total derivatives.

In the present paper we explicitly demonstrated the cancellation of gauge dependent terms in the three-loop $\beta$-function. Also we have verified that all gauge dependent terms vanish in the expression for the two-loop anomalous dimension of the matter superfields (also defined in terms of the bare coupling constant). Taking into account that Eq. (\ref{NSVZ_Bare}) is valid in the Feynman gauge, we see that the NSVZ relation in the considered approximation is also valid in an arbitrary $\xi$-gauge for an arbitrary renormalization prescription supplementing the higher derivative regularization. (Certainly, it is assumed that RGFs are defined in terms of the bare coupling constant.) This exactly agrees with the general results of Refs. \cite{Stepanyantz:2011jy,Stepanyantz:2014ima}.

All results for RGFs defined in terms of the bare coupling constant are also valid for RGFs defined in terms of the renormalized coupling constant in the HD+MSL scheme, because for this renormalization prescription both definitions of RGFs give the same functions up to the renaming of arguments \cite{Kataev:2013eta}.

\section*{Acknowledgements}
\hspace*{\parindent}

The work of M.K., N.M., S.N., I.S., V.Yu.S., and K.S. was supported by Foundation for Advancement of Theoretical Physics and Mathematics ``BASIS'', grants  17-11-120-42 (M.K.), 18-2-6-159-1 (N.M.), 18-2-6-158-1 (S.N.), 19-1-1-45-3 (I.S.), 19-1-1-45-4 (V.Yu.S.), and 19-1-1-45-1 (K.S.).

\appendix

\section{How the gauge dependence is cancelled for the supergraphs with the Pauli--Villars superfields}
\hspace*{\parindent}\label{Appendix_Identities}

\begin{figure}[h]
\begin{picture}(0,8.5)
\put(0.4,6.2){\includegraphics[scale=0.15,clip]{subdiagram1.eps}}
\put(1.75,7.8){$z$}
\put(0.2,7.0){$\bar D_w^2 D_w^2$}
\put(0.35,6){$x$} \put(1.45,6){$w$} \put(2.6,6){$y$}
\put(3.7,7.0){$+$}
\put(4.9,6.2){\includegraphics[scale=0.15,clip]{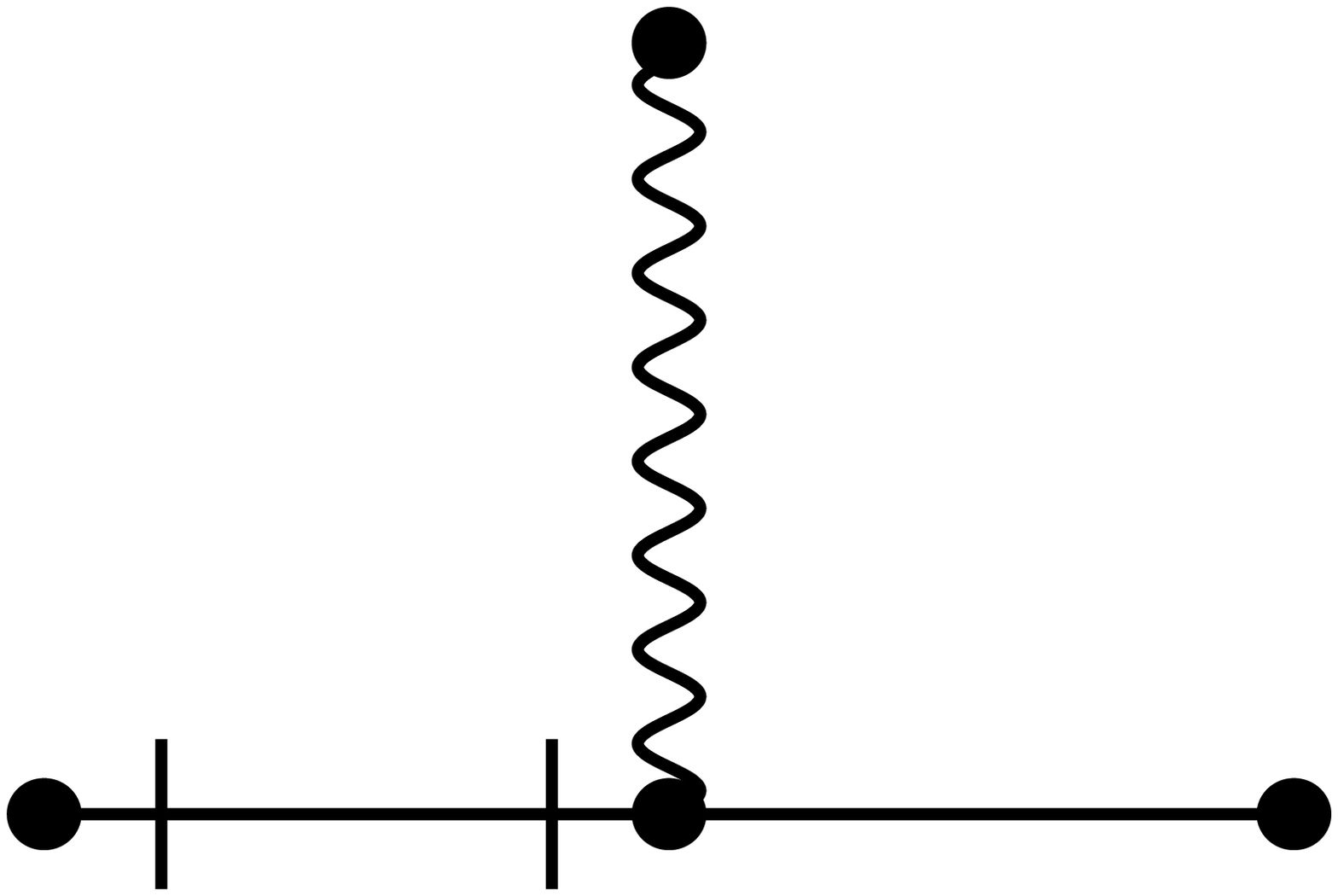}}
\put(6.25,7.8){$z$}
\put(4.7,7.0){$\bar D_w^2 D_w^2$}
\put(4.85,6){$x$} \put(5.95,6){$w$} \put(7.1,6){$y$}
\put(8.1,7.0){\vector(1,0){1}}
\put(9.9,6.2){\includegraphics[scale=0.15,clip]{subdiagram3.eps}}
\put(10.1,7.8){$z$}
\put(10.25,7.0){$\bar D_x^2 D_x^2$}
\put(9.85,6){$x$} \put(11.6,6){$y$}

\put(0.4,3.2){\includegraphics[scale=0.15,clip]{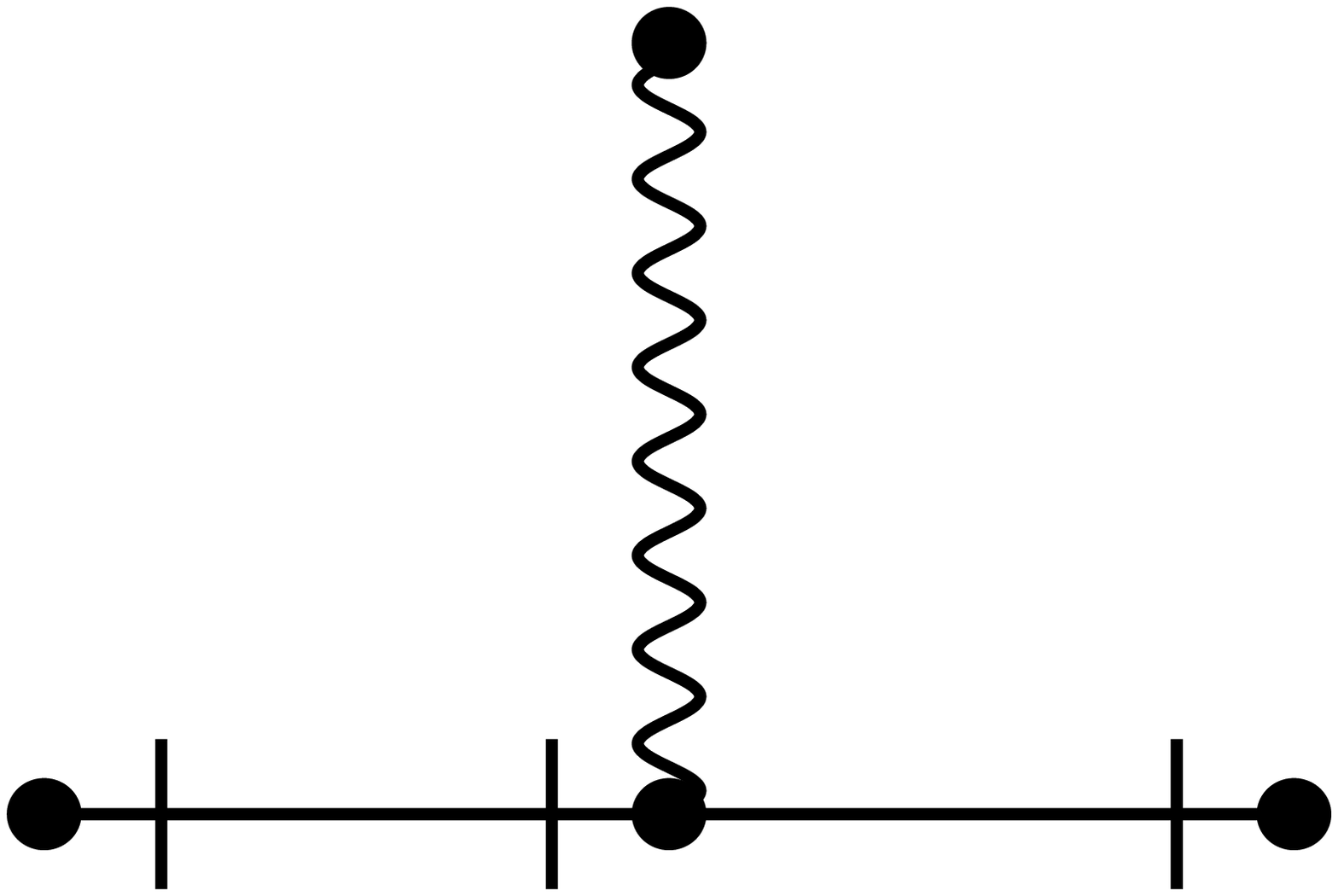}}
\put(1.75,4.8){$z$}
\put(0.2,4.0){$\bar D_w^2 D_w^2$}
\put(0.35,3){$x$} \put(1.45,3){$w$} \put(2.6,3){$y$}
\put(3.7,4.0){$+$}
\put(4.9,3.2){\includegraphics[scale=0.15,clip]{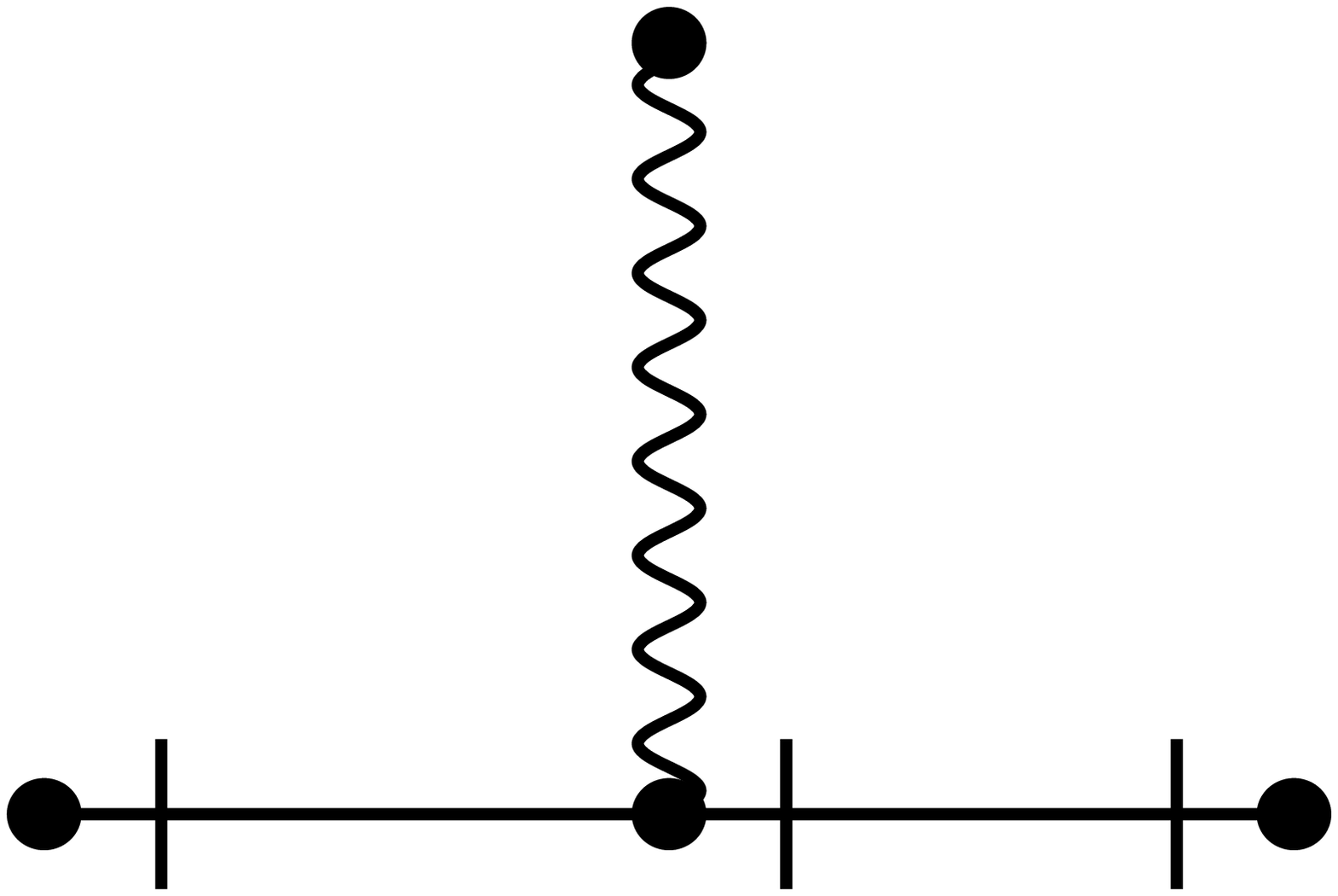}}
\put(6.25,4.8){$z$}
\put(4.7,4.0){$\bar D_w^2 D_w^2$}
\put(4.85,3){$x$} \put(5.95,3){$w$} \put(7.1,3){$y$}
\put(8.1,4.0){\vector(1,0){1}}
\put(9.9,3.2){\includegraphics[scale=0.15,clip]{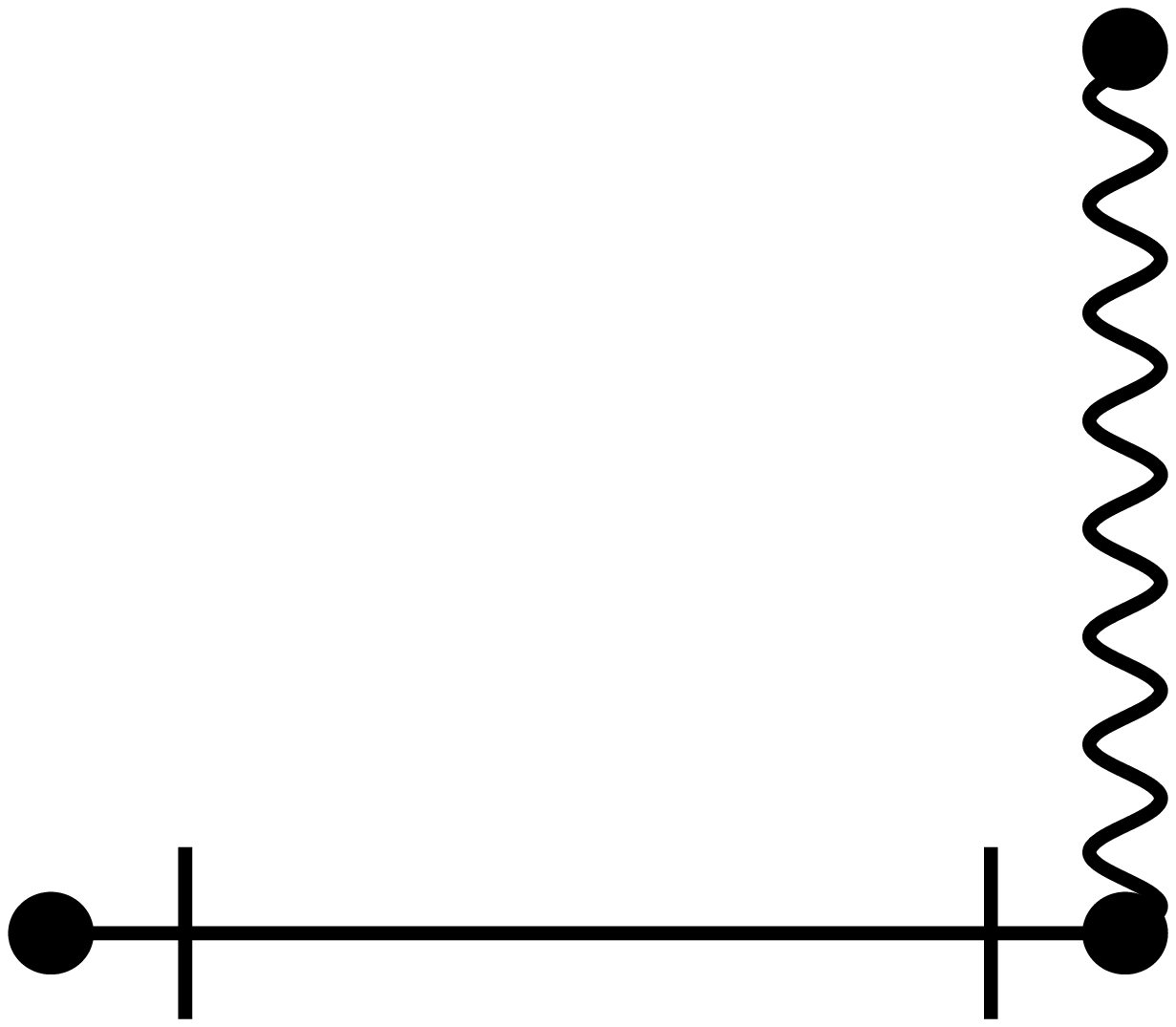}}
\put(9.85,3){$x$} \put(11.6,3){$y$}
\put(11.82,4.8){$z$}
\put(10.4,4.0){$\bar D_y^2 D_y^2$}
\put(12.54,4.0){$+$}
\put(13.7,3.2){\includegraphics[scale=0.15,clip]{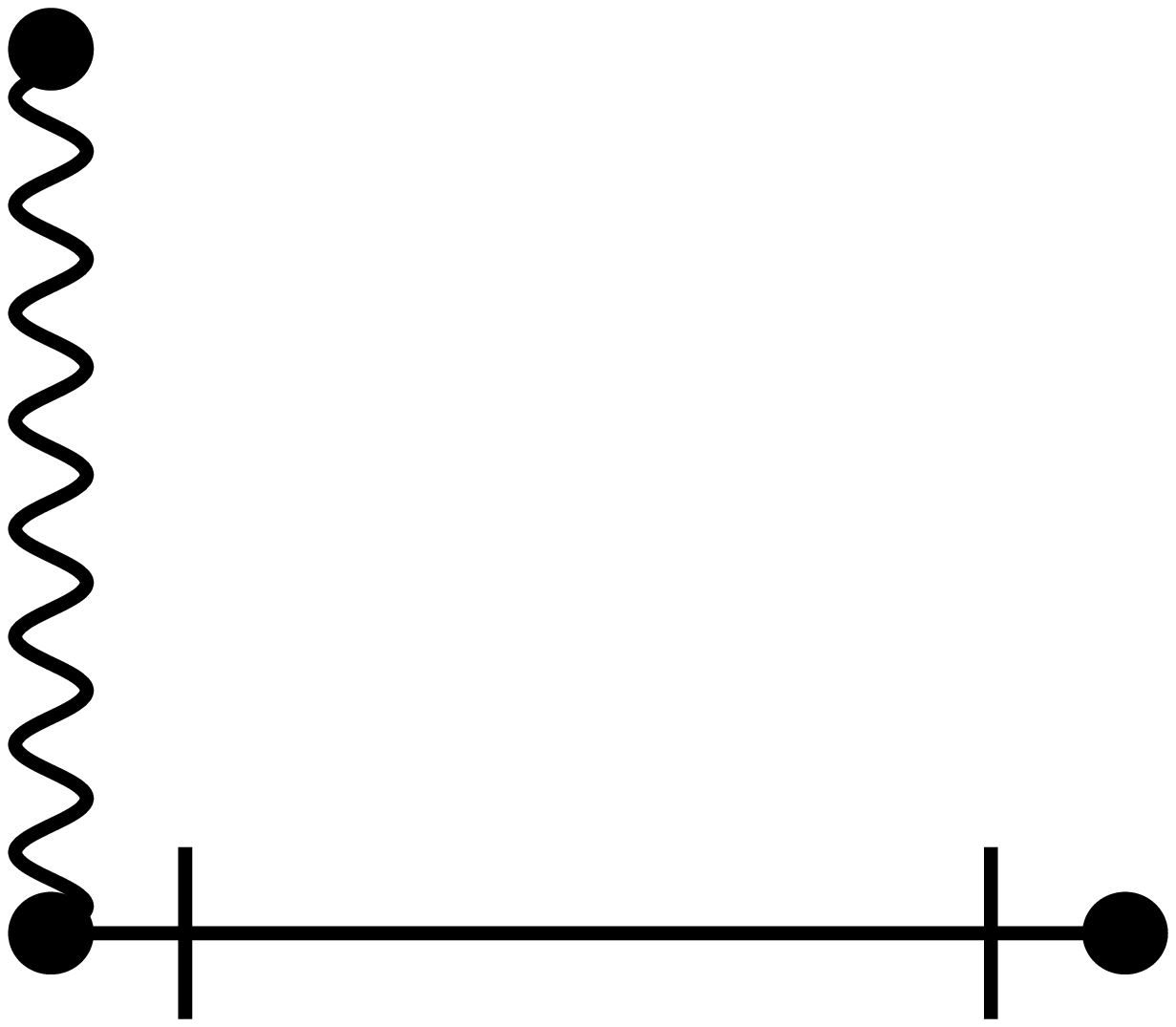}}
\put(13.65,3){$x$} \put(15.4,3){$y$}
\put(13.88,4.8){$z$}
\put(14.0,4.0){$\bar D_x^2 D_x^2$}

\put(0.4,0.2){\includegraphics[scale=0.15,clip]{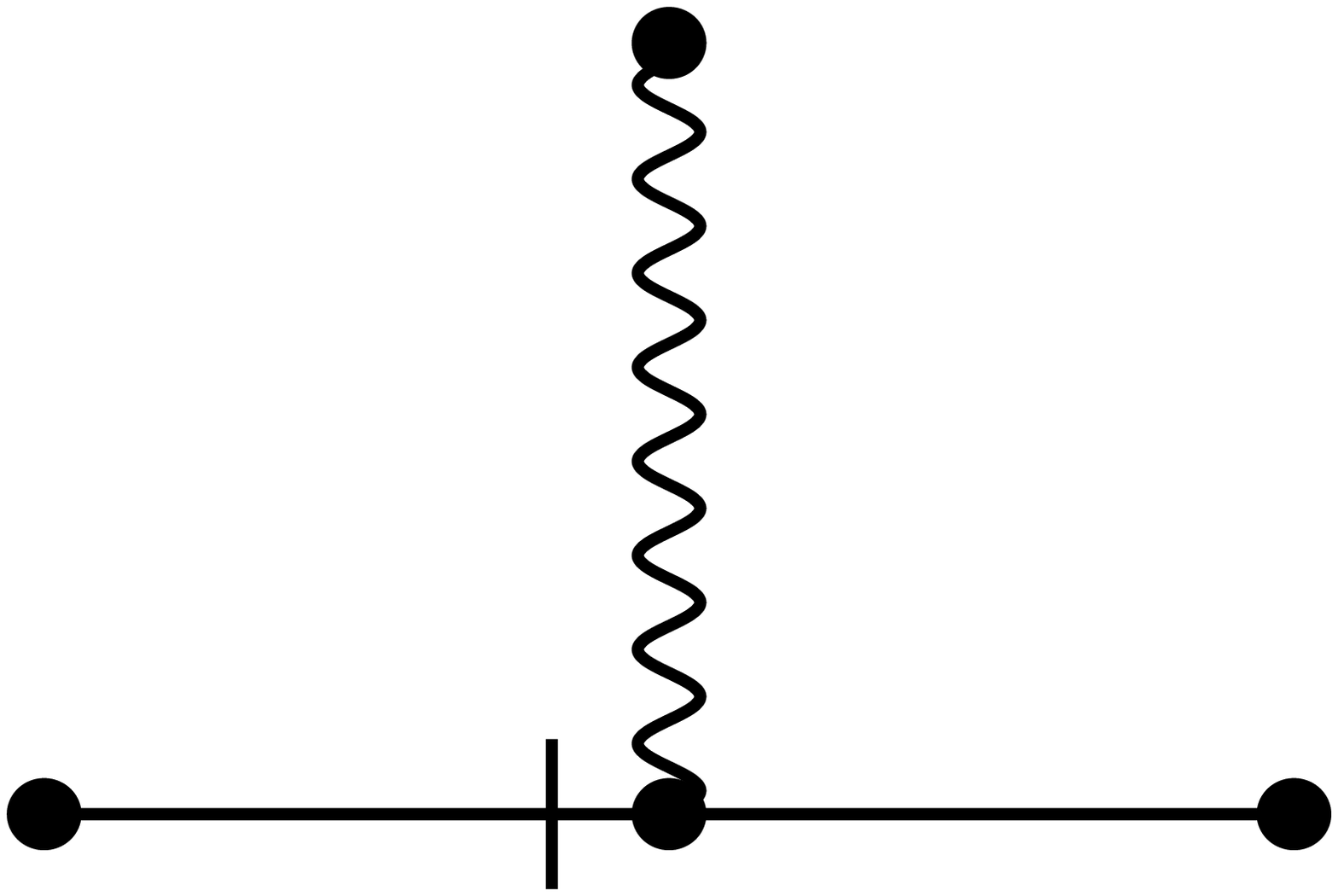}}
\put(1.75,1.8){$z$}
\put(0.2,1.0){$\bar D_w^2 D_w^2$}
\put(0.35,0){$x$} \put(1.45,0){$w$} \put(2.6,0){$y$}
\put(3.7,1.0){$+$}
\put(4.9,0.2){\includegraphics[scale=0.15,clip]{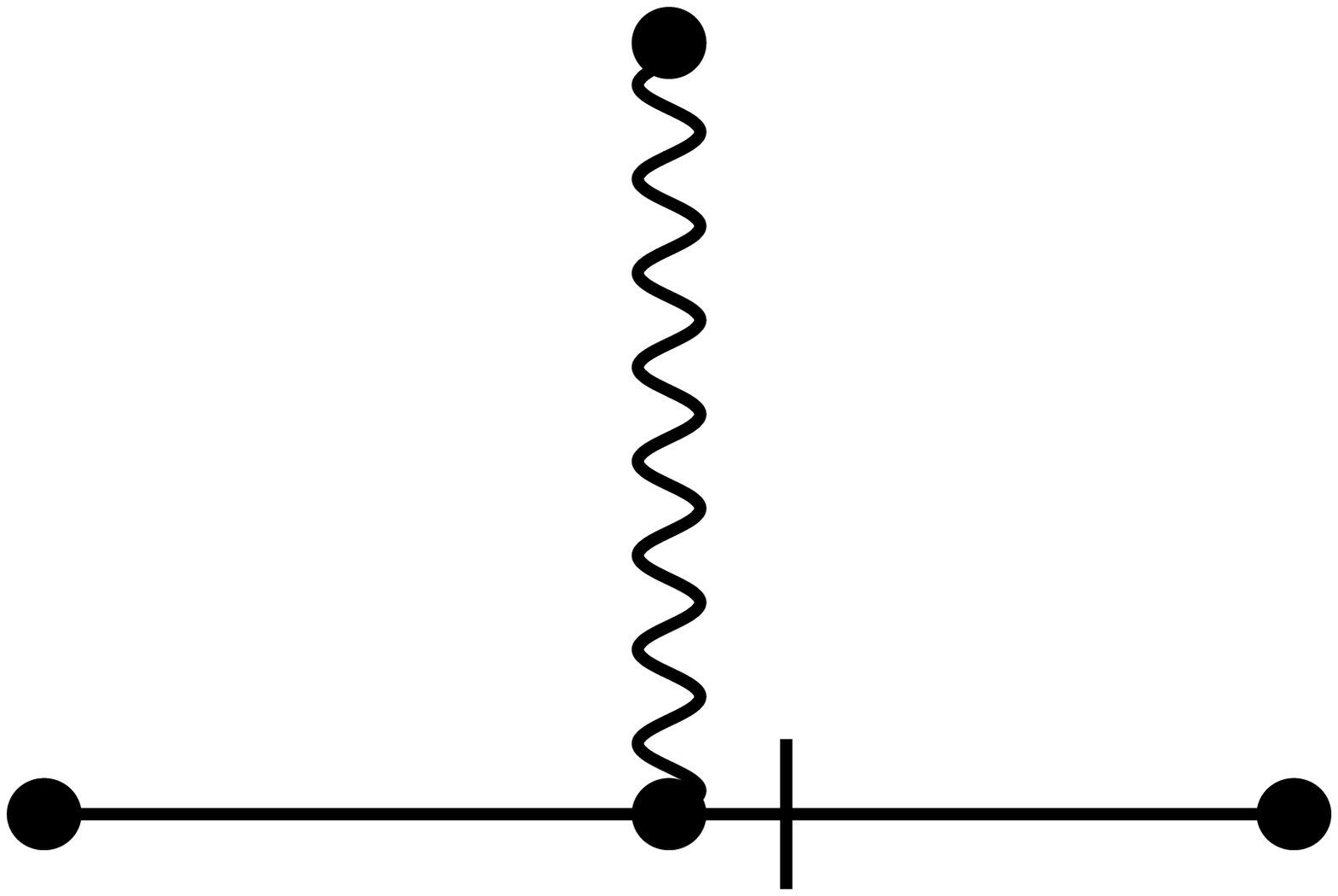}}
\put(6.25,1.8){$z$}
\put(4.7,1.0){$\bar D_w^2 D_w^2$}
\put(4.85,0){$x$} \put(5.95,0){$w$} \put(7.1,0){$y$}
\put(8.1,1.0){\vector(1,0){1}}
\put(9.9,0.9){$0$}
\end{picture}
\caption{This figure illustrates how the gauge dependent terms cancel each other in various supergraphs containing lines of the massive Pauli--Villars superfields.}\label{Figure_Subdiagrams_Massive}
\end{figure}

In this appendix we present identities which can be used for cancelling gauge-dependent terms in supergraphs containing loops of the Pauli--Villars superfields. They generalize Eq. (\ref{Identity2}) presented earlier. The analogs of Eq. (\ref{Identity1}) can be written similarly. In the graphical form these identities are presented in Fig. \ref{Figure_Subdiagrams_Massive}.

The corresponding equations (which can be verified using the standard technique for calculating supergraphs) are written as

\begin{eqnarray}
&& \int d^8w\, \frac{\bar D_w^2 D_w^2}{16\partial^4}\Big(\frac{\xi_0}{K} - \frac{1}{R}\Big)\delta^8_{wz}\cdot \bigg[\frac{\bar D_x^2 D_x^2}{4(\partial^2+M^2)}\delta_{xw}^8\cdot \frac{\bar D_w^2 D_w^2}{4(\partial^2+M^2)}\delta_{wy}^8 \qquad\nonumber\\
&&\qquad\qquad - \frac{M \bar D_x^2}{\partial^2+M^2}\delta_{xw}^8 \cdot \frac{M D_w^2}{\partial^2+M^2}\delta_{wy}^8 \bigg] = - \frac{\bar D_x^2 D_x^2}{16\partial^4}\Big(\frac{\xi_0}{K} - \frac{1}{R}\Big)\delta^8_{xz}\cdot \frac{\bar D_x^2 D_x^2}{\partial^2+M^2}\delta_{xy}^8;\qquad\\
&& \int d^8w\, \frac{\bar D_w^2 D_w^2}{16\partial^4}\Big(\frac{\xi_0}{K} - \frac{1}{R}\Big)\delta^8_{wz}\cdot \bigg[-\frac{M \bar D_x^2}{\partial^2+M^2}\delta_{xw}^8\cdot \frac{D_w^2 \bar D_w^2}{4(\partial^2+M^2)}\delta_{wy}^8 \qquad\nonumber\\
&& \qquad\qquad\qquad\qquad\qquad\qquad\qquad\qquad\qquad\quad\ \ + \frac{\bar D_x^2 D_x^2}{4(\partial^2+M^2)}\delta_{xw}^8\cdot \frac{M \bar D_w^2}{\partial^2+M^2}\delta_{wy}^8 \bigg]\nonumber\\
&& \quad\ \ =  \frac{\bar D_y^2 D_y^2}{4\partial^4} \Big(\frac{\xi_0}{K} - \frac{1}{R}\Big) \delta^8_{yz}\cdot \frac{M \bar D_y^2}{\partial^2 + M^2}\delta^8_{xy} - \frac{\bar D_x^2 D_x^2}{4\partial^4}\Big(\frac{\xi_0}{K} - \frac{1}{R}\Big)\delta^8_{xz} \cdot \frac{M \bar D_x^2}{\partial^2 + M^2} \delta^8_{xy};\qquad\\
&& \int d^8w\, \frac{\bar D_w^2 D_w^2}{16\partial^4}\Big(\frac{\xi_0}{K} - \frac{1}{R}\Big)\delta^8_{wz}\cdot \bigg[
\frac{D_x^2 \bar D_x^2}{4(\partial^2+M^2)}\delta_{xw}^8\cdot \frac{M D_w^2}{\partial^2+M^2}\delta_{wy}^8 \qquad\nonumber\\
&&\qquad\qquad\qquad\qquad\qquad\qquad\qquad\qquad\quad\  - \frac{M D_x^2}{\partial^2+M^2}\delta_{xw}^8\cdot \frac{\bar D_w^2 D_w^2}{4(\partial^2+M^2)}\delta_{wy}^8\bigg] = 0.\qquad
\end{eqnarray}

\noindent
Note that the minus signs appear because the superfields $\Phi$ and $\widetilde\Phi$ have the opposite $U(1)$ charges, see Eq. (\ref{Pauli-Villars_Action}).

\section{Results for the supergraphs presented in Figs. \ref{Figure_Vacuum_Supergraph} and \ref{Figure_3Loop_Supergraphs}}
\hspace*{\parindent}\label{Appendix_Beta_Supergraphs}

In this appendix we present expressions for the supergraphs depicted in Figs. \ref{Figure_Vacuum_Supergraph} and \ref{Figure_3Loop_Supergraphs}, which have been obtained using the method proposed in Ref. \cite{Stepanyantz:2019ihw} and described in Sect. \ref{Section_Three-Loop}. To avoid very large equations, we use the notation

\begin{equation}
\Delta_K \equiv \frac{\xi_0}{K_K}-\frac{1}{R_K}.
\end{equation}

\noindent
By construction, all contributions of the supergraphs under consideration to the function $\beta(\alpha_0)/\alpha_0^2$ are given by integrals of double total derivatives. The results can be written as

\begin{eqnarray}
&&\hspace*{-8mm} \Delta_{\mbox{\scriptsize S1}}\Big(\frac{\beta}{\alpha_0^2}\Big) = 4\pi N_f\, \frac{d}{d\ln\Lambda} \int \frac{d^4Q}{(2\pi)^4} \frac{d^4K}{(2\pi)^4}\, e_0^2\, \frac{\partial^2}{\partial Q^\mu \partial Q_\mu} \bigg\{\frac{1}{K^2 R_K}\bigg(\frac{1}{Q^2(Q+K)^2} - \frac{1}{(Q^2+M^2)}\nonumber\\
&& \times\frac{1}{((Q+K)^2+M^2)}\bigg) +\frac{2\Delta_K}{K^4} \bigg(\frac{1}{Q^2} - \frac{1}{Q^2+M^2}\bigg)\bigg\};\\
&&\hspace*{-8mm} \Delta_{\mbox{\scriptsize S2}}\Big(\frac{\beta}{\alpha_0^2}\Big) = -8\pi N_f\, \frac{d}{d\ln\Lambda} \int \frac{d^4Q}{(2\pi)^4} \frac{d^4K}{(2\pi)^4}\, e_0^2\, \frac{\partial^2}{\partial Q^\mu \partial Q_\mu} \frac{\Delta_K}{K^4} \bigg(\frac{1}{Q^2} - \frac{1}{Q^2+M^2}\bigg);\\
&&\hspace*{-8mm} \Delta_{\mbox{\scriptsize S3}}\Big(\frac{\beta}{\alpha_0^2}\Big) = 4\pi N_f^2\, \frac{d}{d\ln\Lambda} \int \frac{d^4Q}{(2\pi)^4} \frac{d^4K}{(2\pi)^4} \frac{d^4L}{(2\pi)^4}\, e_0^4\, \Big(\frac{\partial^2}{\partial Q^\mu \partial Q_\mu} +\frac{\partial^2}{\partial L^\mu \partial L_\mu}\Big)\bigg\{-\frac{1}{R_K^2 K^2}\nonumber\\
&&\times \bigg(\frac{1}{Q^2(Q+K)^2} - \frac{1}{(Q^2+M^2)((Q+K)^2+M^2)}\bigg) \bigg(\frac{1}{L^2(L+K)^2} - \frac{1}{(L^2+M^2)}\nonumber\\
&& \times \frac{1}{((L+K)^2+M^2)}\bigg) + \frac{4}{R_K^2 K^4} \bigg(\frac{1}{Q^2(Q+K)^2} - \frac{1}{(Q^2+M^2)((Q+K)^2+M^2)}\bigg)\nonumber\\
&& \times \bigg(\frac{1}{L^2} - \frac{1}{L^2+M^2}\bigg) + \frac{4}{K^6}\Big(\frac{\xi_0^2}{K_K^2} - \frac{1}{R_K^2}\Big)\bigg(\frac{1}{Q^2} - \frac{1}{Q^2+M^2}\bigg) \bigg(\frac{1}{L^2} - \frac{1}{L^2+M^2}\bigg)\bigg\};\\
&&\hspace*{-8mm} \Delta_{\mbox{\scriptsize S4}}\Big(\frac{\beta}{\alpha_0^2}\Big) = -16\pi N_f^2\, \frac{d}{d\ln\Lambda} \int \frac{d^4Q}{(2\pi)^4} \frac{d^4K}{(2\pi)^4} \frac{d^4L}{(2\pi)^4}\, e_0^4\, \Big(\frac{\partial^2}{\partial Q^\mu \partial Q_\mu} +\frac{\partial^2}{\partial L^\mu \partial L_\mu}\Big)\bigg(\frac{1}{L^2} - \frac{1}{L^2+M^2}\bigg)\nonumber\\
&&\times \Bigg\{\frac{1}{R_K^2 K^4}
\bigg(\frac{1}{Q^2(Q+K)^2} - \frac{1}{(Q^2+M^2)((Q+K)^2+M^2)}\bigg) + \frac{2}{K^6}\Big(\frac{\xi_0^2}{K_K^2} - \frac{1}{R_K^2}\Big)\bigg(\frac{1}{Q^2}\nonumber\\
&& - \frac{1}{Q^2+M^2}\bigg)\bigg\};\\
&&\hspace*{-8mm} \Delta_{\mbox{\scriptsize S5}}\Big(\frac{\beta}{\alpha_0^2}\Big) = 32\pi N_f^2\, \frac{d}{d\ln\Lambda} \int \frac{d^4Q}{(2\pi)^4} \frac{d^4K}{(2\pi)^4} \frac{d^4L}{(2\pi)^4}\, e_0^4\, \frac{\partial^2}{\partial Q^\mu \partial Q_\mu} \frac{1}{K^6}\bigg( \frac{\xi_0^2}{K_K^2} - \frac{1}{R_K^2} \bigg) \bigg(\frac{1}{Q^2}\nonumber\\
&& - \frac{1}{Q^2+M^2}\bigg) \bigg(\frac{1}{L^2} - \frac{1}{L^2+M^2}\bigg);\\
&&\hspace*{-8mm} \Delta_{\mbox{\scriptsize S6}}\Big(\frac{\beta}{\alpha_0^2}\Big) = -16\pi N_f\, \frac{d}{d\ln\Lambda} \int \frac{d^4Q}{(2\pi)^4} \frac{d^4K}{(2\pi)^4} \frac{d^4L}{(2\pi)^4}\, e_0^4\, \frac{\partial^2}{\partial Q^\mu \partial Q_\mu}\bigg\{\frac{1}{K^2 R_K L^2 R_L}\bigg(\frac{1}{Q^2 (Q+K)^2}\nonumber\\
&& \times \frac{1}{(Q+K+L)^2} - \frac{1}{(Q^2+M^2)((Q+K)^2+M^2)((Q+K+L)^2+M^2)}\bigg)
+\frac{2\Delta_L}{L^4 K^2 R_K}\nonumber\\
&&\times\bigg(\frac{1}{Q^2 (Q+K+L)^2} + \frac{1}{Q^2 (Q+K)^2} - \frac{1}{(Q^2+M^2)((Q+K+L)^2+M^2)} - \frac{1}{(Q^2+M^2)}
\nonumber\\
&&\times \frac{1}{((Q+K)^2+M^2)} \bigg) + \frac{\Delta_K \Delta_L}{K^4 L^4} \bigg(\frac{3}{Q^2} + \frac{(Q+K)^2}{Q^2 (Q+K+L)^2} - \frac{3}{Q^2+M^2} - \frac{1}{(Q^2+M^2)}\nonumber\\
&& \times \frac{(Q+K)^2-3M^2}{((Q+K+L)^2+M^2)}\bigg)\bigg\};\\
&&\hspace*{-8mm} \Delta_{\mbox{\scriptsize S7}}\Big(\frac{\beta}{\alpha_0^2}\Big) = 4 \pi N_f\, \frac{d}{d\ln\Lambda} \int \frac{d^4Q}{(2\pi)^4} \frac{d^4K}{(2\pi)^4} \frac{d^4L}{(2\pi)^4}\, e_0^4\, \frac{\partial^2}{\partial Q^\mu \partial Q_\mu}\bigg\{\frac{1}{K^2 R_K L^2 R_L}\bigg(\frac{(2Q+K+L)^2}{Q^2 (Q+K)^2}\nonumber\\
&& \times \frac{1}{(Q+K+L)^2(Q+L)^2} - \frac{(2Q+K+L)^2+2M^2}{(Q^2+M^2)((Q+K)^2+M^2)((Q+K+L)^2+M^2)}\nonumber\\
&& \times \frac{1}{((Q+L)^2+M^2)} \bigg) +\frac{4\Delta_L}{L^4 K^2 R_K} \bigg( \frac{1}{Q^2 (Q+K)^2} - \frac{1}{(Q^2+M^2)((Q+K)^2+M^2)} \bigg) \nonumber\\
&& + \frac{4 \Delta_K \Delta_L}{K^4 L^4} \bigg(\frac{1}{Q^2} - \frac{1}{Q^2+M^2} + \frac{M^2}{(Q^2+M^2)((Q+K+L)^2+M^2)} \bigg)\bigg\};\\
&&\hspace*{-8mm} \Delta_{\mbox{\scriptsize S8}}\Big(\frac{\beta}{\alpha_0^2}\Big) = 8\pi N_f\, \frac{d}{d\ln\Lambda} \int \frac{d^4Q}{(2\pi)^4} \frac{d^4K}{(2\pi)^4} \frac{d^4L}{(2\pi)^4}\, e_0^4\, \frac{\partial^2}{\partial Q^\mu \partial Q_\mu}\bigg\{\frac{1}{K^2 R_K L^2 R_L}\bigg(\frac{1}{Q^2 (Q+K)^2}\nonumber\\
&& \times \frac{1}{(Q+L)^2} - \frac{Q^2-M^2}{(Q^2+M^2)^2((Q+K)^2+M^2)((Q+L)^2+M^2)}\bigg)
+\frac{2\Delta_L}{L^4 K^2 R_K}\nonumber\\
&&\times\bigg(\frac{1}{(Q+K)^2 (Q+L)^2} + \frac{1}{Q^2 (Q+K)^2} - \frac{1}{((Q+K)^2+M^2)((Q+L)^2+M^2)}
\nonumber\\
&&- \frac{Q^2-M^2}{(Q^2+M^2)^2 ((Q+K)^2+M^2)} \bigg) + \frac{\Delta_K \Delta_L}{K^4 L^4} \bigg(\frac{3}{Q^2} + \frac{Q^2}{(Q+K)^2 (Q+L)^2} - \frac{3}{Q^2+M^2} \nonumber\\
&& + \frac{2M^2}{(Q^2+M^2)^2} - \frac{Q^2-M^2}{((Q+K)^2+M^2)((Q+L)^2+M^2)}\bigg)\bigg\};\\
&&\hspace*{-8mm} \Delta_{\mbox{\scriptsize S9}}\Big(\frac{\beta}{\alpha_0^2}\Big) = 8\pi N_f\, \frac{d}{d\ln\Lambda} \int \frac{d^4Q}{(2\pi)^4} \frac{d^4K}{(2\pi)^4} \frac{d^4L}{(2\pi)^4}\, e_0^4\, \frac{\partial^2}{\partial Q^\mu \partial Q_\mu} \frac{\Delta_K \Delta_L}{K^4 L^4} \bigg(\frac{1}{Q^2} - \frac{Q^2-M^2}{(Q^2+M^2)^2}\bigg);\\
&&\hspace*{-8mm}  \Delta_{\mbox{\scriptsize S10}}\Big(\frac{\beta}{\alpha_0^2}\Big) = 16\pi N_f\, \frac{d}{d\ln\Lambda} \int \frac{d^4Q}{(2\pi)^4} \frac{d^4K}{(2\pi)^4} \frac{d^4L}{(2\pi)^4}\, e_0^4\, \frac{\partial^2}{\partial Q^\mu \partial Q_\mu} \frac{\Delta_L}{L^4} \bigg\{\frac{1}{R_K K^2}\bigg(\frac{1}{Q^2 (Q+K+L)^2}\nonumber\\
&& - \frac{1}{(Q^2+M^2)((Q+K+L)^2+M^2)}\bigg) + \frac{\Delta_K}{2K^4} \bigg(\frac{1}{Q^2} + \frac{(Q+L)^2}{Q^2 (Q+K+L)^2} - \frac{1}{Q^2 + M^2}\nonumber\\
&& -\frac{(Q+L)^2-3M^2}{(Q^2+M^2)((Q+K+L)^2+M^2)}\bigg)\bigg\};\\
&&\hspace*{-8mm}  \Delta_{\mbox{\scriptsize S11}}\Big(\frac{\beta}{\alpha_0^2}\Big) = 16\pi N_f\, \frac{d}{d\ln\Lambda} \int \frac{d^4Q}{(2\pi)^4} \frac{d^4K}{(2\pi)^4} \frac{d^4L}{(2\pi)^4}\, e_0^4\, \frac{\partial^2}{\partial Q^\mu \partial Q_\mu} \frac{\Delta_L}{L^4} \bigg\{\frac{1}{R_K K^2}\bigg(\frac{1}{Q^2 (Q+K)^2}\nonumber\\
&& - \frac{1}{(Q^2+M^2)((Q+K)^2+M^2)}\bigg) + \frac{2\Delta_K}{K^4} \bigg(\frac{1}{Q^2} - \frac{1}{Q^2 + M^2} \bigg)\bigg\};\\
&&\hspace*{-8mm} \Delta_{\mbox{\scriptsize S12}}\Big(\frac{\beta}{\alpha_0^2}\Big) = - 8\pi N_f\, \frac{d}{d\ln\Lambda} \int \frac{d^4Q}{(2\pi)^4} \frac{d^4K}{(2\pi)^4} \frac{d^4L}{(2\pi)^4}\, e_0^4\, \frac{\partial^2}{\partial Q^\mu \partial Q_\mu} \frac{\Delta_K \Delta_L}{K^4 L^4} \bigg(\frac{1}{Q^2} - \frac{1}{Q^2+M^2}\bigg);\\
&&\hspace*{-8mm} \Delta_{\mbox{\scriptsize S13}}\Big(\frac{\beta}{\alpha_0^2}\Big) = -16\pi N_f\, \frac{d}{d\ln\Lambda} \int \frac{d^4Q}{(2\pi)^4} \frac{d^4K}{(2\pi)^4} \frac{d^4L}{(2\pi)^4}\, e_0^4\, \frac{\partial^2}{\partial Q^\mu \partial Q_\mu} \frac{\Delta_L}{L^4} \bigg\{\frac{1}{R_K K^2}\bigg(\frac{1}{Q^2(Q+K)^2}\nonumber\\
&& - \frac{Q^2-M^2}{(Q^2+M^2)^2((Q+K)^2+M^2)}\bigg) + \frac{2\Delta_K}{K^4} \bigg(\frac{1}{Q^2} - \frac{Q^2}{(Q^2+M^2)^2}\bigg)\bigg\}.
\end{eqnarray}

\noindent
The sum of these expressions appears to be gauge-independent and is given by Eq. (\ref{3Loop_Beta}).

\section{Various contributions to the two-loop anomalous dimension of the matter superfields}
\hspace*{\parindent}\label{Appendix_Delta_G}

Here we list the contributions of the superdiagrams presented in Fig. \ref{Figure_Two-Point_Matter} to the function $G$ calculated at the vanishing external momentum $q$:

\begin{eqnarray}
&&\hspace*{-5mm} \Delta_{\mbox{\scriptsize A1}} G\Big|_{q=0} = - 2 e_0^2 \int \frac{d^4K}{(2\pi)^4} \frac{\xi_0}{K^4 K_K};\\
&&\hspace*{-5mm} \Delta_{\mbox{\scriptsize A2}} G\Big|_{q=0} = 2 e_0^2 \int \frac{d^4K}{(2\pi)^4} \frac{\Delta_K}{K^4};\\
&&\hspace*{-5mm} \Delta_{\mbox{\scriptsize A3}} G\Big|_{q=0} = - 8 N_f e_0^4 \int \frac{d^4K}{(2\pi)^4} \frac{d^4L}{(2\pi)^4} \frac{\xi_0^2}{K^6 K_K^2}\bigg(\frac{1}{L^2} - \frac{1}{L^2+M^2}\bigg);\\
&&\hspace*{-5mm} \Delta_{\mbox{\scriptsize A4}} G\Big|_{q=0} = 8 N_f e_0^4 \int \frac{d^4K}{(2\pi)^4} \frac{d^4L}{(2\pi)^4} \frac{\xi_0^2}{K^6 K_K^2}\bigg(\frac{1}{L^2} - \frac{1}{L^2+M^2}\bigg);\\
&&\hspace*{-5mm} \Delta_{\mbox{\scriptsize A5}} G\Big|_{q=0} = 4 N_f e_0^4 \int \frac{d^4K}{(2\pi)^4} \frac{d^4L}{(2\pi)^4}\bigg\{\frac{1}{R_K^2 K^4}\bigg(\frac{1}{L^2 (K+L)^2} - \frac{1}{(L^2+M^2)((K+L)^2+M^2)}\bigg)\nonumber\\
&&\hspace*{-5mm} \qquad + 2\Big(\frac{\xi_0^2}{K_K^2}-\frac{1}{R_K^2}\Big)\bigg(\frac{1}{L^2}-\frac{1}{L^2+M^2}\bigg)\bigg\};\\
&&\hspace*{-5mm} \Delta_{\mbox{\scriptsize A6}} G\Big|_{q=0} = - 8 N_f e_0^4 \int \frac{d^4K}{(2\pi)^4} \frac{d^4L}{(2\pi)^4} \frac{1}{K^6} \bigg(\frac{\xi_0^2}{K_K^2}-\frac{1}{R_K^2}\bigg)\bigg(\frac{1}{L^2} - \frac{1}{L^2+M^2}\bigg);\\
&&\hspace*{-5mm} \Delta_{\mbox{\scriptsize A7}} G\Big|_{q=0} = 4 e_0^4 \int \frac{d^4K}{(2\pi)^4} \frac{d^4L}{(2\pi)^4} \frac{\xi_0^2}{K^4 K_K L^4 K_L};\\
&&\hspace*{-5mm} \Delta_{\mbox{\scriptsize A8}} G\Big|_{q=0} = 4 e_0^4 \int \frac{d^4K}{(2\pi)^4} \frac{d^4L}{(2\pi)^4} \frac{\xi_0}{K^4 K_K}\bigg\{\frac{1}{L^2 R_L (K+L)^2} + \frac{\Delta_L}{L^4}\Big(1+\frac{K^2}{(K+L)^2}\Big)\bigg\};\\
&&\hspace*{-5mm} \Delta_{\mbox{\scriptsize A9}} G\Big|_{q=0} = 4 e_0^4 \int \frac{d^4K}{(2\pi)^4} \frac{d^4L}{(2\pi)^4} \frac{\xi_0}{K^4 K_K}\bigg\{\frac{1}{L^2 R_L (K+L)^2} + \frac{\Delta_L}{L^4}\Big(1+\frac{K^2}{(K+L)^2}\Big)\bigg\};\\
&&\hspace*{-5mm} \Delta_{\mbox{\scriptsize A10}} G\Big|_{q=0} = - 4 e_0^4 \int \frac{d^4K}{(2\pi)^4} \frac{d^4L}{(2\pi)^4} \frac{\xi_0^2}{K^4 K_K L^4 K_L};\\
&&\hspace*{-5mm} \Delta_{\mbox{\scriptsize A11}} G\Big|_{q=0} = - 4 e_0^4 \int \frac{d^4K}{(2\pi)^4} \frac{d^4L}{(2\pi)^4} \frac{\xi_0}{K^4 K_K}\bigg\{\frac{1}{L^2 R_L (K+L)^2} + \frac{\Delta_L}{L^4}\Big(1+\frac{K^2}{(K+L)^2}\Big)\bigg\};\\
&&\hspace*{-5mm} \Delta_{\mbox{\scriptsize A12}} G\Big|_{q=0} = - 4 e_0^4 \int \frac{d^4K}{(2\pi)^4} \frac{d^4L}{(2\pi)^4} \frac{\Delta_L}{L^4}\bigg\{\frac{2}{K^2 R_K (K+L)^2} + \frac{\Delta_K}{2K^4} \Big(1+\frac{2K^2}{(K+L)^2}\Big) \bigg\};\qquad\\
&&\hspace*{-5mm} \Delta_{\mbox{\scriptsize A13}} G\Big|_{q=0} = -4 e_0^4 \int \frac{d^4K}{(2\pi)^4} \frac{d^4L}{(2\pi)^4} \frac{\xi_0 \Delta_L}{K^4 K_K L^4};\\
&&\hspace*{-5mm} \Delta_{\mbox{\scriptsize A14}} G\Big|_{q=0} =  -4 e_0^4 \int \frac{d^4K}{(2\pi)^4} \frac{d^4L}{(2\pi)^4} \frac{\xi_0 \Delta_L}{K^4 K_K L^4};\\
&&\hspace*{-5mm} \Delta_{\mbox{\scriptsize A15}} G\Big|_{q=0} = 2 e_0^4 \int \frac{d^4K}{(2\pi)^4} \frac{d^4L}{(2\pi)^4} \frac{\Delta_K \Delta_L}{K^4 L^4};\\
&&\hspace*{-5mm} \Delta_{\mbox{\scriptsize A16}} G\Big|_{q=0} = 4 e_0^4 \int \frac{d^4K}{(2\pi)^4} \frac{d^4L}{(2\pi)^4} \frac{\xi_0 \Delta_L}{K^4 K_K L^4}.
\end{eqnarray}

\noindent
Note that although these expressions are not well-defined, the total two-loop contribution to the anomalous dimension (defined in terms of the bare couplings)

\begin{eqnarray}\label{Two-Loop_Anomalous_Dimension_Integral}
&& \gamma(\alpha_0) = \frac{d\ln G}{d\ln\Lambda}\bigg|_{q=0} = - \frac{d}{d\ln\Lambda} \int \frac{d^4K}{(2\pi)^4} \frac{2e_0^2}{K^4R_K^2}\bigg\{R_K - 2e_0^2 N_f \int \frac{d^4L}{(2\pi)^4}\bigg(\frac{1}{L^2 (L+K)^2}\qquad\nonumber\\
&& - \frac{1}{(L^2+M^2)((L+K)^2+M^2)}\bigg)\bigg\} + \frac{d}{d\ln\Lambda} \int \frac{d^4K}{(2\pi)^4} \frac{d^4L}{(2\pi)^4}\frac{e_0^4}{R_K R_L}\bigg(\frac{4}{K^2 L^4 (K+L)^2}\nonumber\\
&& - \frac{2}{K^4 L^4}\bigg) + O(e_0^6)
\end{eqnarray}

\noindent
is well-defined, see Ref. \cite{Soloshenko:2003sx} for details. Note that the derivative with respect to $\ln\Lambda$ should be taken at a fixed value of the {\it renormalized} coupling constant before the integrations. We see that all gauge dependent terms cancel each other, and the expression (\ref{Two-Loop_Anomalous_Dimension_Integral}) is gauge independent. This implies that it coincides with the one in the Feynman gauge, which has been calculated in \cite{Soloshenko:2003sx} and is given by Eq. (\ref{Two-Loop_Anomalous_Dimension_Result}) for $R(x)=1+x^n$.

\end{document}